\documentclass[journal]{IEEEtran}
\usepackage{graphicx}
\usepackage{epstopdf}
\usepackage{fancyhdr}
\usepackage{amsmath,amssymb,amstext}
\usepackage{subeqnarray}
\usepackage{cite}
\usepackage{hhline}
\usepackage{bm}
\usepackage{soul}
\usepackage{slashbox}
\usepackage{times,amsmath,amsfonts}
\usepackage{subfigure}
\usepackage{amsmath,amssymb}
\usepackage{graphicx,picture}
\usepackage{cite}
\usepackage{subfigure}
\usepackage{subeqnarray}
\usepackage{cases}
\usepackage{color}
\usepackage{overpic}
\usepackage{multirow}
\usepackage{booktabs}
\usepackage{hyperref}

\pdfoutput=1

\newcounter{saveeqn}%

\allowdisplaybreaks
\DeclareMathSymbol{\Phi}{\mathord}{letters}{8}

\begin{document}

\title{An Overview on Integrated Localization and Communication Towards 6G}

\author{
\IEEEauthorblockN{Zhiqiang Xiao and Yong Zeng,~\IEEEmembership{Member,~IEEE}}

\thanks{
Z. Xiao and Y. Zeng are with the National Mobile Communications Research Laboratory, Southeast University, Nanjing 210096, China. Y. Zeng is also with the Purple Mountain Laboratories, Nanjing 211111, China (e-mail: zhiqiang\_xiao94@163.com, yong\_zeng@seu.edu.cn).
}
}
\maketitle

\begin{abstract}
While the fifth generation (5G) cellular system is being deployed worldwide, researchers have started the investigation of the sixth generation (6G) mobile communication networks.
Although the essential requirements and key usage scenarios of 6G are yet to be defined, it is believed that 6G should be able to provide intelligent and ubiquitous wireless connectivity with Terabits per second (Tbps) data rate and sub-millisecond (sub-ms) latency over three-dimensional (3D) network coverage.
To achieve such goals, acquiring accurate location information of the mobile terminals is becoming extremely useful, not only for location-based services but also for improving wireless communication performance in various ways such as channel estimation, beam alignment, medium access control, routing, and network optimization.
On the other hand, the advancement of communication technologies also brings new opportunities to greatly improve the localization performance, as exemplified by the anticipated centimeter-level localization accuracy in 6G by ultra massive MIMO (multiple-input multiple-output) and millimeter wave (mmWave) technologies.
In this regard, a unified study on integrated localization and communication (ILAC) is necessary to unlock the full potential of wireless networks for the best utilization of network infrastructure and radio resources for dual purposes.
While there are extensive literatures on wireless localization or communications separately, the research on ILAC is still in its infancy.
Therefore, this article aims to give a tutorial overview on ILAC towards 6G wireless networks.
After a holistic survey on wireless localization basics, we present the state-of-the-art results on how wireless localization and communication inter-play with each other in various network layers, together with the main architectures and techniques for localization and communication co-design in current two-dimensional (2D) and future 3D networks with aerial-ground integration.
Finally, we outline some promising future research directions for ILAC.
\end{abstract}


\section{Introduction}
Starting from the second generation (2G), wireless localization has been included as a compulsory feature in the standardization and implementation of cellular networks, with continuous enhancement on the localization accuracy over each generation, e.g., from hundreds of meters accuracy in 2G to tens of meters in the fourth generation (4G).
For the forthcoming fifth generation (5G) mobile networks, localization is regarded as one of the key components, due to its fundamental support for various location-based services, and the requirement on localization accuracy is up to sub-meter level \cite{TS22.071}.
The availability of accurate real-time location information of mobile terminals is expected to play an increasingly important role in future wireless networks.
While the deployment of 5G networks is ongoing, researchers around the world have already started the investigation on the sixth generation (6G) mobile communication targeting for network 2030, with various visions proposed \cite{zong20196g,saad2019vision,strinati20196g,latva2019key,zhang20196gVisions,FG-NET-2030}.
For example, it was envisioned that 6G should achieve ``ubiquitous wireless intelligence'' \cite{latva2019key}, for providing users smart context-aware services through wireless connectivity anywhere in the world.
This renders that acquiring the accurate real-time location information of users becomes more critical than ever before, with potentially centimeter-level localization accuracy for 6G.

However, most current localization services provided by global navigation satellites systems (GNSS), wireless local area networks (WLAN) or cellular networks can at best achieve meter-level localization accuracy in clutter environments.
Such coarse localization services are difficult to meet the centimeter-level localization accuracy requirements of many emerging applications.
For example, the following three promising usage scenarios of 5G-and-beyond networks, namely, \emph{intelligent interactive networks}, \emph{smart city}, and \emph{automatic factory}, all highlight the critical role of accurate localization in future network design.
\paragraph{Intelligent Interactive Networks}
It is believed that the ultimate goal of communication networks is to promote the intelligent interactions across the world, in terms of people-to-people, people-to-machine, and machine-to-machine.
An unprecedented proliferation of new internet-of-things (IoT) services, like multisensory extended reality (XR) encompassing augmented/mixed/virtual reality (AR/MR/VR) \cite{chen2018virtual}, brain-computer interfaces (BCI) \cite{saad2019vision}, as well as tele-presence and tele-control services \cite{latva2019key}, brings excellent opportunities to realize the goal of interaction with everything.
To implement such new applications, it is necessary to achieve the high localization performance, as elaborated in the following.
\begin{itemize}
\item{\textbf{Multisensory XR}:}
XR services will enable users to experience and interact with virtual and immersive environments through first-person view \cite{chen2018virtual}.
To enable truly immersive XR applications, it must deploy XR systems through wireless networks, and thus the tracking accuracy of XR devices is of paramount importance.
For wireless XR applications, a control center collects the tracking information of the XR devices, and sends data to those devices through wireless links.
Therefore, the accuracy of device tracking and the delay of signal measurements will significantly affect the XR information transmission and hence impact the user experience.
For instance, an inaccurate head-tracking may cause cybersickness, like nausea, disorientation, headaches, and eye strain \cite{chang2016performance}.
In general, for XR services, depending on the usage scenarios, the requirement for localization accuracy ranges from 1 centimeter (cm) to 10 cm, and the time delay should be typically less than 20 milliseconds (ms) \cite{chen2018virtual}.
\item{\textbf{Wireless BCI (WBCI)}:}
The forthcoming 5G and future 6G networks bring new opportunities to tailor communication networks into the versatile networks integrated with human-centric communication, wireless sensing, and remote control \cite{saad2019vision}, where people will be enabled to interact with their surrounding environment using various IoT devices connected through the WBCI technology.
It opens the door for people to control their neighboring IoT devices through their brain implants, gestures, empathic as well as haptic messages \cite{chen2018virtual}.
Such a breathtaking technology requires the communication services of extremely high data rate, ultra-low latency, and high reliability, as well as the localization support of high accuracy, e.g., centimeter-level accuracy.
In addition, the cooperative localization among IoT devices is also quite important for WBCI.
\item{\textbf{Tele-presentation and Tele-control}:}
With the advancement of various supporting technologies including high-resolution imaging and sensing, wearable displays, mobile robots and drones, it is expected that the technologies of \emph{tele-presentation} and \emph{tele-control} will become reality in the near future \cite{latva2019key}.
For tele-presentation, a remote environment can be represented through real-time environment capturing, information transmission, and three-dimensional (3D) holographic rendering, which makes the accurate location information critical for 3D mapping.
Furthermore, people may operate the remote IoT devices through wireless networks, just like manipulating them face-to-face, which is referred to as \emph{tele-control} or \emph{tele-operation}.
An exemplary application of tele-presentation and tele-control is the \emph{tele-surgery}, which will enable doctors to perform emergent surgery remotely.
Note that in such use cases, highly-accurate, ultra-reliable, and low-latency localization is vital.
For tele-control, the remote and neighboring localization systems usually have two separate coordinates, perform different localization algorithms, and use different reference nodes, which may cause mismatch errors.
Therefore, for such applications, the real-time infrastructure calibration between the two localization systems is of paramount importance for reducing the mismatch errors.
\end{itemize}

\paragraph{Smart City}
A smart city has the ability to efficiently analyze different requirements from the society, and reasonably manage and optimize public resources, such as electricity, water, transportation, and healthcare, to provide better public services\cite{tariq2019speculative}.
A truly smart city entails many different aspects.
Here, we elaborate two major application scenarios to outline the importance of the accurate localization for smart city:
\begin{itemize}
\item{\textbf{Smart Indoor Services}:}
Over the last decade, IoT technology has developed rapidly, which will flourish the smart indoor services, like smart homes, indoor navigation in shopping malls, and crowd monitoring.
Different from outdoor scenarios, one critical issue of indoor localization is the severe non-light-of-sight (NLoS) signal propagation that may significantly degrade the localization accuracy \cite{witrisal2016high,witrisal2016high1}.
Meanwhile, the privacy protection of location information is another critical issue for public indoor localization services \cite{latva2019key}.
One of the key problems is to identify what kind of location information needs to be protected.
For example, for some public devices, their location information should be accessible to all user devices, while that for user personal devices or some kernel public devices needs to be protected.
\item{\textbf{Smart Transportation}:}
The research on smart transportation is still ongoing, with several standards proposed, like dedicated short-range communications (DSRC) \cite{kenney2011dedicated} and vehicle-to-everything (V2X) \cite{chen2017vehicle}.
The autonomous driving \cite{mao2018deep} and vehicle-to-vehicle (V2V) communications \cite{asadi2014survey} are envisioned as two attractive developing trends of smart transportation, both of which call for advanced localization technologies.
For autonomous driving, the 3D mapping for the real-time scenarios is critical, which requires the accurate relative distances between the vehicle and obstacles to construct the environment model.
The V2V communications also need accurate localization to improve the communication performance.
Compared with other use cases, for smart transportation, the localization systems should be designed not only for high accuracy, but also for wide coverage, as well as for robustness in highly mobile scenarios.
\end{itemize}
\paragraph{Automatic Factory}
The development of connected robotics and autonomous systems (CRAS) like autonomous robotics, drone-delivery systems, etc., promotes the progress of automatic factory \cite{saad2019vision}, such as smart storage, autonomous production, and autonomous delivery.
The accurate localization of various IoT devices is a prerequisite to enable effective cooperation among them.
Different from other applications, for automatic factory, cooperative localization among massive IoT devices is of critical importance, which will require the highly-accurate, low-latency, and highly-reliable location information of the massive devices to build the end-to-end (E2E) communication links.

Table I summarizes the main localization requirements of different applications for 5G/6G networks.
It is observed that achieving highly-accurate, low-latency, and highly-reliable localization will play an important role in future wireless networks.
Furthermore, one promising development trend of future networks is to integrate communication, computing, control, localization, and sensing (3CLS) \cite{saad2019vision}, and further build a self-sustaining networks (SSN) that can maintain the key performance indicators (KPIs) by appropriately managing the radio resources according to the real-time locations of mobile terminals.
\begin{table*}[]
\centering
  \caption{The localization requirements of different applications for 5G/6G networks}\label{The positioning requirements for different applications}
\begin{tabular}{|l|l|l|}
\hline
                  &  Applications& Requirements  \\ \hline
\multirow{3}{*}{Intelligent Interactive Network} &  Multisensory XR&
                  \begin{tabular}[c]{@{}l@{}}
                  Centimeter-level accuracy (i.e. 1-10 cm); \\
                  Low latency (less than 20 ms).
                  \end{tabular}  \\ \cline{2-3}
                  &  WBCI&
                  \begin{tabular}[c]{@{}l@{}}
                  Centimeter-level accuracy; \\
                  Low latency (millisecond-level);\\
                  High requirements on cooperative localization among IoT devices.
                  \end{tabular}\\ \cline{2-3}
                  &  Tele-presentation and Tele-control&
                  \begin{tabular}[c]{@{}l@{}}
                  Centimeter-level accuracy; \\
                  High reliability; \\
                  Calibration between two different localization systems.
                  \end{tabular}  \\ \hline
\multirow{2}{*}{Smart City} &  Smart Indoor Services&
                  \begin{tabular}[c]{@{}l@{}}
                  NLoS-based localization;\\
                  Privacy information classification and protection.\\
                  \end{tabular}  \\ \cline{2-3}
                  &  Smart Transportation&
                  \begin{tabular}[c]{@{}l@{}}
                  Submeter-level localization;\\
                  Wide coverage;\\
                  High mobility tracking;\\
                  Cooperative localization in V2X communication cases.
                  \end{tabular}  \\ \hline
                  Automatic Factory&  CRAS&
                  \begin{tabular}[c]{@{}l@{}}
                  Cooperative localization among massive IoT devices;\\
                  At least submeter-level accuracy;\\
                  High reliability;\\
                  Low latency.
                  \end{tabular}  \\ \hline
\end{tabular}
\end{table*}

To achieve the above goals, the underlying technologies of 5G-and-beyond mobile communication, like millimeter wave (mmWave), massive MIMO (multiple-input multiple-output), and ultra dense networks (UDNs), can be utilized for improving the localization performance.
The mmWave signal with bandwidth up to 2 GHz and center frequencies around 30 GHz and above can provide much higher temporal resolutions for improving time-based localization \cite{Liu2017Prospective}.
Furthermore, the ultra massive antenna arrays consisting of thousands of antenna elements can bring the angular resolution less than 1 degree for angular-based localization \cite{You2020The}.
Besides, the UDNs can increase the likelihood of light-of-sight (LoS) links, which can also be exploited for improving localization performance \cite{Liu2017Prospective}.
On the other hand, the location information of mobile terminals can be used for assisting communications, like location-aided channel estimation, beam alignment, routing, and network optimization.
Furthermore, a unified design on signal waveforms, coding, modulation, and radio resource allocation for the seamless integration of localization and communication can be pursued, and the promising new paradigm is referred to as \emph{integrated localization and communication} (ILAC).

While there are extensive literatures focusing on wireless localizations \cite{HuiSurvey,Yassin2017Recent,zafari2019survey,DardariIndoor,ChristosA,gustafsson2005mobile,sun2005signal,del2017survey}, or communications alone, to our best knowledge, a tutorial overview on ILAC to fully utilize the network infrastructure and radio resources for dual purposes is still missing.
In \cite{HuiSurvey,Yassin2017Recent,zafari2019survey}, the authors provide surveys on indoor localization.
In \cite{HuiSurvey}, the authors focused on the principles of different localization approaches and provided an overview on various localization infrastructures.
In \cite{Yassin2017Recent}, the advanced techniques, such as data fusion, cooperative localization, and game theory, were highlighted to improve the localization performance.
In \cite{zafari2019survey}, the authors provided an up-to-date overview on indoor localization with emphasis on IoT scenarios.
In \cite{DardariIndoor} and \cite{ChristosA}, the surveys of enabling technologies for network localization, tracking and navigation were given.
In \cite{DardariIndoor}, the authors mainly focused on the mathematical theories of different indoor tracking and mapping methods.
In \cite{ChristosA}, a comprehensive review on localization techniques in cellular network, WLAN, and wireless sensor networks (WSNs) was provided.
In \cite{gustafsson2005mobile} and \cite{sun2005signal}, the authors summarized the localization techniques from signal processing perspective.
In \cite{del2017survey}, a survey of cellular-based localization was given, where the evolution of the conventional cellular localization methods were given, together with the envision on the 5G new radio (NR) localization.

In this article, we aim to provide a tutorial overview on ILAC towards 6G.
To this end, we first give a holistic introduction on wireless localization basics, in terms of the main definitions and classifications of localization systems, different localization approaches, fundamental performance analysis and metrics, the major localization infrastructures, and some advanced localization related techniques.
Then we focus on the ILAC targeting for the future wireless network design.
To this end, we first provide an overview on the recent recommendations of the third generation partnership project (3GPP) for 5G localization, and then discuss the enabling technologies of 5G networks towards centimeter-level localization accuracy.
After that, we present the state-of-the-art location-aided communication to expose how wireless localization and communication inter-play with each other in different network layers, and give some initial studies on localization and communication co-design with best utilization of the network infrastructures and radio resources to unlock the full potential of the wireless networks.
Furthermore, a discussion on ILAC for aerial-ground integrated networks will be given, which aims to facilitate the ubiquitous wireless coverage, moving from the conventional two-dimensional (2D) plane to the 3D space.
Finally, we give an architecture of future wireless networks, and discuss the enabling technologies and challenges, attempting to outline some promising future research directions for ILAC.

\section{Wireless Localization Basics}

\subsection{Definition and Classification of Wireless Localization}
\begin{figure}
\centering
\begin{overpic}[width=0.35\textwidth]{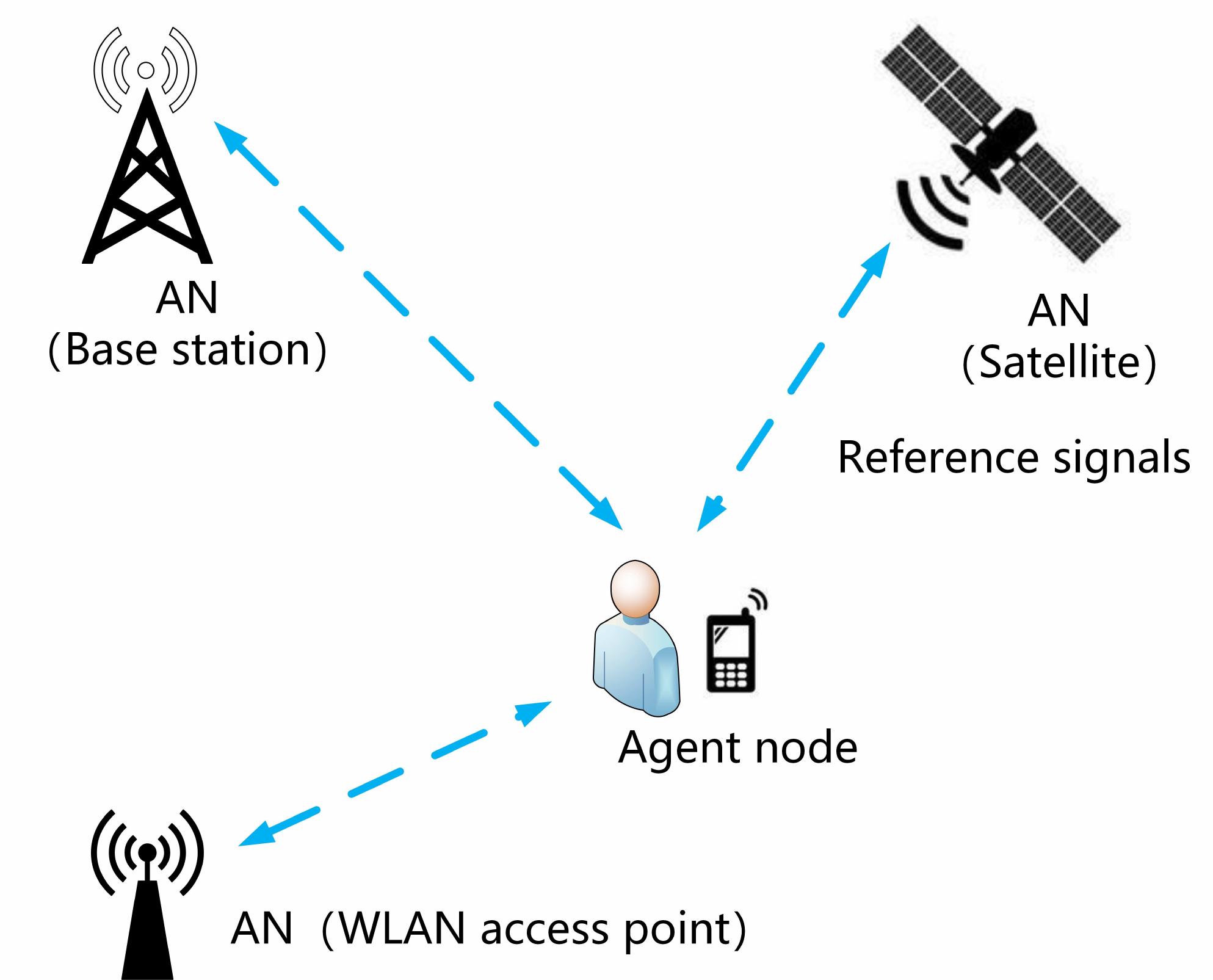}
\end{overpic}
  \caption{A general architecture of wireless localization, with heterogeneous wireless infrastructures including cellular networks, navigation satellites, and WLAN.}\label{Fig general architecture}\vspace{-10pt}
\end{figure}

As illustrated in Fig. \ref{Fig general architecture}, a wireless localization system aims to estimate the location of the targeting object based on a set of wireless reference signals propagated between the reference nodes and the targeting object.
The localization functions can be deployed based on either the existing wireless communication systems, such as cellular networks and WLAN, or dedicated infrastructures, like GNSS.
The targeting object with its unknown location to be localized is often referred to as \emph{agent node} or \emph{mobile user}, and the reference nodes with known locations are often called as \emph{anchor nodes} (ANs) or \emph{landmarks} \cite{DardariIndoor}.
For convenience, throughout this paper, we use the terms \emph{agent node} and ANs to represent the targeting object and reference nodes, respectively, as shown in Fig. \ref{Fig general architecture}.
Compared to a closely related terminology, \emph{wireless positioning}, which estimates the position of the agent node relative to the ANs, wireless localization further locates the estimated position on a coordinate of a map based the locations of the ANs.
Nonetheless, the terminologies \emph{positioning} and \emph{localization} are often used interchangeably \cite{del2017survey}.

Typically, a wireless localization system consists of two major components: a set of ANs and a location estimation unit that can be deployed either on the agent node itself or on a remote site.
The procedures of localization usually include two stages.
In the first stage, specific reference signals are transmitted either by the ANs or the agent node, which are measured by the other end of the link to obtain some location-related information, such as received signal strength (RSS), time of arrival (TOA), time difference of arrival (TDOA), or angle of arrival (AOA) of the received signal.
In the second stage, such measurements are collected at the location estimation unit to estimate the location of the agent node.
Localization systems can be categorized from various perspectives, like based on location estimation algorithms \cite{ChristosA} or localization infrastructures \cite{Yassin2017Recent}.
One popular classification is to consider where the location estimation is performed \cite{Drane1998Positioning}, based on which we have \emph{self-localization} or \emph{remote localization} systems.
\begin{figure} 
  \centering
\subfigure[Self-localization system]{
\label{Fig self-localization}
\begin{overpic}[width=0.35\textwidth]{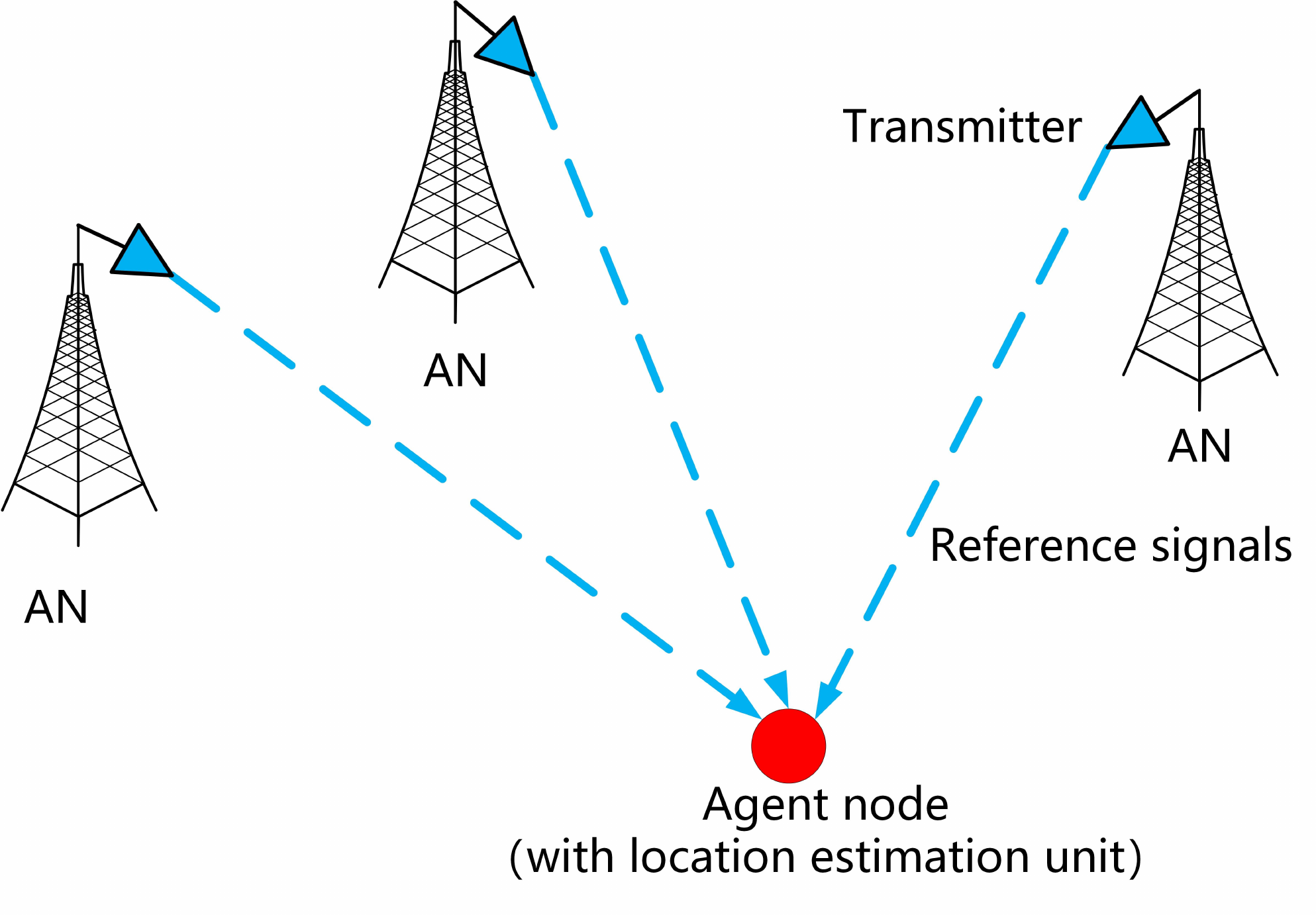}
\end{overpic}
}
\subfigure[Remote localization system]{
\label{Fig remote localization}
\begin{overpic}[width=0.35\textwidth]{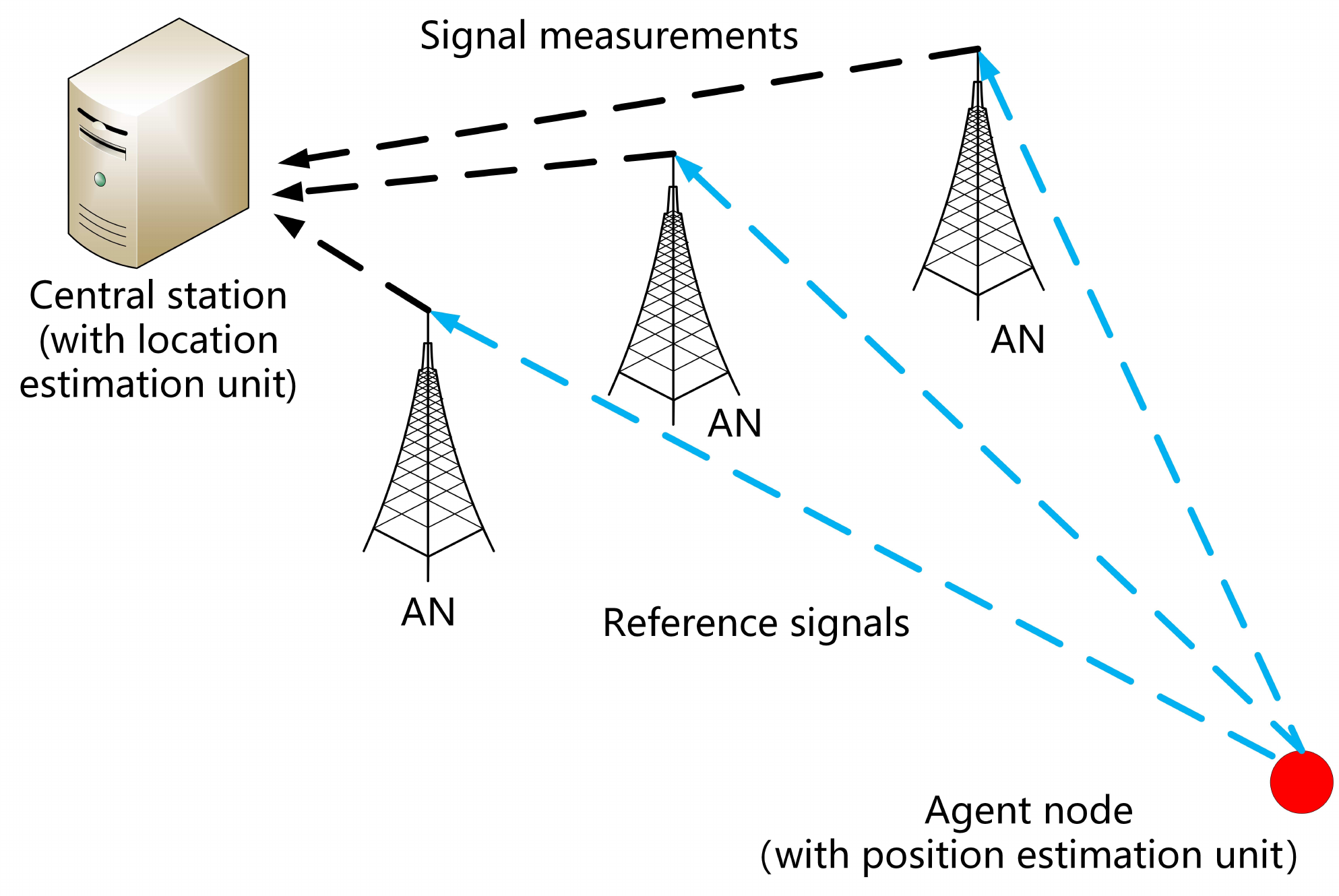}
\end{overpic}
}
\caption{An illustration of self-localization and remote localization systems. (a) For self-localization, the location estimation is performed by the agent node itself. (b) For remote localization, the location estimation is performed by a remote central station.}\label{Preprocess the map and build a field}\label{Fig classification of localization}\vspace{-10pt}
\end{figure}

\subsubsection{Self-Localization}
As illustrated in Fig. \ref{Fig self-localization}, for self-localization systems, a location estimation unit is deployed on the agent node, which receives reference signals transmitted from several ANs.
The agent node has the capability to perform appropriate signal measurements, based on which its own location is estimated.
Self-localization systems have several advantages.
First, since almost all localization-relevant operations are performed locally at the agent node, the localization speed is mainly dependent on the computational capability of the agent node.
Therefore, such systems are easier for performance improvement via updating the computational or measurement units of the agent node, without having to modify the network infrastructure.
Second, self-localization systems have the inherent mechanism for user privacy protection, since the agent node only passively receives signals transmitted from ANs, with little risk of location information leakage from the user side.
Finally, for some dynamic localization scenarios like tracking and navigation, various onboard sensors like inertial measurement units (IMUs) can be handily equipped on the agent node to provide some motion information, which can be utilized to further enhance the localization accuracy \cite{gustafsson2005mobile}.
However, self-localization systems have high hardware requirements on the agent node, such as high caching and computational capability, to accomplish the tasks of signal measurements and location estimation by itself.
The consequence is that such systems, like global positioning system (GPS) or inertial navigation system (INS), can only be deployed on the devices with powerful computational components.

\subsubsection{Remote Localization}
For remote localization systems, as illustrated in Fig. \ref{Fig remote localization}, the reference signals are transmitted from the agent node to ANs.
Upon receiving the reference signals, the ANs would send their respective signal measurements to a remote central station, where the location estimation is performed.
The main advantage of remote localization systems over the self-localization counterparts is the less demanding on the agent node, since almost all time-consuming and complex computing operations are performed at the remote central station, such as cellular base stations (BSs) or computing center.
Therefore, remote localization is especially appealling for resource-limited devices, such as IoT devices and wireless sensor nodes.
In addition, different from self-localization systems where the location information is only acquired by the agent node itself, remote localization systems can preserve locations for all agent nodes in the area of interest, which can be utilized for various purposes, like location-aware communication \cite{di2014location}.
However, since all location information of agent nodes are stored in a remote server, the information privacy and security is a critical issue for remote localization.

\subsection{Basic Localization Techniques}\label{basic localization techniques}
In general, localization techniques can be classified into two main categories: \emph{direct localization} and \emph{two-step localization}.
For direct localization \cite{wax1985decentralized,weiss2004direct,bialer2012maximum}, the received signals are directly used to estimate the location of the agent node, whereas for two-step localization, the location-related information, such as RSS, TOA, TDOA and AOA, is firstly extracted from the received signals, based on which the location of the agent node is estimated.
Note that in principle, direct localization can achieve better performance than two-step localization.
However, by considering the complexity and implementation constraints, \emph{two-step localization} approaches are usually used in practical systems.
Therefore, in this article, we will focus on two-step localization approaches.
Typically, depending on the different principles behind, the two-step localization approaches can be further classified into \emph{geometric-based}, \emph{scene analysis} (also known as \emph{fingerprinting}), and \emph{proximity} approaches\cite{HuiSurvey,zekavat2011handbook}, as discussed in details in the following.

Consider a basic wireless localization system consisting of $N$ ANs and one agent node.
The locations of the ANs are known, denoted by $\mathbf{p}_n, n=1,\cdots,N$, while that of the agent node needs to be estimated, denoted by $\mathbf{w}$.
Regardless of self- or remote localization, a two-step localization approach can be interpreted as a parameters estimation problem.
For the first step for signal measurement, the location-related information is obtained from the received signals, which are in general affected by multi-path effects and NLoS propagation, and a general measurement model can be expressed as
\begin{equation}\label{general measurement model}
r_{n} = h(\mathbf{p}_n,\mathbf{w}) + e_{n}, n=1,2,\cdots,N,
\end{equation}
where $r_n$ denotes the generic signal measurement associated with the $n$th AN, $h(\cdot)$ is a nonlinear function which contains all necessary information to compute the location of the agent node, and $e_n$ represents the measurement error.
Note that the exact expressions of $h\left(\cdot\right)$ for geometric-based methods can be easily established (as given in the subsequent subsections) since there are clear algebraic relationships between ANs and the agent node, while that for fingerprinting-based methods do not exist, and the pre-built fingerprints are usually treated as $h\left(\cdot\right)$.
For the second step for location estimation, the main task is to solve the systems of nonlinear equations in \eqref{general measurement model} to estimate $\mathbf{w}$ based on the obtained signal measurements $\left\{r_n\right\}_{n=1}^N$.
Such nonlinear equations are difficult to solve directly and deserve detailed discussions.
Therefore, in this subsection, we mainly focus on the principles and characteristics of different measurement models, while a comprehensive analysis about the location estimators to solve those nonlinear equations is deferred to Section \ref{location estimators}.

\begin{figure} 
\centering
\subfigure[RSS- or TOA-based localization]{
\label{Fig TOA or RSS}
\begin{overpic}[width=0.3\textwidth]{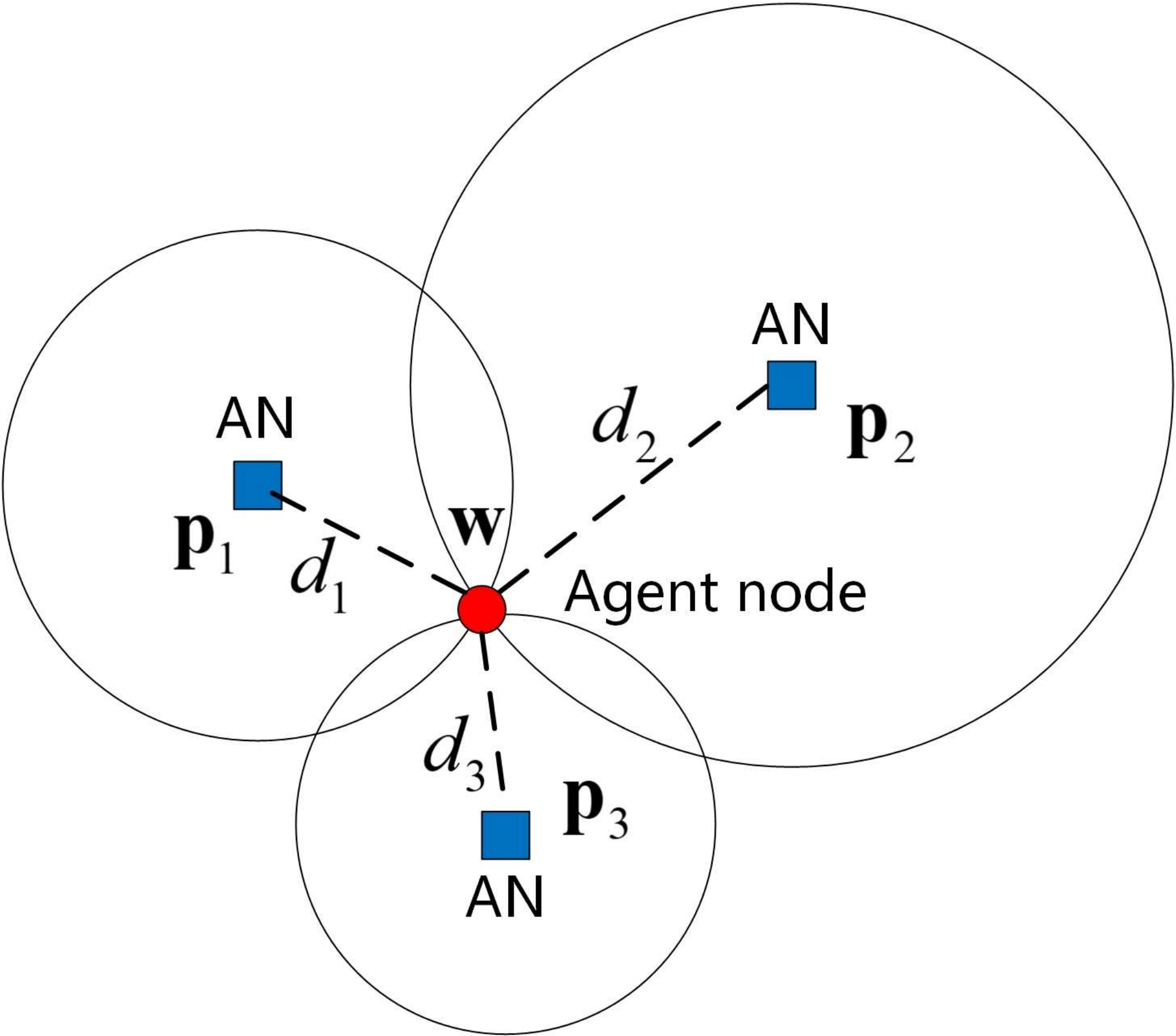}
\end{overpic}
}\quad\quad
\subfigure[TDOA-based localization]{
\label{Fig TDOA}
\begin{overpic}[width=0.3\textwidth]{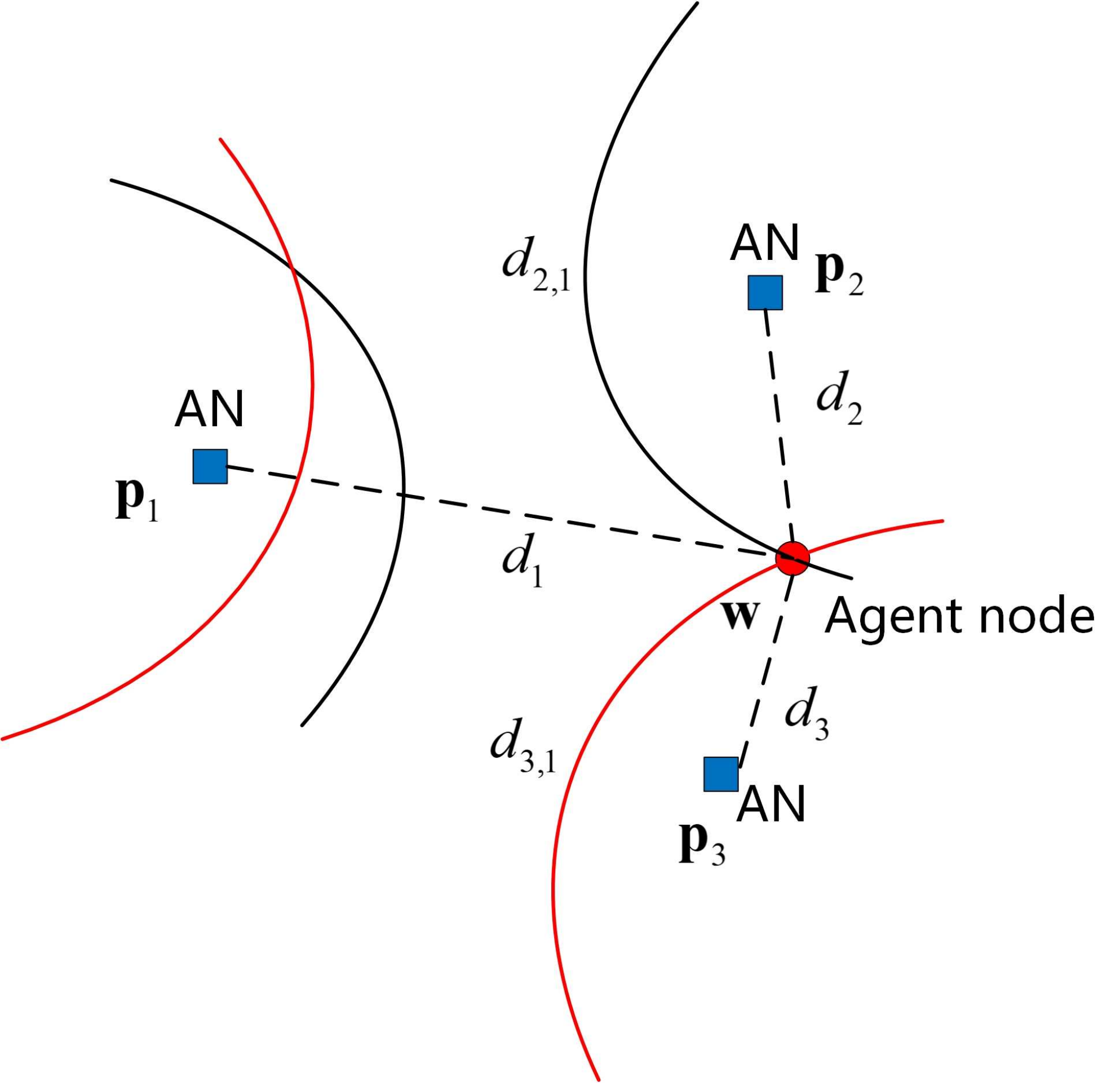}
\end{overpic}
}
\subfigure[AOA-based localization]{
\label{Fig AOA}
\begin{overpic}[width=0.25\textwidth]{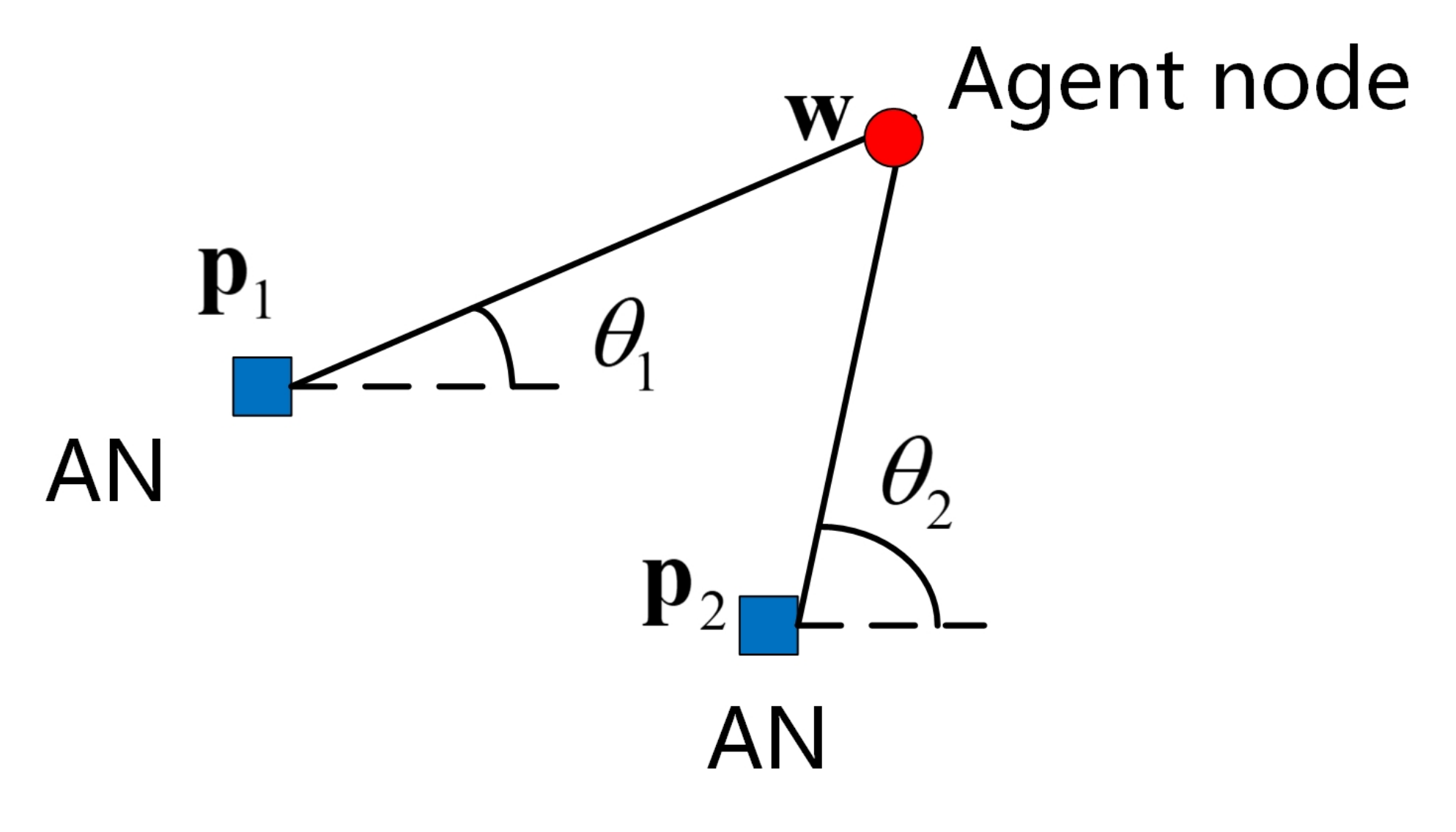}
\end{overpic}
}

\caption{Geometric-based localization in 2D space with perfect signal measurements, where $\textbf{p}_{n}\in \mathbb{R}^{2\times 1}, n=1,2,3$, denote the ANs, $\textbf{w} \in \mathbb{R}^{2\times 1}$ is the agent node to be located. (a) RSS- or TOA-based localization, where $d_n, n=1,2,3$, denote the actual distances between the agent node to the three ANs. (b) TDOA-based localization, where $d_{n,1}, n=2,3$, represent the relative distance between a pair of ANs to the agent node. The two pairs of red and black lines correspond to the hyperbola curves. (c) AOA-based localization, where $\theta_i, i=1,2$ are the angles between the ANs and the agent node.}\label{Fig Triangulation}\vspace{-15pt}
\end{figure}

\subsubsection{Geometric-Based Localization}
As illustrated in Fig. \ref{Fig Triangulation}, geometric-based localization technologies exploit the geometric properties of triangles to locate the agent node.
Typically, geometric-based methods have two variations: \emph{trilateration} and \emph{triangulation}.
Trilateration determines the location of the agent node using the distance-related signal measurements from multiple ANs, so it is also called \emph{ranging}.
For example, in 2D localization scenarios, the agent node would be ideally located at the intersection of at least three circles with centers being the locations of the ANs and radii equal to the distances from the agent node to each of the ANs.
The locations of the ANs are known and their distances to the agent node can be derived from RSSs \cite{yang2013rssi,kumar2009distance,yang2009indoor}, TOAs \cite{chan2006time,ke2009constrained,guvenc2009survey,ledergerber2015robot,wahab2013two} or TDOAs \cite{kaune2011accuracy,jung2011tdoa,kaune2012accuracy} of the received signals.
On the other hand, triangulation usually measures the AOAs of the received signals propagated between the ANs and the agent node, and locates the agent node at the intersection of angle direction lines \cite{kumar2014accurate,wang2015asymptotically}, where the AOAs can be measured with the aid of directional antennas or antenna arrays deployed on the agent node or the ANs.
Different from trilateration which requires at least three ANs, two ANs are sufficient for triangulation to locate the agent node in a 2D space.

\paragraph{RSS}
RSS-based localization approaches use the average power attenuation of the signals propagated between the ANs and the agent node to estimate their distances, based on which a geometric model is formulated to estimate the location of the agent node \cite{yang2013rssi,kumar2009distance,yang2009indoor}.
For example, as illustrated in Fig. \ref{Fig TOA or RSS}, the location of an agent node can be ideally determined in 2D space with the use of three ANs.
In general, the average received power $P_{r,n}$ associated with the $n$th AN can be modelled in dB form as \cite{hata1980empirical}
\begin{equation}\label{lognormal pass loss model}
P_{r,n} = P_0-10\alpha_n\log_{10}d_n+e_{\text{RSS},n},  n=1,2\cdots,N,
\end{equation}
where $P_0$ is the reference received average power at a reference distance of 1 meter (m), $d_n=\left\|\textbf{p}_n-\mathbf{w}\right\|$ is the actual distance between the $n$th AN and the agent node, $\alpha_n$ denotes the path loss exponent, and $e_{\text{RSS,n}}$ represents the error of the RSS measurement.
Assuming that $P_0$ and $\alpha_n, n=1,2,\cdots,N$ are known, the distance between the agent node and each of the ANs can be estimated.
Specifically, let $r_{\text{RSS},n}=P_{r,n}-P_0$ and $h_{\text{RSS}}(\mathbf{p}_n,\mathbf{w})=-10\alpha_n\log_{10}d_n=-10\alpha_n\log_{10}\left(\left\|\mathbf{p}_n-\mathbf{w}\right\|\right)$. The generic model in \eqref{general measurement model} for RSS-based localization can thus be written as
\begin{equation}\label{RSS measurement model}
r_{\text{RSS},n}=h_{\text{RSS}}(\mathbf{p}_n,\mathbf{w})+e_{\text{RSS},n}, n=1,2,\cdots,N.
\end{equation}
The remaining task of the RSS-based localization is to estimate $\mathbf{w}$ based on the obtained $\left\{r_{\text{RSS},n}\right\}_{n=1}^{N}$ in \eqref{RSS measurement model}, which is discussed in Section \ref{location estimators}.

The main advantage of RSS-based localization lies in that time synchronization among different nodes are not needed and RSS measurements are readily available for almost all practical wireless systems.
In addition, different from alternative schemes like TOA, TDOA, or AOA-based approaches, RSS measurements do not rely on LoS signal propagation.
However, the main drawback of RSS-based approaches is the poor localization accuracy, especially in clutter environments, since the signal attenuation in these environments is only weakly correlated with distance, leading to poor accuracy for distance estimation \cite{yang2013rssi}.
Besides, an accurate signal propagation model is necessary for reliable RSS-based distance estimation, which is challenging due to the unpredictable variations of the channel behavior.
Therefore, in practice, RSS-based localization is mostly adopted for those applications with coarse localization accuracy requirements.

\paragraph{TOA}
As illustrated in Fig. \ref{Fig TOA or RSS}, TOA-based approaches first estimate the distances between the agent node and each of the ANs by using the signal propagation delay or time of flight (TOF), denoted by $\tau_{\text{f}}$, and then build the trilateration model to estimate the location of the agent node
\cite{chan2006time,ke2009constrained,guvenc2009survey,ledergerber2015robot,wahab2013two}.
Typically, depending on how $\tau_{\text{f}}$ is defined, TOA-based methods can be further divided into one-way TOA (OW-TOA) \cite{ledergerber2015robot} and two-way TOA (TW-TOA) \cite{wahab2013two}.

For OW-TOA localization, node $A$ (either the AN or the agent node) transmits to node $B$ a packet that contains a timestamp $\tau_{s}$ recording the time when the packet was sent, and node $B$ then measures the TOA of the received signal, denoted by $\tau_{r}$.
The TOA is commonly measured by using matched filtering or correlation techniques, where the TOA measurement is given by the time shift of the reference signal that yields the maximum correlation value with the received signals.
For OW-TOA localization, if the time clock between the ANs and the agent node are perfectly synchronized, it is clear that $\tau_{\text{f}}$ can be determined by node $B$ as $\tau_{\text{f}}=\tau_{r}-\tau_{s}$, and the distance between node $A$ and node $B$ can be calculated as $d=\tau_{\text{f}}\cdot c$, where $c$ is the signal propagation speed, which is typically taken as the light speed.
However,
OW-TOA methods have two main drawbacks.
First, a small time synchronization error between the agent node and ANs can significantly compromise the distance estimation.
Second, the transmitted signal must be labeled with a timestamp to allow the receiving node to calculate $\tau_{\text{f}}$, which increases the complexity of the transmitted signals' structures and may cause additional estimation error.

On the other hand, for TW-TOA localization, node $A$ transmits a packet to node $B$, which responds by sending an acknowledgement packet to node $A$ after a response delay $\tau_{d}$.
Provided that $\tau_d$ is known, node $A$ can calculate its distance to node $B$ based on the signal round-trip time of flight (RTOF), i.e.  $\tau_{\text{RT}}=2\tau_{\text{f}}+\tau_{d}$.
TW-TOA addresses the first drawback of OW-TOA by avoiding the requirement of time synchronization between the two nodes.
However, in practice, it is still difficult for the measurement node, i.e. node $A$, to know the exact response delay $\tau_{d}$.
Although $\tau_{d}$ could be ignored if it is relative small compared with $\tau_{\text{f}}$ in long-range signal propagation, it critically affects the performance for short-range scenarios.
Furthermore, while TW-TOA method eliminates clock synchronization error between the two nodes, relative clock drift could compromise the distance estimation accuracy.
In addition, timestamp is still needed for TW-TOA to compute the RTOF of the transmitted signal.

Mathematically, a general TOA-based measurement model is formulated as \cite{chan2006time,guvenc2009survey}
\begin{equation}\label{TOA model}
c\cdot\tau_{\text{f},n} = d_{n}+e_{\text{TOA},n}, n=1,2,\cdots,N,
\end{equation}
where $\tau_{\text{f},n}$ is the measured TOF of signal propagation between the $n$th AN and the agent node, which is typically affected by positive bias errors introduced in the process of signal measurement, as captured by the additional the measurement error $e_{\text{TOA},n}$.
Following similar notations in \eqref{RSS measurement model}, let $r_{\text{TOA},n}=c\cdot\tau_{\text{f},n}$ and $h_{\text{TOA}}(\mathbf{p}_n,\mathbf{w})=d_n=\left\|\mathbf{p}_n-\mathbf{w}\right\|$,
the generic model in \eqref{general measurement model} for TOA-based localization can be written as
\begin{equation}\label{TOA measurement model}
r_{\text{TOA},n} = h_{\text{TOA}}(\mathbf{p}_n,\mathbf{w}) + e_{\text{TOA},n}, n=1,2,\cdots,N.
\end{equation}
By solving systems of nonlinear equations in \eqref{TOA measurement model}, the location of the agent node can be estimated.

\paragraph{TDOA}
TDOA refers to as the difference on the TOAs of the received signals at two different measurement units \cite{kaune2011accuracy,jung2011tdoa,kaune2012accuracy}.
Compared with TOA-based methods that estimate the absolute distances between each AN and the agent node, TDOA-based methods estimate the relative distance between a pair of ANs to the agent node.
For each TDOA measurements, the agent node would lie on a hyperboloid with a constant distance difference between a pair of ANs to the agent node.
Specifically, the equation of the hyperboloid is given by \cite{HuiSurvey}
\begin{equation}\label{relative distance}
d_{i,j} = \left\|\mathbf{p}_i-\mathbf{w}\right\|-\left\|\mathbf{p}_j-\mathbf{w}\right\|, i\neq j,
\end{equation}
where $d_{i,j}$ denotes the relative distance between the $i$th and $j$th ANs to the agent node, which can be estimated by the TDOA measurements.
For instance, in 2D localization cases, as illustrated in Fig. \ref{Fig TDOA}, the location of the agent node $\mathbf{w}$ could be theoretically estimated from the two intersections of at least two hyperbolas with two pairs of foci, i.e. $\mathbf{p}_1$ versus $\mathbf{p}_2$ and $\mathbf{p}_1$ versus $\mathbf{p}_3$.
Typically, the TDOA measurement schemes used in self- and remote localization are different.
For self-localization, the synchronized ANs broadcast multiple signals to the agent node, which measures the TDOA by itself.
The conventional method for measuring TDOA is based on the cross-correlation techniques,
where the correlation coefficient between a pair of received signals are calculated, and the time delay of the two signals that result in the maximum correlation value is regraded as the TDOA of the two signals.
For remote localization, the ANs first estimate their respective TOAs based on the reference signals transmitted from the agent node, and then exchange measurements with other ANs to compute the TDOAs.

The mathematical model of TDOA measurement is discussed as follows.
Assuming that a reference signal was sent at an unknown time $\tau_0$, which is then received by the $i$th and $j$th measurement units at time $\tau_{i}$ and $\tau_{j}$, respectively.
Therefore, the TDOA between the $i$th and $j$th measurement units is \cite{zekavat2011handbook}
\begin{equation}\label{TDOA}
\tau_{i,j}=(\tau_{i}-\tau_{0})-(\tau_{j}-\tau_{0})=\tau_{i}-\tau_{j}.
\end{equation}
For a localization system consisting of $N$ ANs, there are $N(N-1)/2$ TDOAs from all possible pairs of ANs, but only $N-1$ of them are non-redundant.
For example, for $N=3$, the TDOAs are $\tau_{2,1}$,$\tau_{3,1}$ and $\tau_{3,2}$, but $\tau_{3,2}$ can be obtained by $\tau_{3,2}=\tau_{3,1}-\tau_{2,1}$, which is thus redundant.
Without loss of generality, we consider the first measurement unit as the reference and the non-redundant TDOAs are $\tau_{n,1}, n=2,3,\cdots,N$.
In practice, TDOA measurements suffer from positive bias errors introduced by clock synchronization error among ANs, the multi-path effects, and so on.
Therefore, similar to \eqref{TOA model}, the TDOA measurement model is
\begin{equation}\label{TDOA model}
c\cdot\tau_{n,1} = d_{n,1}+ e_{\text{TDOA},n},  n=2,3,\cdots,N,
\end{equation}
where $d_{n,1}$ is the relative distance between $n$th AN and the 1st AN to the agent node, which is defined in \eqref{relative distance}, and $e_{\text{TDOA},n}$ represents the measurement error.
Following the definition in \eqref{RSS measurement model} and \eqref{TOA measurement model}, the generic model in \eqref{general measurement model} for TDOA-based localization can be expressed as
\begin{equation}\label{TDOA measurement model}
r_{\text{TDOA},n} = h_{\text{TDOA}}(\mathbf{p}_n,\mathbf{w})+e_{\text{TDOA},n}, n=2,3,\cdots,N,
\end{equation}
where $r_{\text{TDOA},n} = c\cdot\tau_{n,1}$ and $h_{\text{TDOA}}(\mathbf{p}_n,\mathbf{w}) = \left\|\mathbf{p}_n-\mathbf{w}\right\|-\left\|\mathbf{p}_1-\mathbf{w}\right\|$.

TDOA-based methods overcome both drawbacks of the TOA-based methods.
First, it only requires time synchronization among ANs, but not between each AN and the agent node, where the complexity of the latter is usually much higher.
This is because that the agent node usually uses quart clocks for timing, which are not as precise as atomic clocks that are generally used at ANs, while the time synchronization among ANs can be more conveniently achieved by using wire backbone networks \cite{zekavat2011handbook}.
Another advantage of TDOA-based localization is that the timestamp is no longer needed, as evident from the TDOA computation in \eqref{TDOA}.
This simplifies the structure of the transmitted signals and avoids the potential sources of error.
However, all time-based positioning methods, like TOA and TDOA, rely heavily on the LoS path to compute TOA or TDOA information, which renders them venerable to NLoS environment.

\paragraph{AOA}
AOA, also called direction of arrival (DOA), refers to the angle of the arriving signal relative to a reference direction at the ANs side (in the uplink), while the terminology angle of departure (AOD) is often used at the agent node side (in the downlink).
Nevertheless, the principles of the angulation methods are similar; that is, using the angles between ANs and the agent node, to determine the location of the agent node \cite{kumar2014accurate,wang2015asymptotically}.
As shown in Fig. \ref{Fig AOA}, for 2D localization, AOA-based methods require only two known ANs with two measured angles to determine the location of the agent node.
To measure the AOAs, the ANs should equip with antenna arrays or directional antennas with spatial resolution capabilities.

For 2D localization, the AOA between the agent node $\mathbf{w}=\left[w_x,w_y\right]^{T}$ and the $n$th AN $\mathbf{p}_n=\left[p_{x,n},p_{y,n}\right]^{T}$ can be expressed as \cite{zekavat2011handbook}
\begin{equation}\label{azimuth}
\theta_n = \tan^{-1}\left(\frac{w_y-p_{y,n}}{w_x-p_{x,n}}\right), n=1,2,\cdots,N,
\end{equation}
where $\theta_n\in\left(-\pi,\pi\right)$ represents the azimuth angle in a 2D plane.
In the presence of angle estimation errors, the AOA measurements can be modeled as
\begin{equation}\label{AOA 2D measurement model}
r_{\text{AOA},n} = h_{\text{AOA}}(\mathbf{p}_n,\mathbf{w}) + e_{\text{AOA},n}, n =1,2,\cdots,N,
\end{equation}
where $r_{\text{AOA},n}$ is the measured AOA with the error $e_{\text{AOA},n}$, and $h_{\text{AOA}}(\mathbf{p}_n,\mathbf{w})=\theta_n = \tan^{-1}\left(\frac{w_y-p_{y,n}}{w_x-p_{x,n}}\right)$.
Compared with 2D localization, AOA-based 3D localization is more challenging, since the AOAs are represented by pairs of azimuth and elevation angles in a 3D space.
As the two angles are coupled with each other in their respective nonlinear measurement equations, one cannot apply the 2D AOA methods to the azimuth and elevation angles separately.
Specifically, considering the agent node at $\mathbf{w}=\left[w_{x},w_{y},w_{z}\right]^{T}$ and the ANs at $\mathbf{p}_n=\left[p_{x,n},p_{y,n},p_{z,n}\right]^{T}, n=1,2,\cdots,N$, their azimuth and elevation angles can be expressed as \cite{wang2015asymptotically}
\begin{equation}\label{azimuth and elevation}
\left[\begin{array}{l}
\theta_n \\
\phi_n
\end{array}\right]
=
\left[\begin{array}{l}
\tan^{-1}\left(\frac{w_y-p_{y,n}}{w_x-p_{x,n}}\right) \\
\tan^{-1}\left(\frac{w_z-p_{z,n}}{\left(w_x-p_{x,n}\right)\cos\theta_n+\left(w_y-p_{y,n}\right)\sin\theta_n}\right)
\end{array}\right],
\end{equation}
where $\phi_{n}\in\left(-\pi/2,\pi/2\right)$ represents the elevation angle.
Therefore, similar to \eqref{AOA 2D measurement model}, the AOA measurement model in 3D space is
\begin{equation}\label{AOA 3D measurement model}
\mathbf{r}_{\text{AOA},n} = \mathbf{h}_{\text{AOA}}(\mathbf{p}_n,\mathbf{w}) + \mathbf{e}_{\text{AOA},n}, n=1,2,\cdots,N,
\end{equation}
where $\mathbf{r}_{\text{AOA},n}=\left[\hat{\theta}_n,\hat{\phi}_n\right]^{T}$ is a vector consisting of the measured azimuth $\hat{\theta}_n$ and elevation $\hat{\phi}_n$ with measurement errors $\mathbf{e}_{\text{AOA},n}=\left[e_{\theta_n},e_{\phi_n}\right]^{T}$, and $\mathbf{h}_{\text{AOA}}(\mathbf{p}_n,\mathbf{w})=\left[\theta_n, \phi_n\right]^{T}$ is defined in \eqref{azimuth and elevation}.

The advantage of AOA-based localization over the TOA/TDOA counterparts lies in that it does not require time synchronization between the measuring units, and the 2D localization can be achieved by using two ANs, as opposed to three ANs in TOA-, TDOA-, or RSS-based methods.
However, the ANs for AOA-based localization need to equip with highly directional antennas or large antenna arrays for highly accurate angular resolution.
Besides, in complex multi-path environment, like urban areas or indoor scenarios, the AOA estimation is subject to significant errors due to the NLoS propagation.
In this case, it is important for AOA measuring units to distinguish the AOA of the LoS path from those NLoS paths.

\subsubsection{Scene Analysis/Fingerprinting-based Localization}
The ever-increasing number of sensors on smart devices has led to a rapid advancement of wireless localization.
Since the performance of geometric-based localization approaches degrades significantly in complex environments, alternative approaches based on \emph{scene analysis} or \emph{fingerprinting} have been proposed \cite{feng2011received,honkavirta2009comparative,yang2015wifi,vo2015survey,he2015wi,khalajmehrabadi2017modern}.
Such methods first exploit the data collected by the sensors, like cameras, accelerometer, or specific WiFi access points (APs), to extract unique geotagged signatures, i.e. \emph{fingerprints}, and then pinpoint the location of the agent node by matching the online signal measurements against the pre-recorded geotagged fingerprints.

Various information can be sensed and used as fingerprints for localization.
Depending on the types of fingerprints, such localization approaches can be classified into visual, motion, and signal fingerprint-based methods \cite{vo2015survey}.
For visual fingerprint-based localization, a set of geotagged images are pre-recorded in a database.
During the localization process, a query image captured by the agent node equipped with camera is used to find the best-matched image from the database, and the location associated with the best-matched image is returned as the location of the agent node.
For motion fingerprint-based localization, the measurements collected from the accelerometer and gyroscopes on a mobile agent node are combined and matched with a map of the area to estimate the location of the agent node.
Note that the map is characterized by motion fingerprints (e.g. traveled distance and orientation) of the mobile agent node.
For signal fingerprint-based localization, the location-related signal measurements are sensed and stored in a database as signal fingerprints.
In many applications, signal fingerprints usually correspond to the RSS indicators (RSSIs) instead of TOA, TDOA or AOA \cite{he2015wi}.
The main reason is that the time- or angle-based signal measurements rely heavily on LoS geometric assumption, which is hardly satisfied in complex environments.
Compared with visual fingerprint-based approaches, signal fingerprint-based approaches can achieve higher localization accuracy, since RSSIs generally have stronger correlation with location than images \cite{he2015wi}.
On the other hand, while motion fingerprint-based methods can achieve higher accuracy than signal fingerprint-based methods, the motion sensors like accelerometer and gyroscopes are costly hardware.
By contrast, RSS measurements utilizing the existing infrastructure can avoid the cost of using additional hardware.
Therefore, a variety of radio frequency (RF) signal fingerprint-based localization systems have been developed, like RADAR system \cite{bahl2000radar} that employs WiFi signals for indoor localization.
In this article, we mainly focus on RF signal fingerprint-based methods.

\begin{figure} 
\centering
\subfigure[Offline training phase]{
\label{Fig Offline}
\begin{overpic}[width=0.4\textwidth]{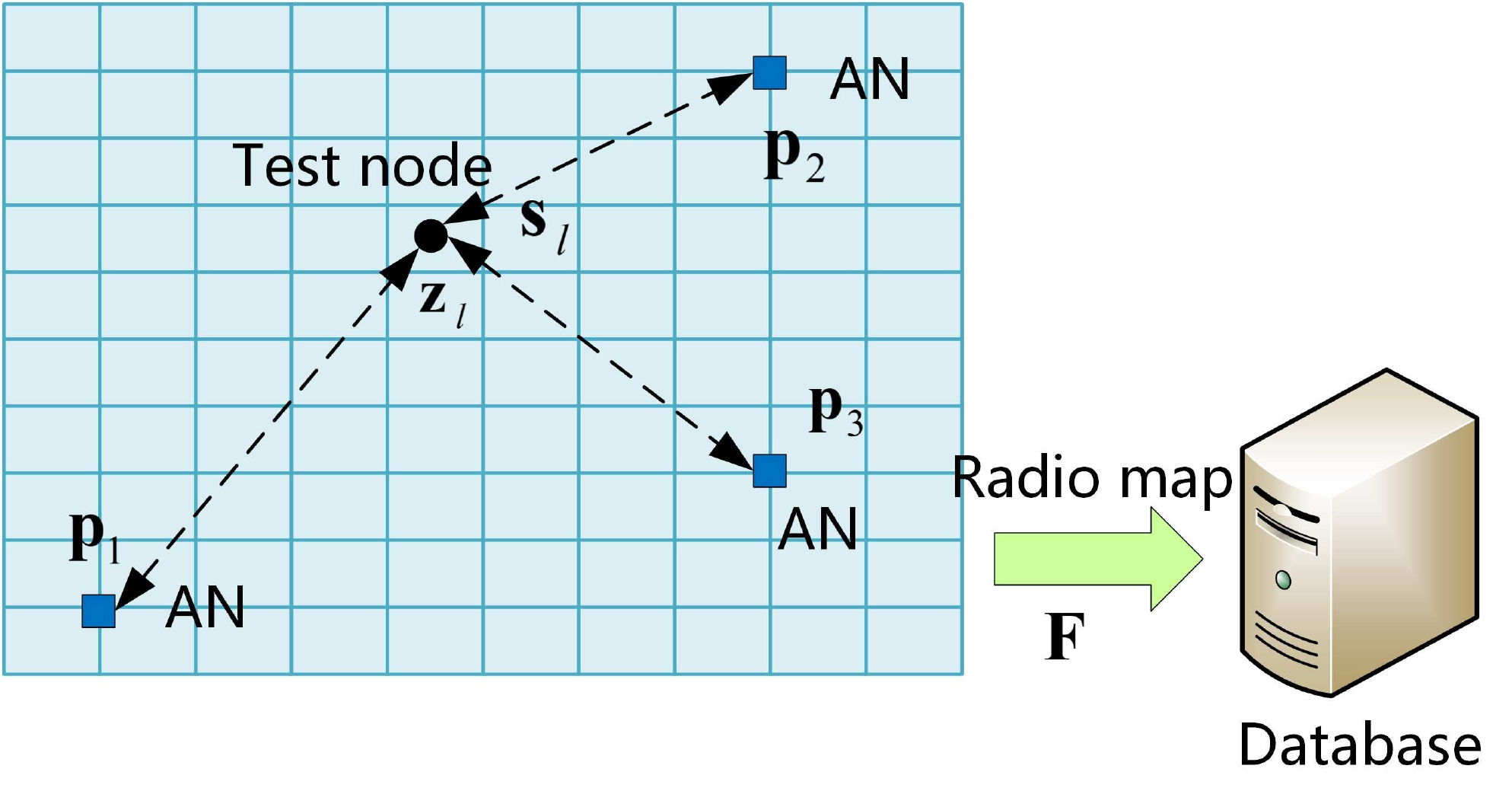}
\end{overpic}
}

\subfigure[Online localization phase]{
\label{Fig Online}
\begin{overpic}[width=0.4\textwidth]{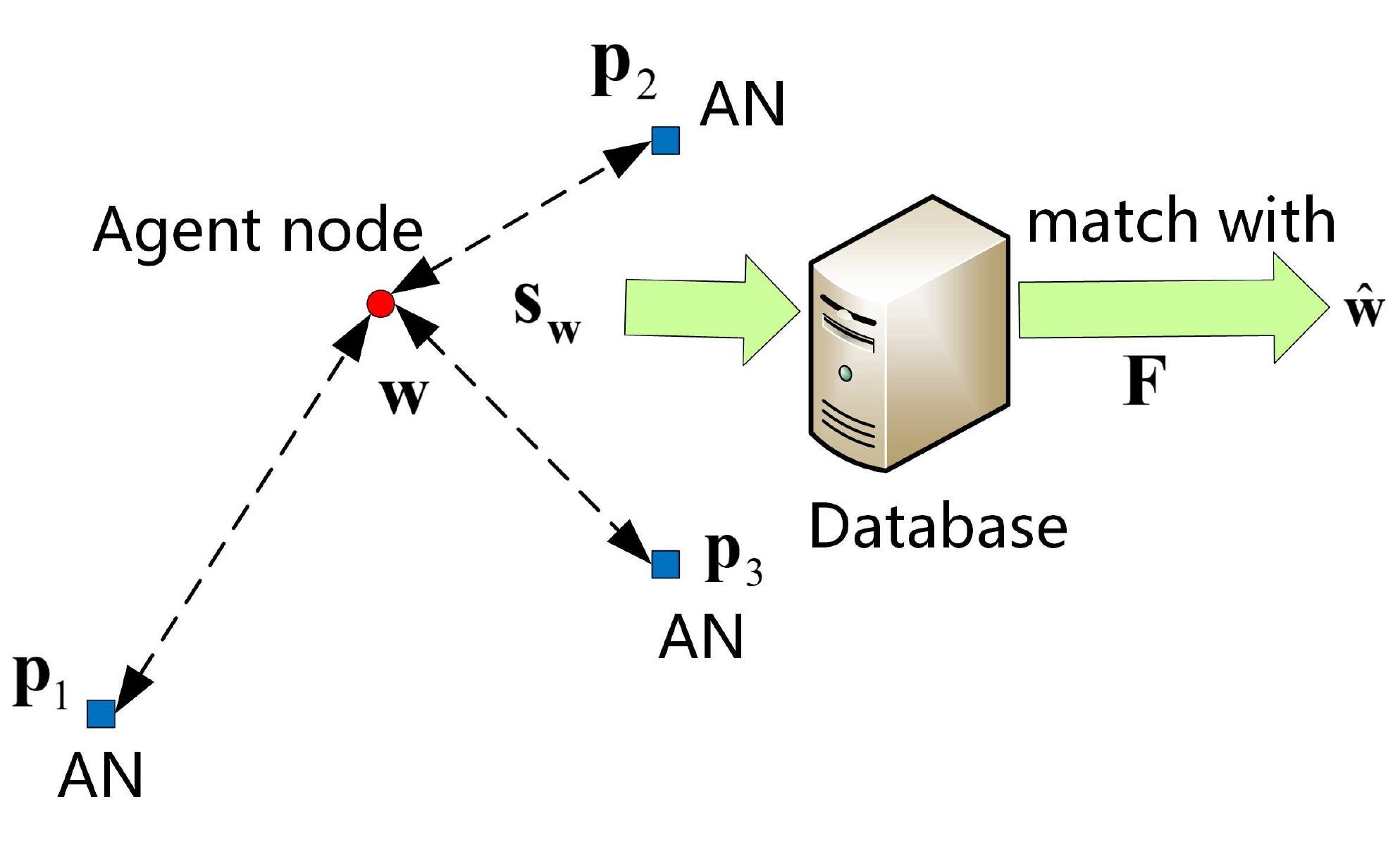}
\end{overpic}
}

\caption{The procedures of fingerprint-based localization, where $\mathbf{p}_n, n=1,2,3$ denote the ANs. (a) Offline training phase, where $\mathbf{s}_l$ is the RF fingerprint at the cell $\mathbf{z}_l$. After the test node travels the whole area of interest, the radio map $\mathbf{F}$ consisting of all fingerprints in the area will be stored in a database. (b) Online localization phase, where the online RSS measurements $\mathbf{s}_{\mathbf{w}}$ of the agent node is sent to the database. After pattern matching $\mathbf{s}_{\mathbf{w}}$ against with the radio map $\mathbf{F}$, the location of the agent node can be estimated as $\hat{\mathbf{w}}$.}\label{Fig Fingerprint}\vspace{-10pt}
\end{figure}

In general, the RF fingerprinting-based localization includes \emph{offline training} and \emph{online localization} phases.
As illustrated in Fig. \ref{Fig Fingerprint}, taking the 2D localization as an example, the area of interest is firstly divided into $L$ cells, where the location of the $l$th cell is known, denoted by $\mathbf{z}_l=\left[z_{x,l},z_{y,l}\right]^{T}, l=1,2,\cdots,L$.
For offline training, a mobile test node travels to the $L$ cells and communicates with the $N$ ANs to measure the RSS fingerprints, which are denoted as $\mathbf{s}_l=\left[s_{l,1},s_{l,2},\cdots,s_{l,N}\right]^{T}$ for measurements at $\mathbf{z}_l, l=1,2,\cdots,L$.
Then, the entire \emph{radio map} of the area of interest is obtained as
$\mathbf{F}=\left[\mathbf{s}_1,\mathbf{s}_2,\cdots,\mathbf{s}_L\right]^{T}$,
which would be stored in a database for online localization.
For online localization, a real-time RSS measurements of the agent node at the location of $\mathbf{w}=\left[w_{x},w_{y}\right]^{T}$ are measured as $\mathbf{s}_\mathbf{w}=\left[s_{\mathbf{w},1},s_{\mathbf{w},2},\cdots,s_{\mathbf{w},N}\right]^{T}$,
and the location of the agent node can be estimated as
$\hat{\mathbf{w}}=\left[\hat{w}_{x},\hat{w}_{y}\right]^{T}$ based on a rule $g(\cdot)$ that compares the received online measurements $\mathbf{s}_\mathbf{w}$ against the radio map $\mathbf{F}$ \cite{khalajmehrabadi2017modern,honkavirta2009comparative},
\begin{equation}\label{fingerprint comparision}
\hat{\mathbf{w}} = g\left(\mathbf{F},\mathbf{s}_\mathbf{w}\right).
\end{equation}
Depending on the mathematical model of RSS fingerprints, there are two main conventional localization approaches: \emph{deterministic} and \emph{probabilistic}.

\paragraph{Deterministic}
For deterministic approaches, the measured RSSs are assumed to be static.
In general, the location of the agent node is estimated to be the location of the cell, whose fingerprint is the closest to the online RSS measurements \cite{khalajmehrabadi2017modern}, i.e.,
\begin{equation}\label{deterministic}
\hat{\mathbf{w}}=\arg\min\limits_{l=1,\cdots,L} d\left(\mathbf{s}_l,\mathbf{s}_{\mathbf{w}}\right),
\end{equation}
where $d(\cdot)$ denotes a certain distance metric.
For example, applying the Euclidean distance metric in \eqref{deterministic}, we can obtain
\begin{equation}\label{NN}
\hat{\mathbf{w}}=\arg\min\limits_{l=1,\cdots,L} \left\|\mathbf{s}_l-\mathbf{s}_\mathbf{w}\right\|.
\end{equation}
Solving \eqref{NN} to estimate the location of the agent node is known as the nearest neighbor (NN) method.
Another well-known deterministic method is the $K$-nearest neighborhood (KNN) \cite{HuiSurvey,yang2015wifi}, where the location estimation is obtained by averaging the locations of the $K$ cells in the radio map that have the nearest distances.
The weighted KNN (WKNN) is a variant of the KNN method \cite{khalajmehrabadi2017modern,honkavirta2009comparative}, where the selected locations of the closest cells are combined by assigning weights to estimate the location of the agent node, where the weights are usually proportional to the inverse of their corresponding $d\left(\mathbf{s}_l,\mathbf{s}_{\mathbf{w}}\right)$.
In general, KNN and WKNN methods can achieve better performance than NN method.
However, as the density of the radio map increases, NN method may achieve comparable performance as KNN or WKNN methods.
Technically, RF fingerprinting methods originate from machine learning classification, so other machine learning methods such as support vector machine (SVM) and linear discriminant analysis (LDA) can be also used for location fingerprinting \cite{HuiSurvey}.
Such methods can achieve better localization accuracy compared with KNN, WKNN, or NN, but with higher computational complexity.
The main advantage of the deterministic approaches is their simplicity.
However, a single static RSS fingerprint of a cell may not be sufficient to uniquely represent the feature of the cell due to the time-varying nature of wireless signal propagation, so probabilistic approaches are developed.

\paragraph{Probabilistic}
Probabilistic methods use statistical inference between online signal measurement $\mathbf{s}_{\mathbf{w}}$ and the stored radio map $\mathbf{F}$ to estimate the location of the agent node, where the RSS fingerprint at a cell is treated as a random vector and the knowledge of RSS distribution in the area of interests is acquired through offline training.
The underlying principle of probabilistic localization is the maximum a posteriori (MAP) estimation, which estimates the location of the agent node by maximizing the conditional probability of the location given the online RSS measurements \cite{seco2009survey}, i.e.,
\begin{equation}\label{MAP}
\hat{\mathbf{w}}=\arg\max\limits_{\mathbf{z}_l, l=1,\cdots,L}P\left(\mathbf{z}_{l}|\mathbf{s}_{\mathbf{w}}\right),
\end{equation}
where $P\left(\mathbf{z}_{l}|\mathbf{s}_{\mathbf{w}}\right)$ is the conditional probability of the agent node at location $\mathbf{z}_{l}$ given the online RSS measurements $\mathbf{s}_{\mathbf{w}}$.
In the absence of a priori knowledge about the location of the agent node, the probabilities of the agent node at the each cell of the radio map is equal.
Then by using the Bayes' formula, equation \eqref{MAP} can be further transformed into
\begin{equation}\label{ML}
\hat{\mathbf{w}}=\arg\max\limits_{\mathbf{z}_l, l=1,\cdots,L}P\left(\mathbf{s}_\mathbf{w}|\mathbf{z}_l\right),
\end{equation}
which is known as maximum likelihood (ML) estimation, where $P\left(\mathbf{s}_\mathbf{w}|\mathbf{z}_l\right)$ is the probability of RSS distribution at the given location $\mathbf{z}_{l}$.
Therefore, the probabilistic localization methods rely on the estimation of the conditional probability $P\left(\mathbf{s}_\mathbf{w}|\mathbf{z}_l\right)$.
There are two main approaches to approximate $P\left(\mathbf{s}_\mathbf{w}|\mathbf{z}_l\right)$, namely, parametric and non-parametric estimation \cite{HuiSurvey,khalajmehrabadi2017modern}.
For parametric estimation methods, the known analytical distribution functions, such as Gaussian, lognormal, and kernel functions, are used to approximate temporal RSS characteristics.
However, these parametric estimation methods usually require some probabilistic assumptions (like the probabilistic independence), which makes it challenging to apply in some practical situations.
Unlike parametric estimation methods, non-parametric density estimation methods do not make any assumption about the RSS fingerprint distribution.
For non-parametric estimation, the fingerprint distributions are proportional to the current centralized histogram, which is known as \emph{histogram matching} \cite{khalajmehrabadi2017modern}.
However, for such methods, a large number of time samples are needed for each cell to generate a histogram.

Compared with geometric-based methods, the main advantage of fingerprinting methods is their robustness to signal measurement errors introduced by multi-path effects.
This is because that fingerprinting methods transform localization problems into the problems of pattern matching by dividing the process of localization into offline training and online matching phases.
In fingerprinting methods, a location is characterized by its detected signal patterns.
Therefore, without having to know the exact locations of ANs, fingerprinting requires neither distance nor angle measurement, rendering it especially feasible in clutter environments, like indoor or urban scenarios.
However, such methods also have some drawbacks.
First, for offline training, it is labor intensive and time-consuming to extract the RF fingerprints and construct the radio map.
The selected RF fingerprints must uniquely correspond to a given location and should have low variability during a certain time interval.
The process of training RF fingerprints is time-consuming since the distribution of RSS fingerprints is usually non-Gaussian, skewed, and multimodal.
In addition, since the signal propagation environment is inherently time-varying, the radio map needs to be updated regularly.
Furthermore, for online localization, it is necessary to limit the search size and exploit efficient pattern matching algorithms for reducing the cache consumption and computational complexity.
Therefore, compared with geometric-based localization, fingerprinting is more suitable for small-size environments.

\subsubsection{Proximity-Based Localization}
\begin{figure} 
\centering
{
\begin{overpic}[width=0.3\textwidth]{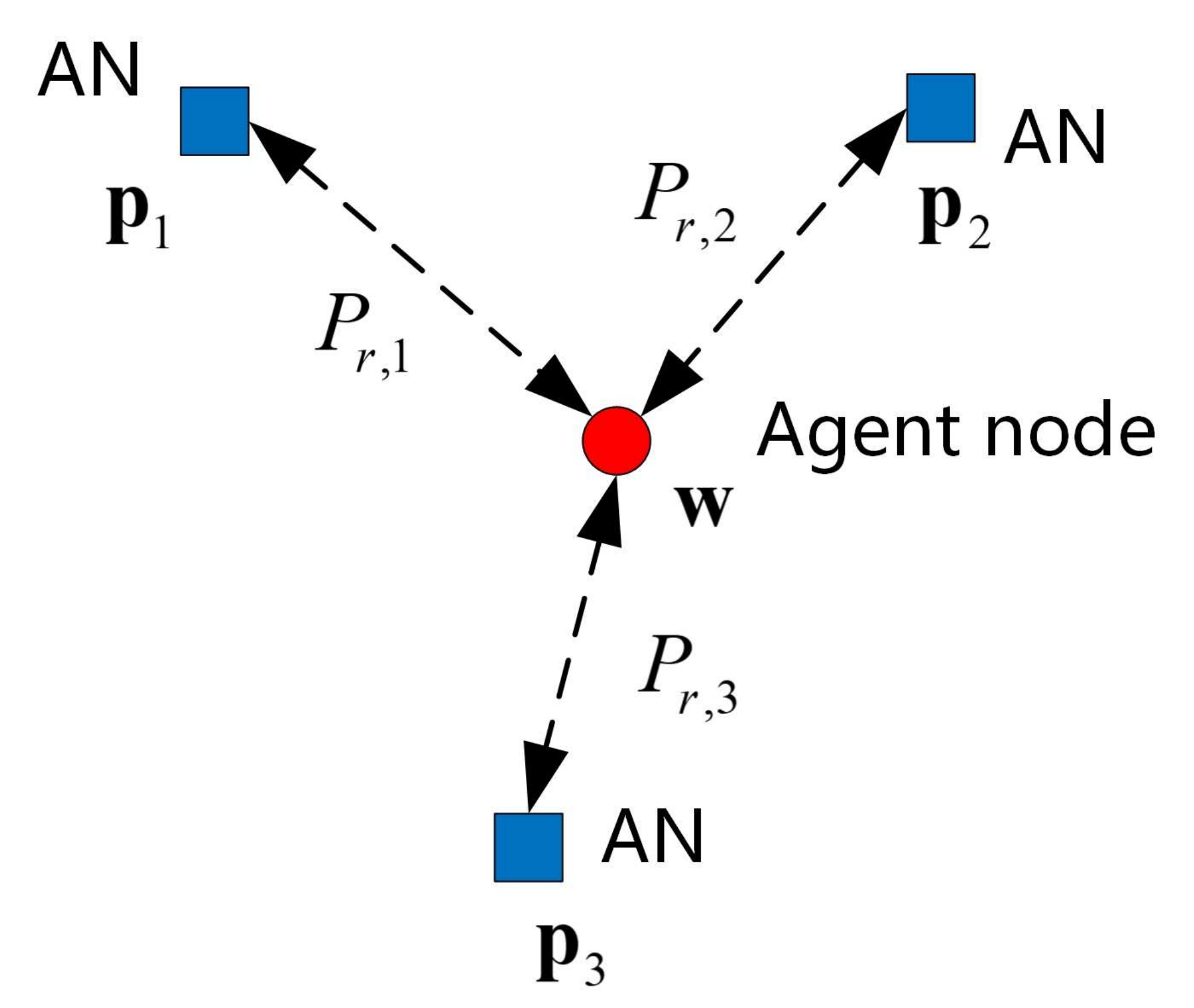}
\end{overpic}
}
\caption{Proximity-based localization using RSS measurements, where $\mathbf{p}_1$, $\mathbf{p}_2$ and $\mathbf{p}_3$ denote the ANs, $\mathbf{w}$ denotes the location of the targeting agent node, and $P_{r,n}, n=1,2,3$ represent the RSS measurements. The location of the agent node would be determined by comparing the RSSs under certain proximity constraints.}\label{FIG proximity}\vspace{-10pt}
\end{figure}
The principle of proximity-based localization is that the location of the agent node is determined according to proximity constraints \cite{brida2007accuracy,he2003range,li2009rendered}.
The mathematical model behind such proximity methods is similar to that of the deterministic approaches in fingerprinting-based localization.
As illustrated in Fig. \ref{FIG proximity}, such methods usually use RSS measurements to detect the agent node in a dense grid of ANs, and the location of the AN which has the strongest RSS is treated as the location of the agent node.
However, different from fingerprinting methods, proximity location estimation depends on the locations of the actual ANs, thus the localization accuracy relies on the density of ANs, which can only be improved by increasing the number of ANs.
In general, proximity method is simpler to implement than other localization techniques.
However, this method can only provide very coarse localization service, so it is usually used in systems with low requirements on the location accuracy.
The representative application for this method include Cell-ID (CID) \cite{del2017survey}, RFID \cite{Yassin2017Recent} and bluetooth-based localization systems \cite{ChristosA}, which are often used in cellular and IoT networks.
Another use case of proximity-based localization is to reduce the search size of fingerprint-based localization before fingerprints pattern matching is performed.

A summary of the aforementioned the above localization approaches is provided in Table \ref{Summary of Different Localization Approaches}, which compares their differences on measurement models, advantages, and disadvantages.

\begin{table*}[]
\centering
  \caption{Summary of Different Localization Techniques}\label{Summary of Different Localization Approaches}
  \setlength{\tabcolsep}{1mm}
\begin{tabular}{|l|l|l|l|l|}
\hline
                  Model&  Location-Related Information&  Advantages&  Disadvantages \\ \hline
\multirow{4}{*}{Geometric-Based}
                  &  RSS&
                  \begin{tabular}[c]{@{}l@{}}
                  Time synchronization is not required;\\
                  LoS path is not necessary;\\
                  Available for almost all wireless systems.
                  \end{tabular}&
                  \begin{tabular}[c]{@{}l@{}}
                  Low accuracy;\\
                  Vulnerable to complex environments.
                  \end{tabular} \\ \cline{2-4}
                  &  TOA&
                  \begin{tabular}[c]{@{}l@{}}
                  High accuracy for LoS scenarios.\\
                  \end{tabular}&
                  \begin{tabular}[c]{@{}l@{}}
                  Time synchronization between
                  ANs and \\the agent node is necessary;\\
                  Timestamp is required;\\
                  LoS path is assumed.
                  \end{tabular}  \\ \cline{2-4}
                  &  TDOA&
                  \begin{tabular}[c]{@{}l@{}}
                  High accuracy for LoS scenarios;\\
                  Time synchronization between ANs and\\
                  agent node is not needed;\\
                  Timestamp is not needed.\\
                  \end{tabular}&
                  LoS path is assumed.
                   \\ \cline{2-4}
                  &  AOA&
                  \begin{tabular}[c]{@{}l@{}}
                  Time synchronization is not required;\\
                  Two ANs are sufficient for 2D localization.\\
                  \end{tabular}&
                  \begin{tabular}[c]{@{}l@{}}
                  Directional antennas or arrays are needed;\\
                  LoS path is assumed in general.
                  \\(Some advanced methods can work on \\NLoS condition, which will be discussed \\in Section \ref{5G Localization}).
                  \end{tabular}
                  \\ \hline
                  \begin{tabular}[c]{@{}l@{}}
                  Scene Analysis\\ (or Fingerprinting)
                  \end{tabular}&  Fingerprints&
                  \begin{tabular}[c]{@{}l@{}}
                  Accuracy can be very high;\\
                  Robust to clutter environments;\\
                  LoS path is not required.
                  \end{tabular}&
                  \begin{tabular}[c]{@{}l@{}}
                  Training phase is labour intensive and \\time-consuming;\\
                  Accurate mapping from locations to \\fingerprints is needed;\\
                  Radio map needs to be updated regularly.
                  \end{tabular}    \\ \hline
                  Proximity&  RSS&
                  \begin{tabular}[c]{@{}l@{}}
                  Simple and inexpensive; \\
                  Easy to implement.
                  \end{tabular}
                  & Very low accuracy.    \\ \hline
\end{tabular}
\end{table*}

\subsection{Location Estimators}\label{location estimators}
In general, there are two categories of location estimators, namely, nonlinear and linear, to solve the localization problems defined in \eqref{general measurement model}.
The nonlinear estimators directly solve the problems by minimizing a cost function constructed from \eqref{general measurement model}.
Such nonlinear estimators usually result in high localization accuracy.
However, sometimes the global solution of such schemes may not be guaranteed as their cost functions are usually multi-modal, and nonlinear estimators usually have high time complexity if grid or random search is involved.
By contrast, linear estimators which convert the nonlinear equations into a set of linear equations can find efficient solutions quickly, with degraded localization accuracy compared to nonlinear estimators.

\subsubsection{Nonlinear Estimators}
Typical nonlinear estimators include the nonlinear least squares (NLS), weighted nonlinear least squares (WNLS) and maximum likelihood (ML) estimators \cite{zekavat2011handbook}.
Base on the generic model \eqref{general measurement model}, the general cost function of the NLS estimator is defined as \cite{guvenc2009survey}
\begin{equation}\label{NLS}
\begin{split}
\mathrm{V}_{\text{NLS}}\left(\mathbf{w}\right)&=\sum\limits_{n=1}^{N}\left(r_n-h(\mathbf{p}_n,\mathbf{w})\right)^2\\
&=\left(\mathbf{r}-\mathbf{h}\left(\mathbf{w}\right)\right)^{T}\left(\mathbf{r}-\mathbf{h}\left(\mathbf{w}\right)\right),
\end{split}
\end{equation}
where $\mathbf{r}=\left[r_1,\cdots,r_N\right]^{T}$ and $\mathbf{h}(\mathbf{w})=\left[h(\mathbf{p}_1,\mathbf{w}),\cdots,h(\mathbf{p}_N,\mathbf{w})\right]^{T}$ are $N$ dimensional vectors.
The solution of NLS estimator corresponds to the estimated location $\hat{\mathbf{w}}$ that minimizes the cost function \eqref{NLS}, i.e.,
\begin{equation}\label{NLS estimator}
\hat{\mathbf{w}}=\arg\min\limits_{\mathbf{w}}\mathrm{V}_{\text{NLS}}(\mathbf{w}).
\end{equation}
The NLS estimator does not rely on any assumption about the error statistics.
However, when the covariance of the error vector $\mathbf{e}=\left[e_1,\cdots,e_N\right]^{T}$ is available, we can obtain the WNLS estimator, which is defined as \cite{seco2009survey}
\begin{equation}\label{WNLS}
\begin{split}
\hat{\mathbf{w}}&=\arg\min\limits_{\mathbf{w}}\mathrm{V}_{\text{WNLS}}(\mathbf{w})\\
&=\arg\min\limits_{\mathbf{w}}\left(\mathbf{r}-\mathbf{h}\left(\mathbf{w}\right)\right)^{T}\mathbf{C}^{-1}(\mathbf{e})\left(\mathbf{r}-\mathbf{h}\left(\mathbf{w}\right)\right),
\end{split}
\end{equation}
where $\textbf{C}(\mathbf{e})=\mathbb{E}\left[\mathbf{e}\mathbf{e}^T\right]$ denotes the covariance of $\mathbf{e}$, and $\mathbb{E}\left[\cdot\right]$ represents the expectation operation.
Furthermore, when the error probability distribution $P_e(\mathbf{e})$ is known, the ML estimator can be used for location estimation \cite{chan2006exact,chang2004estimation}
\begin{equation}\label{ML}
\begin{split}
\hat{\mathbf{w}}&=\arg\min\limits_{\mathbf{w}}\mathrm{V}_{\text{ML}}(\mathbf{w})\\
&=\arg\min\limits_{\mathbf{w}}\log P_e(\mathbf{r}-\mathbf{h}(\mathbf{w})).
\end{split}
\end{equation}
Note that when the errors satisfy the zero-mean Gaussian distribution, the ML and WNLS estimators have the same performance.
In summary, the NLS estimator is simpler than WNLS or ML estimators and can be a practical choice if the noise information is unavailable, while when the error covariance matrix is available, WNLS can perform better than NLS, and the ML is optimal, since it can attain the the Cramer-Rao lower bound (CRLB) \cite{shen2010fundamental,shen2010fundamental2}, which is discussed in Section \ref{Performance Metrics}.

In general, there are two ways to solve the optimization problem in \eqref{NLS estimator}, \eqref{WNLS} and \eqref{ML}.
The first one is to perform a global exploration by using grid or random search techniques, such as genetic algorithm \cite{Marks2007Two-phase} or particle swarm optimization \cite{ChuangAnEffective}.
However, although such methods can achieve high localization accuracy, they are time-consuming and the global convergence may not be always guaranteed.
The other way is the iterative search algorithm, which requires a good initialization to avoid trapping at the undesired local minima.
There are commonly three iterative search schemes, namely, Newton-Raphson \cite{saab2016application}, Gauss-Newton \cite{bejar2010distributed}, and steep descent methods \cite{torrieri1984statistical}.
Such methods will start with the initial estimate $\hat{\mathbf{w}}^0$ and iterates until the $k$th iteration satisfying a certain criterion like
$\left\|\hat{\mathbf{w}}^{k}-\hat{\mathbf{w}}^{k-1}\right\|<\epsilon$, where $\epsilon$ is a sufficiently small positive constant \cite{torrieri1984statistical}.
In general, the Newton-Raphson and Gauss-Newton methods are more effective than the steepest descent method, while the steepest descent method is more stable than the former, since the inverse of Hessian matrix may not exist in the Netwon-Raphson or Gauss-Newton methods in some cases \cite{zekavat2011handbook}.

\subsubsection{Linear Estimators}
For geometric-based localization, since the function $h(\cdot)$ has a clear expression given by the algebraic relationships between ANs and the agent node, as discussed in Section \ref{basic localization techniques}, the corresponding localization problems may be also solved in closed-form through linear estimators.
The linear estimators mainly include the linear least squares (LLS) and weighted LLS (WLLS).
The aim of linearization is to covert the nonlinear equations in \eqref{general measurement model} into linear forms, based on which the location of the agent node can be estimated through ordinary least squares (LS) techniques.
Taking the 2D TOA-based measurement model in \eqref{TOA measurement model} as an example, by taking squares on both sides of \eqref{TOA measurement model}, we have
\begin{equation}\label{square TOA}
r_{\text{TOA},n}^2=\left\|\mathbf{p}_n-\mathbf{w}\right\|^2+e_{\text{TOA},n}^2+2e_{\text{TOA},n}\left\|\mathbf{p}_n-\mathbf{w}\right\|,
\end{equation}
where $n=1,2,\cdots,N$.
Let the error terms be $m_{\text{TOA},n}=e_{\text{TOA},n}^2+2e_{\text{TOA},n}\left\|\mathbf{p}_n-\mathbf{w}\right\|$, so \eqref{square TOA} can be further simplified into
\begin{equation}\label{TOA simplify}
\begin{array}{l}
r_{\text{TOA},n}^2=\left\|\mathbf{p}_n-\mathbf{w}\right\|^2+m_{\text{TOA},n}
\\=(w_x-p_{x,n})^2+(w_y-p_{y,n})^2+m_{\text{TOA},n}\\
=p_{x,n}^2+p_{y,n}^2+w_x^2+w_y^2-2w_xp_{x,n}-2w_yp_{y,n}+m_{\text{TOA},n}\\
=p_{x,n}^2+p_{y,n}^2+\xi-2w_xp_{x,n}-2w_yp_{y,n}+m_{\text{TOA},n}
\end{array}
\end{equation}
where $\xi=w_x^2+w_y^2$ is a dummy variable in the third simplification step.
According to \cite{seco2009survey}, with the assumption that the errors are relatively small, we can eliminate $m_{\text{TOA},n}$ and  linearize \eqref{TOA simplify} into the following compact form
\begin{equation}\label{Linearization I}
\mathbf{A}\boldsymbol \theta=\mathbf{b},
\end{equation}
where 
\begin{equation}\label{A}\nonumber
\mathbf{A}=
\left[\begin{array}{l l l}
-2p_{x,1} & -2p_{y,1} & 1\\
\vdots & \vdots & \vdots\\
-2p_{x,N} & -2p_{y,N} & 1
\end{array}\right],
\boldsymbol \theta=\left[w_x,w_y,\xi\right]^T,
\end{equation}
and
\begin{equation}\nonumber
\mathbf{b}=\left[
\begin{array}{l}
r_{\text{TOA},1}^2-p_{x,1}^2-p_{y,1}^2 \\
\vdots \\
r_{\text{TOA},N}^2-p_{x,N}^2-p_{y,N}^2
\end{array}
\right].
\end{equation}
Then the LS solution of \eqref{Linearization I} is found by
\begin{equation}\label{LLS I}
\hat{\boldsymbol\theta}=\left(\mathbf{A}^{T}\mathbf{A}\right)^{-1}\mathbf{A}^{T}\mathbf{b},
\end{equation}
where the first and second entries of $\hat{\boldsymbol \theta}$ is the estimated location of the agent node.
An alternative way for LLS TOA-based localization is proposed in \cite{Jr2000A} , which subtracts the first equation of \eqref{Linearization I} from the remaining equations. Assuming that the noise is sufficiently small, the linearization form of \eqref{TOA measurement model} can be obtained by \eqref{Linearization I} with
\begin{equation}\label{B}\nonumber
\mathbf{A}=
\left[\begin{array}{l l l}
-2(p_{x,2}-p_{x,1}) & -2(p_{y,2}-p_{y,1}) & 1\\
\vdots & \vdots & \vdots\\
-2(p_{x,N}-p_{x,1}) & -2(p_{y,N}-p_{y,1}) & 1
\end{array}\right],
\end{equation}
and
\begin{equation}\nonumber
\mathbf{b}=\left[
\begin{array}{l}
r_{\text{TOA},2}^2-r_{\text{TOA},1}^2+\left\|\mathbf{p}_1\right\|^2-\left\|\mathbf{p}_2\right\|^2 \\
\vdots \\
r_{\text{TOA},N}^2-r_{\text{TOA},1}^2+\left\|\mathbf{p}_1\right\|^2-\left\|\mathbf{p}_N\right\|^2
\end{array}
\right].
\end{equation}
Similar to TOA, the equivalent closed-form solutions for TDOA \cite{ChanAsimple1994,Fang1990Simple}, AOA\cite{Nardone1997AcLoSed,NGavish1992Performance} and RSS \cite{Chen2002Maximum} based localization can also be obtained.
Although the LLS estimators provide a closed-form solution, the solutions are sub-optimal in general, due to the discarding of information in the linearization process, and it only performs well when the noise is relatively small.
The WLLS estimator \cite{Navidi1998Statistical} is more generic than LLS scheme by making the use of the mean and covariance information of the measurement errors, which can provide higher localization accuracy.

A comparison of different location estimators, in terms of their advantages, drawbacks, and their corresponding measurement models, are provided in Table \ref{Comparison of Different Location Estimators}.
\begin{table*}[]
\centering
  \caption{Comparison of Different Location Estimators}\label{Comparison of Different Location Estimators}
\begin{tabular}{|l|l|l|l|}
\hline
 Estimators& Measurement Model&  Advantages& Disadvantages  \\ \hline
 NLS& \multirow{3}{*}{\begin{tabular}[c]{@{}l@{}}
                  \\
                  \\Geometric-based;\\
                  Fingerprinting;\\
                  Proximity
                  \end{tabular}}&
                  \begin{tabular}[c]{@{}l@{}}
                  High accuracy;\\
                  Error statistics is not required.
                  \end{tabular}&
                  \begin{tabular}[c]{@{}l@{}}
                  Global optimal solution cannot be guaranteed;\\
                  High complexity.
                  \end{tabular}\\ \cline{1-1} \cline{3-4}
 WNLS&                   &
                  \begin{tabular}[c]{@{}l@{}}
                  Higher accuracy than NLS.\\
                  \end{tabular}&
                  \begin{tabular}[c]{@{}l@{}}
                  Error covariance is needed;\\
                  Global optimal solution cannot be guaranteed;\\
                  High complexity.
                  \end{tabular}\\ \cline{1-1} \cline{3-4}
 ML&                   &
                  \begin{tabular}[c]{@{}l@{}}
                  Highest accuracy compared to other estimators;\\
                  Can achieve the theoretical CRLB.
                  \end{tabular}&
                  \begin{tabular}[c]{@{}l@{}}
                  Requires error probability distribution information;\\
                  Global optimal solution cannot be guaranteed;\\
                  High complexity.
                  \end{tabular} \\ \hline
 LLS& \multirow{2}{*}{Geometric-based} &
                  \begin{tabular}[c]{@{}l@{}}
                  Closed-form solution is guaranteed;\\
                  Computationally efficient;\\
                  Error statistics is not required.
                  \end{tabular}  &
                  \begin{tabular}[c]{@{}l@{}}
                  Low accuracy especially for clutter environments.
                  \end{tabular}  \\ \cline{1-1} \cline{3-4}
 WLLS&                   &
                  \begin{tabular}[c]{@{}l@{}}
                  Higher accuracy than LLS; \\
                  Computationally efficient.\\
                  \end{tabular}&
                  \begin{tabular}[c]{@{}l@{}}
                  Error statistics are needed;\\
                  May require iterative computation.
                  \end{tabular}\\ \hline
\end{tabular}
\end{table*}

\subsection{Main Sources of Error and Mitigation Techniques}
Localization performance is fundamentally limited by various estimation biases and measurement errors.
Therefore, it is important to analyze their source of error to improve the robustness of localization systems.
Here, we discuss three main sources of error, together with their corresponding mitigation techniques.

\subsubsection{Multi-path Fading}
Multi-path fading commonly exists in wireless channels, which can considerably degrade the localization performance.
In particular, for narrowband localization systems in clutter environments, the signals that arrive at the receiver via different paths are superimposed with each other, resulting them unresolvable at the receiver.
Moreover, the multi-path effect varies with signal propagation environments, making the signal detection more difficult.
To mitigate this effect, some diversity combining techniques are proposed, and for the ultrawide bandwith (UWB) systems, the multi-path components are usually resolvable temporally without resorting to complex algorithms \cite{lee2002ranging,denis2003impact}.
However, in harsh environments, the large number of mulitpath components still degrade the localization performance, especially for the geometric-based algorithms which need to distinguish the LoS path from the large number of NLoS paths to obtain the location information of the agent node.
Recently, a new line of research direction is the multi-path assisted localization by using the advanced tracking algorithms or by considering the signal reflectors as virtual transmitters to achieve high localization accuracy \cite{witrisal2016high,witrisal2016high2,gentner2016multipath}.

\subsubsection{NLoS Propagation}
The adversary impact of NLoS propagation lies in that the received NLoS signals weaken the correlation between signal measurements and link distance, since it will introduce a positive bias to the range estimate.
In general, there are three methods to cope with the NLoS condition.
The first method is based on the statistical information of the NLoS error.
By assuming a scattering model of the environment, the statistics of signal measurements can be obtained, and then the well-known techniques, like MAP or ML, can be used to mitigate the effects of NLoS errors.
However, the difficulty of such methods is to obtain an accurate model, which may change with terrain and/or the construction/demolition of buildings.
The second method uses all NLoS and LoS measurements with appropriate weights to minimize the effects of the NLoS contributions, where the weights are generated from the localization geometry and the ANs layout.
Although this method is effective even in the cases without LoS measurements, its solution is unreliable because the NLoS errors are always present.
The third method is to identify and discard those NLoS measurements, and perform localization only based on the LoS measurements.
In essence, the problem of NLoS identification in this method is converted into a statistical detection problem, where the NLoS and LoS conditions are considered as two hypotheses, and the goal of the problem is to figure out a metric to differentiate the NLoS and LoS hypotheses.
For instance, we can identify the NLoS path based on the statistics of range measurements.
Usually, the NLoS range measurements which are positively biased with non-Gaussian distribution tend to have a larger variance compared to the LoS counterpart with Gaussian distribution.
However, in some harsh environments, almost all measurements come from NLoS paths, so there are insufficient LoS measurements for localization.
For such cases, the localization methods using NLoS measurements and geometrical information are proposed \cite{zekavat2011handbook}.
In general, these NLoS localization techniques can be divided into two categories.
One is NLoS localization using signal measurements combining with the priori knowledge of the environment map.
The other is the localization using the measurements from scatters. Note that in the latter method, the NLoS measurements are first identified, and then the geometrical relationship among the ANs, the agent node and the scatters are used to locate the agent node.

\subsubsection{Systematic Error}
The systematic errors refer to the errors originated from the localization system itself, such as the imperfect signal measurements and radio miscalibration.
For instance, in time-based localization systems, the ANs are equipped with the oscillators for time synchronization.
However, the oscillators often experience independent frequency drifts, resulting in clock drift and offset, that may degrade the localization accuracy.
Systematic errors often bias the location estimators, making the mean of the estimator larger than the true value.
These errors are usually constant with respect to the targeting location and cannot be eliminated by averaging over a multiple repeated measurements.
Nevertheless, some techniques can effectively mitigate these errors.
For example, real-time infrastructure calibration can mitigate the localization performance degradation, where wireless links among ANs are made periodically to calibrate the parameters of the localization system.
Alternatively, some techniques like clock offset correction and recursive Bayesian approach have been proposed to tackle systematic errors \cite{gustafsson2010statistical,gunnarsson2014particle}.
\vspace{-5pt}
\subsection{Performance Metrics}\label{Performance Metrics}
The performance of localization systems can be evaluated from various aspects.
In this subsection, we outline the main performance evaluation metrics, including \emph{accuracy}, \emph{precision}, \emph{complexity}, \emph{coverage} and \emph{scalability}, while a performance comparison across several localization infrastructures is given in Section \ref{Localization Infrastructures}.

\subsubsection{Accuracy}
Accuracy (or location error) is usually measured as the Euclidean distance between the estimated location $\hat{\mathbf{w}}$ and the actual location $\mathbf{w}$ of the agent node, which is typically the most important performance metric to analyze the overall system performance.
In practice, various statistics can be adopted for this evaluation criterion, such as the mean square error (MSE) of the location estimates, which is defined as
\begin{equation}\label{MSE}
\begin{split}
e_{\text{MSE}}(\hat{\mathbf{w}})&=\mathbb{E}\left[\left\|\hat{\mathbf{w}}-\mathbf{w}\right\|^2\right]\\
&=\mathrm{tr}\left(\mathbf{C}\left(\hat{\mathbf{w}},\mathbf{w}\right)\right)
+\left\|\mathbb{E}\left(\hat{\mathbf{w}}\right)-\mathbf{w}\right\|^2,
\end{split}
\end{equation}
where $\mathrm{tr}(\cdot)$ indicates the matrix trace, and $\mathbf{C}\left(\hat{\mathbf{w}},\mathbf{w}\right)$ denotes the covariance matrix of $\hat{\mathbf{w}}$ and $\mathbf{w}$, which is defined as
\begin{equation}\label{COV}
\mathbf{C}\left(\hat{\mathbf{w}},\mathbf{w}\right)=\mathbb{E}\left[(\hat{\mathbf{w}}-\mathbf{w})(\hat{\mathbf{w}}-\mathbf{w})^{T}\right].
\end{equation}
The first and second terms of \eqref{MSE} represent the \emph{variance} and \emph{bias} of the estimated location, respectively.
Note that the bias is usually a constant unknown error introduced by the signal measurement, which can be mitigated through appropriate methods.
For unbiased cases, the CRLB gives the lower bound on the variance of $\hat{\mathbf{w}}$ \cite{shen2010fundamental2,shen2010fundamental}, i.e.,
\begin{equation}\label{CRLB}
\mathbf{C}\left(\hat{\mathbf{w}},\mathbf{w}\right) \succeq \mathbf{I}^{-1}\left(\mathbf{w}\right),
\end{equation}
where $\succeq$ denotes the matrix $\mathbf{C}\left(\hat{\mathbf{w}},\mathbf{w}\right)- \mathbf{I}^{-1}\left(\mathbf{w}\right)$ is positive semidefinite, and $\mathbf{I}(\mathbf{w})$ is the Fisher information matrix (FIM) of $\mathbf{w}$, given by \cite{guvenc2009survey}
\begin{equation}\label{FIM}
\mathbf{I}\left(\mathbf{w}\right)=-\mathrm{E}\left[\frac{\partial^2\ln P\left(\mathbf{r}\mid\mathbf{w}\right)}{\partial\mathbf{w}\partial\mathbf{w}^{T}}\right],
\end{equation}
where $P\left(\mathbf{r}\mid\mathbf{w}\right)$ denotes the conditional probability density function (PDF) of the measurement vector $\mathbf{r}$.
Taking the 2D TOA-based localization as an example, the measurement vector can be generated by \eqref{TOA measurement model}, which is a vector of the measured distances between the agent node and each of ANs, denoted by
\begin{equation}\label{distance vector}
\mathbf{r}=\hat{\mathbf{d}}=\left[\hat{d}_1,\hat{d}_2,\cdots,\hat{d}_N\right]^{T}.
\end{equation}
For convenience, here we consider the measurement errors $e_{\text{TOA},n}, n=1,\cdots,N$ that are zero-mean Gausssian distributed, and the conditional PDF of measured distance $\hat{\mathbf{d}}$ is
\begin{equation}\label{TOA PDF}
P\left(\hat{\mathbf{d}}\mid\mathbf{w}\right)=\prod_{n=1}^N \frac{1}{\sqrt{2\pi\sigma_n^2}}\exp\left(-\frac{\left(\hat{d}_n-d_n\right)^2}{2\sigma_n^2}\right),
\end{equation}
where $d_n, n=1,\cdots,N$ denote the actual distances between each AN and the agent node.
By substituting \eqref{TOA PDF} into \eqref{FIM}, the FIM can be calculated as \cite{chang2004estimation}
\begin{equation}\label{FIM compeleted}
\mathbf{I}\left(\mathbf{w}\right)=\left[
\begin{array}{l l}
\sum\limits_{i=1}^{N}\frac{\left(w_x-p_{x,i}\right)^2}{\sigma_i^2 d_i^2} &
\sum\limits_{i=1}^{N}\frac{\left(w_x-p_{x,i}\right)\left(w_y-p_{y,i}\right)}{\sigma_i^2 d_i^2} \\
\sum\limits_{i=1}^{N}\frac{\left(w_x-p_{x,i}\right)\left(w_y-p_{y,i}\right)}{\sigma_i^2 d_i^2} &
\sum\limits_{i=1}^{N}\frac{\left(w_y-p_{y,i}\right)^2}{\sigma_i^2 d_i^2}
\end{array}
\right].
\end{equation}
Then substituting \eqref{FIM compeleted} into \eqref{CRLB}, the CRLB for the TOA-based localization method can be obtained.
In a similar manner, the FIMs for TDOA, RSS, and AOA measurements can also be obtained.
The CRLB implies that the MSE of location estimates satisfies the following bound
\begin{equation}\label{MSE bound}
\begin{split}
e_{\text{MSE}}(\hat{\mathbf{w}})&=\mathbb{E}\left[\left\|\hat{\mathbf{w}}-\mathbf{w}\right\|^2\right]\\
&\ge\mathrm{tr}\left(\mathbf{C}\left(\hat{\mathbf{w}},\mathbf{w}\right)\right)
\ge\mathrm{tr}\left(\mathbf{I}^{-1}\left(\mathbf{w}\right)\right).
\end{split}
\end{equation}
Another useful evaluation criterion is the root mean square error (RMSE), which is the root of the MSE with the following bound \cite{gustafsson2005mobile}
\begin{equation}\label{RMSE}
\begin{split}
e_{\text{RMSE}}(\hat{\mathbf{w}})&=\sqrt{\mathbb{E}\left[\left\|\hat{\mathbf{w}}-\mathbf{w}\right\|^2\right]}\\
&\ge\sqrt{\mathrm{tr}\left(\mathrm{Cov}\left(\hat{\mathbf{w}}\right)\right)}
\ge\sqrt{\mathrm{tr}\left(\mathbf{I}^{-1}\left(\mathbf{w}\right)\right)}.
\end{split}
\end{equation}
The CRLB determines the attainable location accuracy of the unbiased system. However, many practical estimators are biased because of signal NLoS propagation and other factors, so in practice the system performance may not achieve the CRLB.
Other bounds like the Bayesian Cramer Rao bound \cite{dauwels2005computing}, Weiss-Weinstein bound \cite{weinstein1988general}, and extended Zik-Zakai bound \cite{zeira1994realizable} are tighter but require more complicated evaluations compared with CRLB.

\subsubsection{Precision}
Precision reveals the variation of location estimation with respect to the localization accuracy \cite{HuiSurvey}.
Specifically, precision measures the statistical characterization of the accuracy which varies over many localization trials.
In some works, the geometrical dilution of precision (GDOP) is also used to measure the variation of localization errors \cite{HuiSurvey}.
Taking the TOA-based localization as an example, the GDOP is defined as \cite{guvenc2009survey}
\begin{equation}\label{GDOP}
\text{GDOP}=\frac{e_\text{RMSE}(\hat{\mathbf{w}})}{e_\text{RMSE}(\hat{\mathbf{d}})},
\end{equation}
where the numerator and denominator are the RMSE of the location estimate and the range estimate, respectively.
The smaller GDOP value means the better performance on localization precision.
In addition, the GDOP also reveals the relation between the achievable localization accuracy and the geometry distribution of the ANs, which can be adopted as a criterion for ANs placement and selection to minimize the GDOP value.

Another evaluation metric is the localization error outage (LEO), which is defined as the probability when the localization error exceeds a certain threshold $e_\text{th}$ \cite{DardariIndoor}
\begin{equation}\label{LEO}
\text{LEO}(e_{\text{th}}) = \Pr\left\{\left\|\hat{\mathbf{w}}-\mathbf{w}\right\|\ge e_\text{th}\right\}.
\end{equation}
An equivalent expression of \eqref{LEO} is the cumulative distribution function (CDF) of the localization error defined by
\begin{equation}\label{CDF}
\text{CDF}(e_\text{th})=1-\text{LEO}(e_{\text{th}}),
\end{equation}
which denotes the success probability of location estimations with respect to a predefined accuracy.
In practice, the LEO or CDF reveals the probability of confidence in the location estimate.
When the accuracies of two localization algorithms are the same, the algorithm that gives lower LEO or higher CDF values has better precision \cite{DardariIndoor}.
For example, a localization system with $\text{CDF}(1.5)=0.9$ (a precision of 90\% within 1.5 m) performs better than that with $\text{CDF}(1.5)=0.5$ (a precision of 50\% within 1.5 m).

\subsubsection{Complexity}
The complexity of a localization system depends on the hardware, process of signal measurement, and computational complexity of the localization algorithm \cite{HuiSurvey}.
In general, it is difficult to analytically derive the complexity formula of different localization techniques.
Therefore, the computational complexity of the location estimators is usually treated as the complexity of the localization system.
There is always a trade-off between accuracy and complexity in the sense that more accurate localization usually requires the higher computational complexity.
On the other hand, the location update rate or \emph{latency} can be also used as a criterion to evaluate the system complexity, which reflects the time delay between two consecutive location updates for the same agent node and is very important for navigation.

\subsubsection{Coverage and Scalability}
In general, the localization performance degrades as the distance between each AN and the agent node increases.
The coverage refers to the maximum area where the localization system can provide effective localization services with guaranteed performance in terms of accuracy, precision, latency, and so on.
In general, the coverage can be roughly classified into global, local, and indoor coverage depending on different localization infrastructures.
One the other hand, the scalability reflects the adaptive capability of the localization system when the localization scope gets large \cite{HuiSurvey}.
As the localization coverage increases, the wireless channels may become congested and the localization system need to perform more signal measurement and calculation operations.

\subsection{Localization Infrastructures}\label{Localization Infrastructures}
There are two basic ways to deploy the localization systems.
The first one is to build a dedicated localization infrastructure, like the GNSS.
The second way is to reuse the existing wireless network infrastructures with the signals of opportunity (SoOP), like cellular networks, WLAN, etc., to provide wireless localization services, in addition to communication services.
For the first approach, the main advantage is that it can achieve high localization performance by using specific reference signal and professional hardware, while the drawback is the cost of the hardware and the limitation on the system scalability.
For the second approach, it avoids the expensive and time-consuming deployment of infrastructure, but such systems usually rely on sophisticated algorithms to improve the performance.
In this subsection, we mainly discuss the most popular localization infrastructures, including GNSS, cellular networks, WiFi, and UWB based localization systems, and a detailed comparison on their performance is given in Table \ref{Comparison of Different Localization Infrastructures}.
\begin{table*}[]
\centering
  \caption{Comparison of Different Localization Infrastructures}\label{Comparison of Different Localization Infrastructures}
\begin{tabular}{|l|l|l|l|l|}
\hline
                  &                   Example&  Techniques&  Type &Performance  \\ \hline
                  GNSS&                   GPS&  OW-TOA/TDOA& Trilateration&
                  \begin{tabular}[c]{@{}l@{}}
                  10-20 m accuracy;  \\
                  Global coverage;\\
                  About 30 s latency.
                  \end{tabular}
                    \\ \hline
\multirow{8}{*}{Cellular Networks} & \multirow{2}{*}{2G} &  CID + TA&
                  Proximity &\begin{tabular}[c]{@{}l@{}}
                  About 550 m accuracy;\\
                  Local area coverage;\\
                  Low latency.
                  \end{tabular}
                  \\ \cline{3-5}
                  &                   &
                  E-OTD& Trilateration&
                  \begin{tabular}[c]{@{}l@{}}
                  50-300 m accuracy;\\
                  Local area coverage;\\
                  Medium latency.
                  \end{tabular}
                  \\ \cline{2-5}
                  & \multirow{2}{*}{3G} &  OTDOA/UTDOA& Trilateration&
                  \begin{tabular}[c]{@{}l@{}}
                  50-200 m accuracy;\\
                  Local area coverage;\\
                  Medium latency.
                  \end{tabular}
                  \\ \cline{3-5}
                  &                   &  RFPM& Scene analysis &
                  \begin{tabular}[c]{@{}l@{}}
                  Over 50 m accuracy;\\
                  Local area coverage.\\
                  \end{tabular}   \\ \cline{3-5}
                  &                   &  A-GPS&  Trilateration& \begin{tabular}[c]{@{}l@{}}
                  10-50 m accuracy;\\
                  Global coverage;\\
                  High latency.\\
                  \end{tabular}   \\ \cline{2-5}
                  & \multirow{3}{*}{4G} &  E-CID&  Angulation + Proximity& \begin{tabular}[c]{@{}l@{}}
                  About 150 m accuracy;\\
                  Local area coverage;\\
                  Low latency.
                  \end{tabular}    \\ \cline{3-5}
                  &                   &  OTDOA/UTDOA&  Trilateration&  \begin{tabular}[c]{@{}l@{}}
                  25-200 m accuracy;\\
                  Local area coverage;\\
                  Medium latency.
                  \end{tabular}  \\ \cline{3-5}
                  &                   &  A-GNSS&  Trilateration& \begin{tabular}[c]{@{}l@{}}
                  Less than 10 m accuracy;\\
                  Global coverage;\\
                  High latency.
                  \end{tabular}   \\ \hline
                  WiFi&                   IEEE 802.11&  RFPM&  Scene analysis& \begin{tabular}[c]{@{}l@{}}
                  1-5 m accuracy;\\
                  50-100 m  coverage in general;\\
                  Medium latency.
                  \end{tabular}   \\ \hline
                  UWB&                    IEEE 802.15.4a&  TOA/TDOA&  Trilateration& \begin{tabular}[c]{@{}l@{}}
                  0.1-1 m accuracy;\\
                  Indoor area coverage;\\
                  Medium latency.
                  \end{tabular}   \\ \hline
\end{tabular}
\end{table*}

\subsubsection{GNSS}
Several countries in the world have already developed and launched various GNNSs, including the GPS by U.S., the Galileo by Europe, the Beidou by China, the GLONASS by Russia, as well as other regional systems like the Japanese Quasi-Zenith satellite system (QZSS) and the Indian regional navigational satellite system (IRNNS).
Although all of these GNNSs introduce different innovation at the system and signal levels, they share common theoretical and functional principles.
Taking the most popular GPS as an example, it is a self-localization system based on a man-made constellation of 27 earth-orbiting satellites.
The localization techniques behind GPS are the OW-TOA or TDOA methods, where at least four clearly visible satellites are required for the agent node to localize itself in terms of latitude, longitude, and altitude.
In general, the localization accuracy of GPS ranges from 10 m to 20 m \cite{gioia2015stand}, and the precision performance is highly dependent on the geometric distribution and selection of satellites, which can be measured by the GDOP value.
The stand-alone GPS suffers from the problem of time to first fix (TTFF) or \emph{cold start}; that is, when a GPS receiver is first turned on, it needs a long time (about 30 s or even a few minutes) to acquire the satellite signals.
To this end, an assisted-GPS (A-GPS) technique has been developed \cite{gioia2015stand,Duffett2005EGPS}, which uses a location server equipped with the GPS receiver that can simultaneously detect the satellites to help the user equipment (UE) to acquire GPS signals more quickly.
Although GPS can provide localization service globally, its performance is degraded in severe scattering environments, like indoor or urban areas.

\subsubsection{Cellular Networks}
Initially, cellular networks was designed for communications, and all UEs in the networks that need location services resorted to the GNSS.
However, due to the poor performance of the GNSS in urban and indoor environments, the cellular-based localization was proposed as a good complementary of GNSS to enhance its performance and robustness.
The first cellular-based localization system is the E-911 \cite{Rappaport1996Position} introduced by the Federal Communication Commission (FCC) of the U.S. to provide emergency services.
Moreover, since the location information within the cellular networks can be exploited for commercial services and network optimization, in the subsequent cellular networks standards from 2G to 4G, more advanced cellular-based localization techniques were proposed.
For cellular networks, both of the uplink and downlink between the BSs and UEs can be exploited for localization.
In 2G cellular standard, the localization schemes in global system
for mobile communications (GSM) and code-division multiple access (CDMA) network include CID, timing advance (TA), and enhanced observed time difference (E-OTD)
\cite{Drane1998Positioning,spirito2001experimental,lin2005mobile,borkowski2004enhanced},
which are mainly based on the uplink time-based measurements that required strict time synchronization, leading to coarse localization accuracy, ranging from 50 m to 550 m.
In 3G networks, the specific location measurement units (LMUs) were introduced in wideband CDMA (WCDMA) systems, which can be integrated with the BSs to improve the signal measurement performance.
The uplink localization methods in 3G networks are similar to those in 2G networks, where the enhanced CID (E-CID) and uplink TDOA (UTDOA) are the improved versions of CID and E-OTD, respectively \cite{del2017survey}.
Different from 2G localization, the localization method based on the downlink signals was specified in 3G networks, i.e. observed TDOA (OTDOA).
The OTDOA information is measured as the reference signal time difference (RSTD) by the UE, which then reports the RSTD to cellular networks to calculate its location.
Moreover, a RF pattern matching (RFPM) method was proposed in the 3GPP Release 10 \cite{ghosh2010lte}.
In addition, the A-GPS method is adopted by 3G networks, which can provide location service in GPS-denied areas.
In general, the localization accuracy of 3G networks ranges from 50 m to 200 m \cite{sadowski2014tdoa,del2014joint,driusso2015estimation}.
The location services in 4G long term evolution (LTE) was firstly defined in Release 9 \cite{Cherian2013LTE}.
Different from 2G or 3G networks, the 4G networks specify the LTE positioning protocol (LPP) \cite{TS36.355} to exchange information between UE and the remote location server.
Furthermore, a dedicated positioning reference signal (PRS) was introduced in LTE standards, which has high configurability in terms of power, time, and frequency allocation, and can improve localization performance \cite{EricssonWhitePaper}.
In the forthcoming 5G networks, the location information is becoming more critical for various application \cite{Liu2017Prospective}, including content prefetching, radio environment mapping, proactive radio resource management, etc..
All of the prospective applications require the localization services with higher accuracy.
Moreover, the availability, scalability, security, as well as privacy are also new challenges for localization systems \cite{Yassin2017Recent}.
Fortunately, the 5G communication systems enabled by higher carrier frequencies, wider signal bandwidths, denser networks, and Massive MIMO technologies can bring new opportunities for localization, which will be discussed in the Section III.

\subsubsection{WiFi}
WiFi technology can be used as a promising indoor localization scheme, thanks to its ubiquitous availability and handy for short-range RSS measurements.
It usually operates in two unlicensed bands, i.e. 2.4 GHz (IEEE 802.11b/g) and 5 GHz (IEEE 802.11a), with a range of 50-100 m in general \cite{he2015wi}, and has now increased to about 1 kilometer (km) in IEEE 802.11ah \cite{centenaro2016long}.
The main advantage of WiFi-based localization is its almost ubiquitous availability since most smart devices today are WiFi enabled.
However, WiFi signals transmit on the unlicensed industrial, scientific and medical (ISM) band which are vulnerable to interference \cite{he2015wi}.
The most commonly known WiFi-based localization approach is the RSS-based fingerprinting, while TOA-, TDOA- and AOA-based methods are relative less used since angular and time delay measurements are complex.
Usually, the accuracy of typical WiFi localization systems is approximately 1 to 5 m with a few seconds update rate \cite{liu2012push,liu2011wifi,davidson2016survey}.
Recently, the accuracy of WiFi-based localization systems has achieved the decimeter-level in the certain scenarios \cite{zou2014online,kotaru2015spotfi}.
There are many survey papers providing reviews about WiFi-based localization, e.g., \cite{he2015wi,vo2015survey}, and \cite{davidson2016survey}.
In \cite{he2015wi}, the authors provided a comprehensive overview about WiFi-based indoor localization techniques, while in \cite{vo2015survey} the authors focused on the outdoor fingerprinting-based localization with WiFi signals. In \cite{davidson2016survey}, the localization methods with the use of available measurements performed on smartphone are reviewed.

\subsubsection{UWB}
The UWB signal refers to the signal whose spectrum is lager than 500 MHZ in the frequency range from 3.1 GHZ to 10.6 GHz \cite{oppermann2005uwb,sahinoglu2008ultra}.
In general, the UWB spectrum can be acquired either by generating a series of extremely short duration pulses less than 1 nanosecond (ns) or by aggregating a number of narrowband subcarriers.
The first UWB standard is IEEE 802.15.4a, which was designed for low-rate wireless personal area networks (WPAN) in short range \cite{gezici2005localization}.
Different from conventional narrowband signals, UWB signals have higher temporal resolutions, lower transmission power consumption, and the ability to resolve multi-path and penetrate obstacles, which makes it quite promising to provide the centimeter-level localization services \cite{mekonnen2010constrained,alarifi2016ultra}.
The UWB technology can be implemented on self- or remote localization systems, and it can be incorporated with different localization techniques to improve the performance.
For fingerprinting localization, UWB fingerprinting enables the small ambiguity region even with a single AN, so that high localization accuracy can be achieved for both LoS and NLoS scenarios.
For geometric-based localization, due to the high temporal resolution of UWB signals, the time-based localization methods can achieve very high accuracy \cite{mekonnen2010constrained,gezici2005localization}.
However, UWB is not suitable for RSS- or AOA-based localization methods, since the RSS and AOA measurements cannot benefit from the huge bandwidth of UWB.
In general, the localization accuracy of UWB-based systems ranges from 0.1 m to 1 m, but the main drawback is its short coverage, which renders it only suitable for indoor environments \cite{alarifi2016ultra,mekonnen2010constrained}.

\subsection{Advanced Localization Techniques}
In the above subsections, we mainly focus on the static localization problems for one agent node.
In this subsection, we discuss some advanced techniques for localization, including \emph{tracking}, \emph{simultaneous localization and mapping} (SLAM), \emph{cooperative localization} and \emph{data fusion}.

\subsubsection{Tracking}
The problems of tracking can be viewed as a sequence of independent localization problems, but in the more general sense, besides location estimation, it further involves the estimation of velocity, acceleration, and all past states of the mobile agent node.
Compared with static localization, tracking entails mobility modelling to describe the agent node's movement.
Mathematically, the problems of tracking can be formulated as
\begin{subequations}\label{Tracking}
\begin{equation}\label{dynamic}
\mathbf{x}_t=\zeta(\mathbf{x}_{t-1})+\mathbf{z}_t
\end{equation}
\begin{equation}
\mathbf{y}_t=h(\mathbf{x}_t)+\mathbf{v}_t,
\end{equation}
\end{subequations}
where $\mathbf{x}_t$ denotes the global state of the agent node at time step $t$, including its location, velocity, acceleration, etc., while $\mathbf{y}_t$ is the measurement at step $t$; the function $\zeta(\cdot)$ models dynamics of the agent node, and the function $h(\cdot)$ is the measurement model; $\mathbf{z}_t$ and $\mathbf{v}_t$ represent the additive random noise of the state dynamics and measurement, respectively.
The goal of tracking is to obtain a time succession of the agent node's states $\mathbf{x}=\left\{\mathbf{x}_1,\cdots,\mathbf{x}_T\right\}$ from a set of measurements $\mathbf{y}=\left\{\mathbf{y}_1,\cdots,\mathbf{y}_T\right\}$.
Therefore, the tracking problem can be viewed as an estimation problem.
To operate in real-time, various filtering-based techniques have been proposed, such as the Bayesian filtering methodology, like Kalman filer (KF) \cite{gustafsson2010statistical}, extended KF (EKF), and particle filter (PF)\cite{gunnarsson2014particle}.
For more comprehensive review on tracking, readers may refer to \cite{DardariIndoor} and \cite{ChristosA}.

\subsubsection{SLAM}
Different from the localization problems, for SLAM, the locations of the ANs may not always be prior known, and the goal is to locate a set of fixed ANs and construct a map of the surrounding environment when a mobile node navigates through a predetermined path.
Typically, for the simplest SLAM problem, only one mobile node performs the environment surveying and locates the ANs, while for more sophisticated cases cooperative SLAM is involved.
Different from tracking, for SLAM, the navigation scheme of the mobile node may also affect its state, thus the dynamics equation in \eqref{dynamic} can be revised as \cite{DardariIndoor}
\begin{equation}\label{SLAM}
\mathbf{x}_t = \zeta(\mathbf{x}_{t-1})+\mathbf{u}_t+\mathbf{z}_t,
\end{equation}
where $\mathbf{u}_t$ is a control parameter used to guide the mobile node following the predetermined path.
Note that since the location of ANs are unknown, the state $\mathbf{x}_t$ should contain the mobile's state for the classical tracking as well as the ANs' locations.
In practice, the mathematical tools to solve the tracking problems are also suitable for the SLAM problem, especially the EKF which is widely used for SLAM.
More comprehensive overviews on this topic are given in \cite{Durrant2006Sim,Bailey2006Sim,cadena2016past}.
\subsubsection{Cooperative Localization}
In many application scenarios, due to the presence of NLoS propagation, some agent nodes are difficult to directly communicate with a sufficient number of ANs for localization, which may degrade the localization accuracy.
To this end, the cooperative localization techniques have been proposed \cite{patwari2005locating,mao2007wireless,wymeersch2009cooperative}, which can improve the localization performance, particulary in complex environments.
For noncooperative localization, all agent nodes need to communicate with ANs, thus a high density of ANs or long-range ANs transmission coverage is required.
Compared with noncooperative localization, cooperative localization allows the inter-communication among agent nodes, and the agent node can obtain information from both ANs and other agent nodes, so cooperative localization can not only improve accuracy but also extend the localization coverage.
The cooperative localization is also a parameters estimation problem, which can be solved through two kinds of methods.
One is the deterministic method, which includes the classical LS, multidimensional scaling, multilateration, and so on \cite{niculescu2001ad,shang2004localization,vivekanandan2007concentric}.
However, such methods rely on the assumption of a Gaussian model for all measurements uncertainties, which may not be effective in some practical scenarios.
The other category is the probabilistic method, which is known as \emph{belief propagation} (BP) \cite{ihler2005nonparametric,wymeersch2009cooperative}.
Such methods can not only obtain the location estimations but also measure the uncertainty of these estimates.
A detailed fundamental analysis about cooperative localization can be found in \cite{shen2010fundamental}, where the equivalent Fisher information (EFI) for cooperative networks was derived.
In \cite{wymeersch2009cooperative}, the authors discussed the main cooperative localization approaches from the perspective of estimation theory and factor graphs.

\subsubsection{Data Fusion}
Due to the multi-path effects, the performance of localization systems that only use a single type of measurements is severely limited in complex environments.
To this end, the study on fusing different types of information to improve the localization performance has gained a momentum \cite{gustafsson2010statistical,gunnarsson2014particle,radnosrati2015new,de2015mobile}.
So far, the wireless network has become heterogeneous with various wireless technologies, such as cellular network, WLAN, RFID, and Bluetooth.
Therefore, hybrid data fusion (HDF) has attracted research interest for unlocking the full potential of localization systems \cite{ChristosA}.
An example of data fusion localization is the \emph{SiRstarV} \cite{ChristosA}, which combines real-time data from GNSS satellites, WLAN, cellular, as well as multiple IMU sensors to improve the localization performance.
Numerical results show that its positioning error is within 9 m for 68\% cases and 13.1 m for 95\% cases over several tests \cite{bullock2012continuous}.
A generic framework of fusion technologies used for tracking is provided in \cite{radnosrati2015new}, where fusion exists across all stages of localization systems.
In the signal measurement stage, different signal measurements like TOA, TDOA, AOA, and RSS are combined.
In the position estimation stage, different localization techniques such as trilateration, triangulation, and fingerprinting are performed.
Finally, a temporal filter like EKF is applied to smooth the estimated user trajectory with the help of information gathered from multiple IMU sensors.

\section{Integrated Localization and Communication for 5G and Beyond}
The future mobile communication networks are expected to realize the vision of Internet of everything (IoE) with a versatile networks not only for ubiquitous communications, but also for seamless localization and intelligent automatic control with high accuracy \cite{saad2019vision,series2015imt}.
To fulfill this magnificent goal, it is necessary to develop high performance localization techniques, so as to not only meet the requirements of various emerging commercial and industrial on location-based services, but also to improve the communication performance in various aspects at different network layers.
In this section, we first overview the recent standardizations of 3GPP on localization technologies in 5G NR from Release 15 to Release 17, and discuss the developing trends of the localization systems.
Then we discuss the enabling technologies of 5G networks towards centimeter-level localization.
Since the accurate location information is beneficial for communications, we overview the location-aware communication techniques in different network layers.
After that, due to the correlation between communication and localization, we discuss the co-design of localization and communication systems and attempt to give some insights for future network design.
Finally, as a promising vision of 6G, we discuss the development trends of localization and communication techniques in future aerial-and-ground integrated networks.
\subsection{3GPP Standardizations for 5G Localization}
Compared with 4G-LTE, 5G networks enabled by higher carrier frequencies, wider bandwidth, and massive antenna arrays is expected to achieve enhanced localization performance in terms of accuracy, reliability, coverage and latency \cite{keating2019overview}.
In general, the localization accuracy of LTE is between 25 m and 200 m, while as reported by 3GPP, the localization systems in 5G networks need to achieve submeter-level or even centimeter-level accuracy with low latency.
In Release 15, a general description of location services (LCS) and the corresponding requirements are given in TS 22.071 \cite{TS22.071}.
Release 15 specifies the  CID and radio access technology (RAT)-independent positioning methods by reusing LPP \cite{TS36.355}, but the RAT-dependent positioning methods are excluded.
In Release 16, the recent new use cases of localization are identified in TR 22.872 \cite{TR22.872}, which mainly includes wearables, advertisement push, flow monitoring and control, as well as emergency call.
The accuracy requirements for such applications range from 1 m to 3 m in indoor scenarios, and below 50 m horizontal and 3 m vertical in outdoor scenarios, with less than 10 seconds (s) TTFF and 1 s latency.
Such requirements cannot be satisfied with the current cellular networks, e.g. LTE,
so a study item was concluded in March 2019 to investigate the NR positioning support, and the technical report is summarized in TR 38.855 of Release 16 \cite{TR38.855}.
In Release 16, the regulatory requirements on positioning are listed as follows:
\begin{itemize}
\item{Horizontal positioning error less than 50 m for 80\% of UEs.}
\item{Vertical positioning error less than 5 m for 80\% of UEs.}
\item{E2E latency and TTFF less than 30 s.}
\end{itemize}
These are regraded as the minimum performance targets for NR positioning studies.
Furthermore, for commercial use cases in indoor and outdoor scenarios, the localization technologies should meet the following requirements:
\begin{itemize}
\item{For indoor deployment scenarios, the horizontal and vertical positioning error less than 3 m for 80\% of UEs.}
\item{For outdoor deployment scenarios, the horizontal positioning error is less than 10 m, and vertical positioning error is less than 3 m for 80\% of UEs.}
\item{E2E latency is less than 1 s for both indoor and outdoor deployment scenarios.}
\end{itemize}
In order to fulfill the above requirements, Release 16 recommends the following RAT-dependent localization methods, including downlink TDOA (DL-TDOA), downlink AOD (DL-AOD), uplink TDOA (UL-TDOA), uplink AOA (UL-AOA), multi-cell RTT (Multi-RTT), and E-CID.
In addition, the combination of RAT-dependent and RAT-independent techniques like GNSS, Bluetooth, WLAN, and sensors are also considered for NR positioning.
The simulation results of these proposed NR positioning methods are presented in \cite{TR38.855}, which considers three different scenarios, i.e. urban macro (UMa), urban micro (UMi), and indoor office (InH).
The simulation results show that DL-TDOA can meet the regulatory requirements in all scenarios, while under some specific evaluation assumptions, some other techniques can meet the requirements of commercial performance \cite{keating2019overview}.
To satisfy the increasing requirements on localization accuracy resulting from new applications and industrial IoT (IIoT) use cases, the studies on NR positioning enhancements is ongoing in Release 17 (TS 38.857).
As the recommendation of TS 38.857, the NR positioning in Release 17 should meet the following exemplary performance targets:
\begin{itemize}
\item{For general commercial use cases (e.g. TS 22.261 \cite{TS36.355}), the submeter-level positioning accuracy should be guaranteed.}
\item{For IIoT use cases (e.g. TR 22.804 \cite{TR22.804}), positioning error should be less than 0.2 m.}
\item{The latency requirement is less than 100 ms generally, while for some IIoT use cases, the 10 ms latency is desired.}
\end{itemize}
Two main goals of TS 38.857 are: i) study enhancements to support high accuracy, low latency, network efficiency (or scalability), and device efficiency;
ii) study solutions to support integrity and reliability of assistance data and position information.
Furthermore, the enhancements on the combination of diverse positioning techniques and the flexibility of the networks are further emphasized.
To meet the requirements of the emerging autonomous applications, the future localization technology should not only guarantee the high accuracy, but also be able to determine the reliability and uncertainty or confidence level of the location-related data.

\subsection{Towards Centimeter Localization for 5G and Beyond}\label{5G Localization}
Massive MIMO, mmWave, UDNs and device-to-device (D2D) communication are four underlying technologies of 5G networks, which can not only improve the communication performance, but also potentially benefit for localization \cite{Liu2017Prospective}.
For massive MIMO, the BS equipped with a large antenna array can steer highly directional beams and provide high angular resolution, which can be utilized for angle-based localization \cite{heath2016overview}.
The mmWave technology can provide large bandwidth on the order of GHz, which offers high temporal resolution, thereby ensuring more accurate time-based localization \cite{lemic2016localization}.
In addition, since mmWave signals suffer from high path loss than the sub-6 GHz counterparts, the cell size in 5G networks will shrink, which renders UDN a promising technology for 5G networks.
The high density of BSs in UDN will increase the LoS probability, which may also improve the localization performance.
On the other hand, the promising D2D communication, where two neighboring devices directly communicate with each other while bypassing the BS, may flourish the device-centric cooperative localization.
\subsubsection{mmWave Massive MIMO Localization}
The main advantage of massive MIMO is its unprecedented potential of high spectral efficiency \cite{larsson2014massive,rusek2012scaling,lu2014overview}.
In massive MIMO systems, the BS equipped with a large number of antennas can serve a large number of UEs simultaneously with high data rates in the same frequency through dense spatial multiplexing.
This is achieved via beamforming/prcoding by large antenna arrays with high angular resolution, where the channels among different UEs are asymptotically orthogonal \cite{jungnickel2014role,larsson2014massive}.
In general, the large antenna array can be deployed as one-dimensional linear array or two-dimensional planar array, e.g. uniform linear arrays (ULAs) \cite{shahmansoori20155g} versus uniform rectangular arrays (URAs) \cite{abu2018error}.
For ULA with $N$ antenna elements, the beam can steer in 1D angles $\theta$, and the unit-norm array response vector is \cite{shahmansoori20155g}
\begin{equation}\label{ULA}
\textbf{a}(\theta)=\frac{1}{\sqrt{N}}\left[1,e^{j\frac{2\pi}{\lambda}d\sin(\theta)},
\cdots,e^{j(N-1)\frac{2\pi}{\lambda}d\sin(\theta)}\right]^{T},
\end{equation}
where $\lambda$ is the signal wavelength and $d$ is the distance between the adjacent antenna elements.
For URA with $N=N_x\times N_y$ antenna elements, the operation of beamforming lies in 2D angles referred to azimuth and elevation angles $(\theta,\phi)$, and the unit-norm array response vector is \cite{abu2018error}
\begin{equation}\label{URA}
\begin{array}{l}
\left[\textbf{a}(\theta,\phi)\right]_{n_x,n_y}=\frac{1}{\sqrt{N}}e^{j\frac{2\pi}{\lambda}d\sin(\phi)[(n_x-1)\cos(\theta)+(n_y-1)\sin(\theta)]},
\\
n_x\in\left\{1,\cdots,N_x\right\},n_y\in\left\{1,\cdots,N_y\right\}.
\end{array}
\end{equation}

Another promising technology of 5G networks is mmWave communication, which can achieve high data rate with low latency, due to its availability of the large bandwidth on the order of GHz \cite{sur201560}.
In particular, mmWave communication operates at a carrier frequency range from around 30 GHz to 300 GHz.
The mmWave communication at 60 GHz with bandwidth up to 7 GHz has been standardized for WPANs, e.g., IEEE 802.11ad \cite{perahia2010ieee} and IEEE 802.15.3c \cite{baykas2011ieee}.
In \cite{lemic2016localization}, the authors compared the raw resolution of time-based localization across different frequency bands, where the raw resolution is defined as the ratio of the the speed of light and the available bandwidth.
They revealed that mmWave signal specified in IEEE 802.11ad with bandwidth over 2 GHz can achieve raw resolution of roughly 15 cm, while that of the UWB signal with bandwidth over 500 MHz is only about 60 cm.
Compared with conventional location reference signals, which are usually transmitted on the ISM frequency bands, and suffer from severe interference and multi-path effects, mmWave signals can reduce the probability of interference and have the capability to resolve the multi-path components thanks to the very large bandwidth \cite{singh2011interference}.
Another important feature of mmWave transmission is the channel sparsity; that is, only a limited number of propagation paths can reach the receiver due to the short signal wavelength, which can be also exploited to enhance the localization performance \cite{deng2014mm,saloranta2016utilization}.
However, to overcome the high path loss associated with mmWave signals due to the short wavelength \cite{qingling2006rain}, mmWave transmission is usually combined with massive MIMO for directional beamforming, for which the accurate angular information can be extracted and utilized for localization.
Therefore, mmWave massive MIMO-based localization may significantly outperform the conventional angle-based localization at lower frequencies, which suffer from rich scattering and poor multi-path separability \cite{shahmansoori20155g}.
\begin{figure} 
\centering
\subfigure[Sub-6 GHz system with omnidirectional antennas]{
\label{mmWaveA}
\begin{overpic}[width=0.35\textwidth]{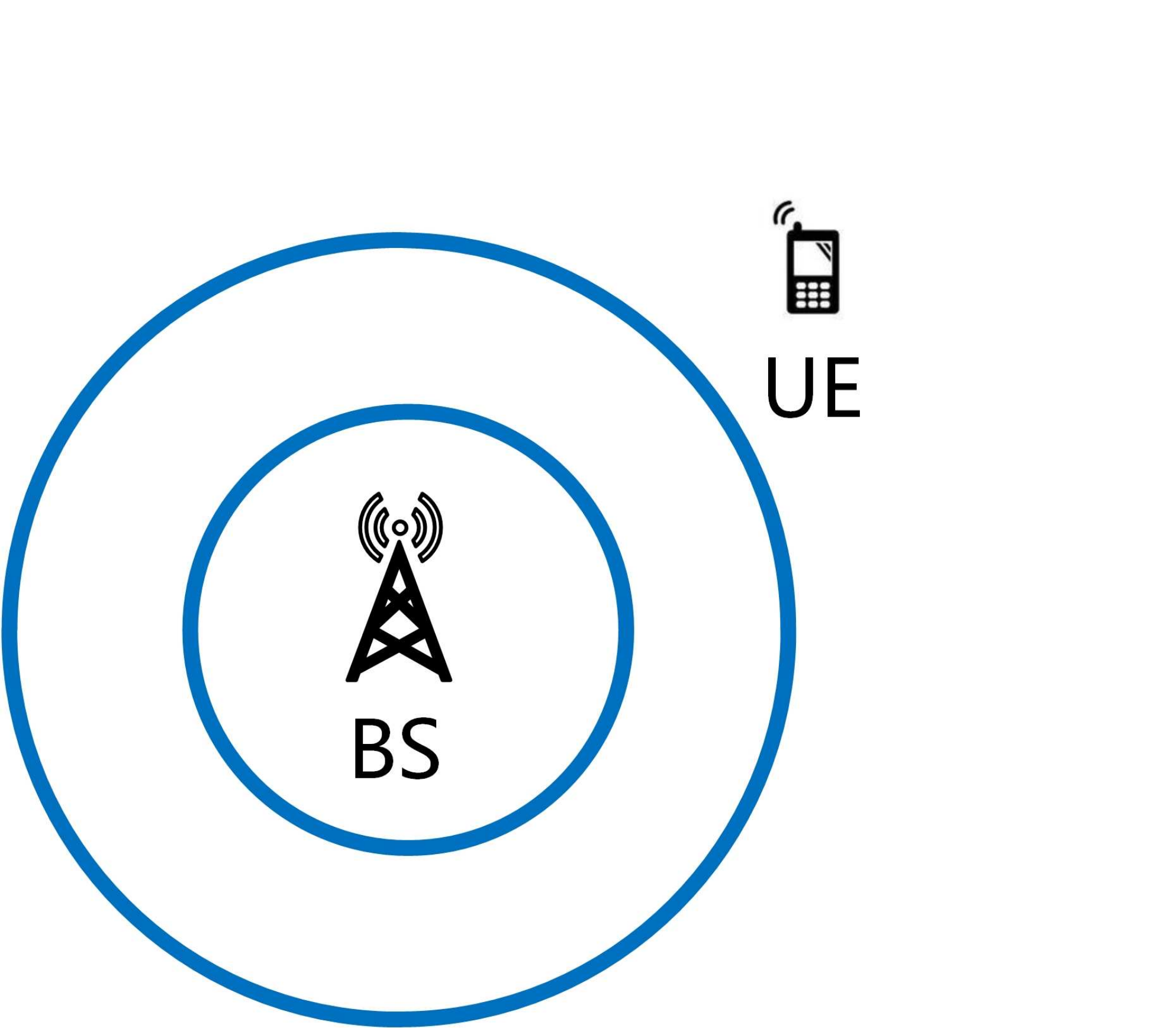}
\end{overpic}
}\quad\quad
\subfigure[mmWave system with directional antennas]{
\label{mmWaveB}
\begin{overpic}[width=0.35\textwidth]{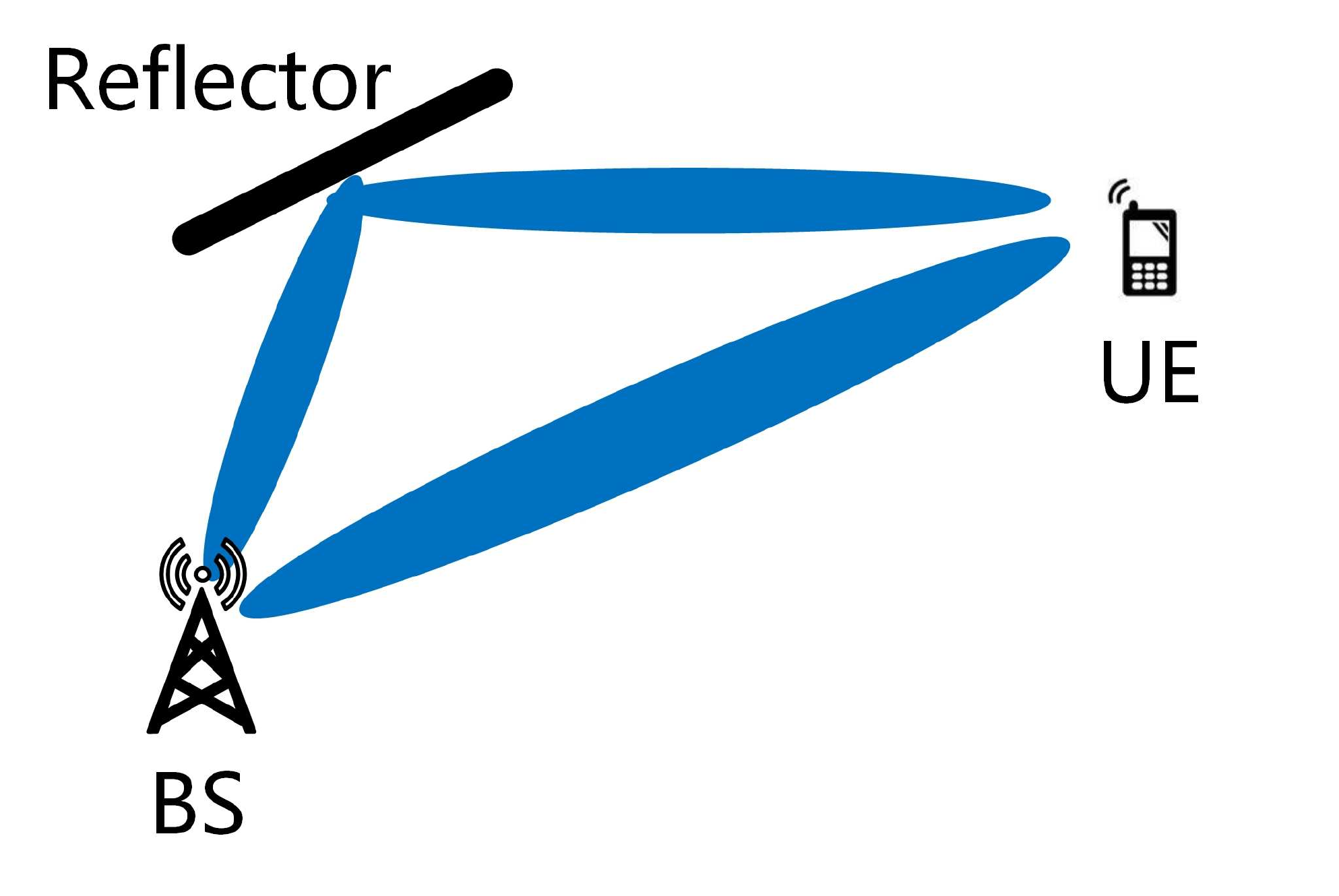}
\end{overpic}
}
\caption{Signal propagation in sub-6 GHz and mmWave systems. (a) For sub-6 GHz system with omnidirectional antennas, signals suffer from rich scattering and poor multi-path separation capability. (b) For mmWave system with directional antennas, signals will arrive at the receiver with very few paths due to the high carrier frequency and high antenna direction. }\label{mmWave}\vspace{-10pt}
\end{figure}
As illustrated in Fig. \ref{mmWave}, compared with traditional sub-6 GHz system,
the limited scattering and high directivity are unique characteristic of mmWave massive MIMO communications, which bring new opportunities for localization.
In the following contents, we discuss the mmWave massive MIMO localization in terms of channel model, parameter estimation, and localization approaches.

\paragraph{Channel Model}
For convenience, we consider a MIMO system equipped with ULAs with $N_t$ antennas at the BS and $N_r$ antennas at the UE.
The corresponding MIMO channel can be modelled as \cite{zeng2016millimeter}
\begin{equation}\label{MIMO}
\textbf{H}(t)=\sum\limits_{l=1}^{L}\alpha_l\textbf{a}(\theta_r,l)\textbf{a}(\theta_t,l)^{{H}}\delta(t-\tau_l),
\end{equation}
where $L$ denotes the total number of multi-paths, which is usually small due to the multi-path sparsity in mmWave regime \cite{zeng2016millimeter}; $\alpha_l$ and $\tau_l$ represent the complex path gain and the delay of the $l$th path, respectively; $\mathbf{a}(\theta_{r,l})$ and $\mathbf{a}(\theta_{t,l})$ are the antenna array response vectors for angles $\theta_{r,l}$ and $\theta_{t,l}$ seen by the UE and BS, respectively.
The sparsity of mmWave channel can simplify the estimation of channel parameters \cite{deng2014mm}.
This method uses a virtual representation for \eqref{MIMO} by sampling in time, frequency, and/or space with the aid of the discrete fourier transform (DFT).
For instance, the beamspace MIMO channel can be represented as
\begin{equation}\label{sparsity}
\mathbf{H}_b(t)=\mathbf{U}_r^{H}\mathbf{H}(t)\mathbf{U}_t,
\end{equation}
where $\mathbf{H}_b(t)$ is the beamspace channel matrix, which is a linear equivalent representation of the antenna domain channel matrix $\mathbf{H}(t)$, $\mathbf{U}_r$ and $\mathbf{U}_t$ are the DFT matrices whose columns are response and steering vectors, respectively, which are orthogonal.
Estimating parameters from $\mathbf{H}_b(t)$ is much easier than the parameter estimation based on $\mathbf{H}(t)$, since the latter determines the parameters in a nonlinear manner.

\paragraph{Parameter Estimation}
For MIMO systems, the source localization involves the AOAs and AODs estimation, which can be classified into localization for point sources and distributed sources \cite{lee1997distributed,astely1999effects}.
For point source localization, the signal of each source emitted from a single AOA or the AOAs of different sources are distinguishable.
On the other hand, for the distributed sources localization, the signal of each source is emitted from an angular region.
In general, point sources usually correspond to the LoS propagation scenarios \cite{lee1997distributed}, while distributed sources are commonly used for the multi-path scenarios \cite{astely1999effects}.
For distributed sources, they can be further classified into coherently distributed (CD) sources \cite{wan2015doa} and incoherently distributed (ID) sources \cite{hu2014esprit}, depending on whether they are slowly time-varying channels or rapidly time-varying channels.

In general, the parameters for estimation include the azimuth and elevation angles of signal departure and arrival, signal propagation delay, Doppler shift, and the corresponding uncertainty of these parameters.
The channel parameter estimation approaches can be categorized based on the source type.
For point sources, the authors in \cite{wen2019survey} categorized the channel parameter estimation methods into subspace-based and compressive sensing methods.
The subspace-based methods treat the parameters into tensors, and use the tensor decomposition method to reduce the dimensionality.
Such methods can achieve a good balance between estimation accuracy and computational complexity \cite{haardt2014subspace}.
Compressive sensing is a promising method to recover the sparse signals, which is particularly suitable for mmWave massive MIMO system due to its sparsity in angular and delay domain \cite{tsai2018efficient}.
For distributed sources, the parameter estimation can be classified into CD sources and ID sources estimation.
The classical estimation approaches for point sources have been successfully applied into CD sources estimation \cite{valaee1995parametric,shahbazpanahi2001distributed,lee2003low}, and the multiple signal classification (MUSIC) algorithms were employed to estimate the AOA of CD sources \cite{wan2015doa}.
However, the parameter estimation for ID sources is more complicated.
These methods can be generally divided into parametric methods and non-parametric methods.
The parametric approaches include ML \cite{trump1996estimation}, covariance matching \cite{ottersten1998covariance} and pseudo-subspace \cite{zoubir2008efficient}.
The ML estimation is optimal \cite{trump1996estimation}, but it suffers from high complexity.
The LS estimator are proposed by using the covariance matrix matching techniques to reduce the computational complexity \cite{ottersten1998covariance}.
The non-parametric approaches like beamforming approaches \cite{tapio2002direction} have lower the computational complexity but with compromised performance compared with parametric approaches.

\paragraph{Localization Approaches}
The conventional localization approaches can be also used in mmWave MIMO systems for improving localization performance.
In \cite{yang20193} and \cite{yang20193D}, a 3D localization method was proposed by using hybrid signal measurements in mmWave systems, which can joint estimate the position and velocity of the UEs and construct the environment maps.
In \cite{vari2014mmwaves}, the method based on RSS measurements of mmWave signals was proposed, which can achieve localization accuracy of one meter.
The mmWave-based object tracking was proposed in \cite{wei2015mtrack}, where the RSS and signal phase were used as the features for tracking an object.
In \cite{lemic2016localization}, a set of feasible localization approaches in mmWave bands were discussed, and the results demonstrated that mmWave localization can achieve decimeter level localization accuracy.
In \cite{shahmansoori20155g} and \cite{shahmansoori2017position}, the methods for estimating the object's position and orientation by using mmWave MIMO were proposed, where the CRLB on position and rotation angle estimation was derived.
Moreover, due to the highly directional narrow beamforming of mmWave signals, the beam training protocols that has been standardized in IEEE 802.11ad \cite{perahia2010ieee} can be used for improving localization.
With beam training protocol, the strongest signals are selected for AOA, AOD, and TOA estimation.
Furthermore, the mmWave localization can turn multi-path from a foe into a friend \cite{witrisal2016high}.
By considering the signal reflectors as virtual transmitters \cite{witrisal2016high,witrisal2016high2}, the high accurate localization is possible even without the LoS link.
A hybrid localization approach for massive MIMO systems combining with TDOA, AOA, and AOD measurements was proposed in \cite{li2007position}.
In \cite{savic2015fingerprinting}, a fingerprinting-based localization method in massive MIMO systems was proposed, where the uplink RSS measurements were used as the fingerprints.
In \cite{garcia2017direct}, the direct source localization (DiSouL) was proposed by jointly processing the observations at the distributed massive MIMO BSs.
In \cite{koivisto2017joint} and \cite{koivisto2017high}, the combination of TDOA and AOA measurements by using the EKF was proposed.
A comprehensive discussion on massive MIMO localization is given in \cite{wen2019survey}.
An overview about the 5G mmWave localization for vehicular networks is given in \cite{wymeersch20175g}.

\subsubsection{D2D Communication and Cooperative Localization}
To meet the ever-increasing throughput demand of various applications on mobile devices, one promising method is to shift the current cell-centric architecture to device-centric architecture \cite{boccardi2014five}.
In traditional cellular networks, all communications must go through the BSs even if both communication entities are very close with each other, while D2D communication enables that two devices can communicate without traversing the BSs.
The related use cases include multi-hop relaying \cite{lin2000multihop}, peer-to-peer (P2P) communication \cite{lei2012operator}, machine-to-machine (M2M) communication \cite{pratas2013low}, cellular offloading \cite{bao2010dataspotting}, and so on.
A detailed survey of D2D communication is given in \cite{asadi2014survey}.
In general, D2D communication combining with densely deployed small cells can achieve high data rate with high spectral efficiency and low latency, which paves the way for cooperative localization.
In this subsection, we outline the localization schemes based on different types of D2D communications in cellular networks.

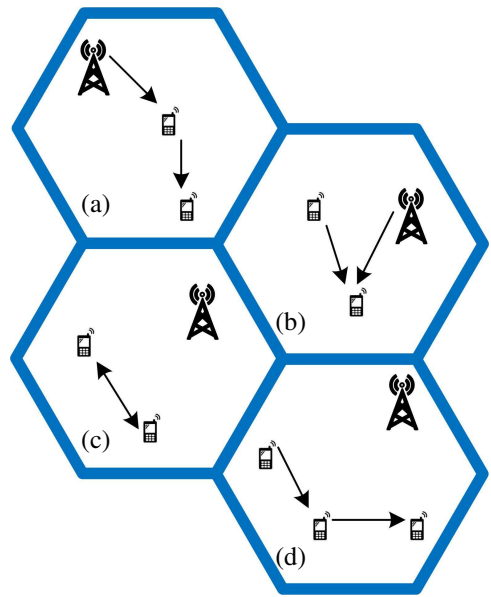
\begin{figure}
\centering
\begin{overpic}[width=0.35\textwidth]{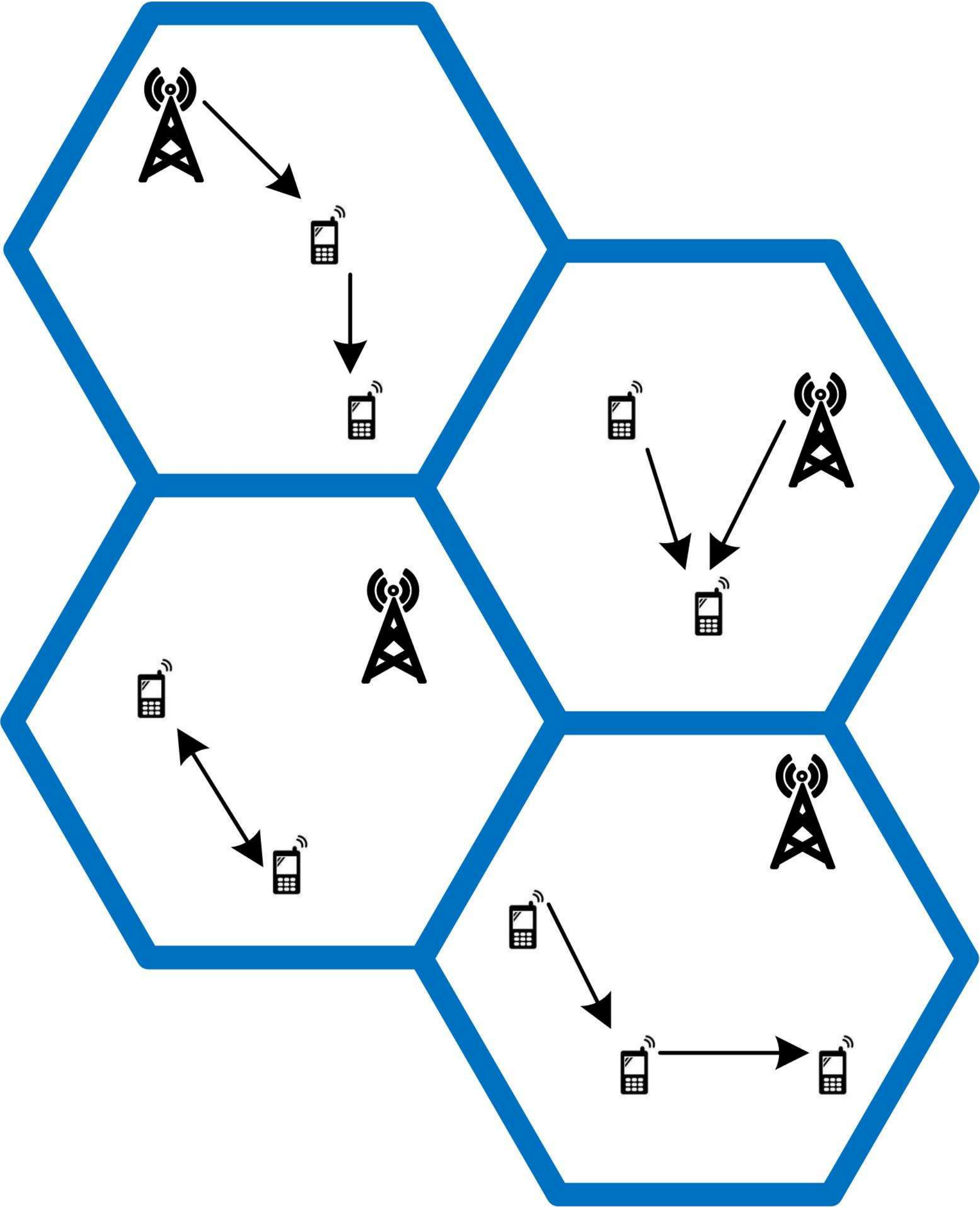}
\put(12,65){(a)}
\put(45,45){(b)}
\put(12,25){(c)}
\put(45,5){(d)}
\end{overpic}
  \caption{An illustration of the device-centric networks and four types of D2D communication. (a) Device relaying from BS. (b) BS-aided D2D communication. (c) Direct D2D communication. (d) Device relaying from other device.}\label{DeviceCentric}\vspace{-12pt}
\end{figure}

Depending on the role that the BSs played in different communication schemes, D2D communication can be classified into four categories \cite{tehrani2014device}, namely, \emph{device relaying from BS}, \emph{BS-aided D2D communication}, \emph{direct D2D communication}, and \emph{device relaying from other device}, as illustrated in Fig. \ref{DeviceCentric}.
For device relaying to BS, a device located at the edge of a cell can only receive weak signals transmitted from the BS, which is assisted by other devices via relaying.
In this case, device localization for the cell edge users can be achieved by utilizing the assisting devices as the pseudo BSs with known location, and the signal measurements among these devices can be used for localization.
However, the location information of the assisting devices might be inaccurate, which will degrade the localization accuracy of the target device.
Therefore, how to mitigate the effect of error propagation is a critical issue.
For BS-aided D2D communication, two devices communicate with each other via their direct link, together with the assisted information provided by the BS.
In this case, a device can measure the signals from both the BS and the other device, which can be utilized for localization.
Therefore, the assisting device act like an additional AN to provide extra information to improve the localization performance.
By contrast, in the architecture of direct D2D communication or device relaying to other device, two devices communicate with each other directly or indirectly via information relaying by other devices without traversing the BS.
In such architectures, for the target device, the relative distance and/or angles to other devices can be obtained, and if the locations of other devices are known, cooperative localization can be achieved.
For the multi-hop based cooperative localization, the key problem is to perform location estimation using multiple signal measurements from different devices.
Based on the spectrum usage, D2D communication can be classified into \emph{inband} D2D communication, which reuses the cellular spectrum for D2D data transmission, and \emph{outband} D2D communication, which usually uses unlicensed spectrum.
From localization perspective, the main advantage of inband D2D lies in the better spectrum utilization, which avoids the consumption of extra hardware for other unlicensed spectrum, while the main drawback is the potential high interference, which may severely degrade the localization performance.
On the other hand, the outband D2D communication transmit signals by using different wireless technologies, and the location-related information are measured from heterogeneous networks.
In this case, how to coordinate the signal transmission over different bands and data fusion for improving localization accuracy is a key problem.
Furthermore, for all cooperative localization schemes, it is critical to protect the location privacy of the users and avoid the significant performance degradation simultaneously.
\subsubsection{Localization in UDNs}
Network densification is a promising technology to meet the ever-increasing demands on area coverage and capacity enhancement in 6G mobile networks.
In the extreme case, we would have UDNs, which refers to such networks with more cells than users \cite{baldemair2015ultra,kamel2016ultra,ding2015will}.
In \cite{ding2015will}, An quantitative definition of the UDN was given, where a network can be considered as ultra-dense if the density of cells is no smaller than $10^3$ cells/$\text{km}^2$.
In general, the densely deployed cells in UDNs can be classified into full-functioning BSs (picocells and femtocells) and macro-extension APs (relays and remote radio heads) \cite{kamel2016ultra}.
The coverage area of all these cells are typically small, ranging from a few meters up to 100 meters.
The fundamental difference of UDNs from traditional networks lies in that it not only enables higher data rates with less energy consumption for communication, but also brings new opportunities for localization.
First, since more small cells are in the close vicinity of users, the localization error of CID-based methods can be reduced.
Furthermore, as the cell shrinks in UDNs, the probability of LoS link increases, which is beneficial for accurate location-related measurements \cite{koivisto2017high}.
In \cite{werner2015joint}, the authors proposed a joint localization method combining TOA and DOA measurements, as well as a real-time UE tracking method by using an EKF.
The simulation results showed that the localization accuracy is below 1 m for 95\% of the case with the signal bandwidth below 10 MHz using one or two base stations in the 5G UDN.
In \cite{koivisto2018joint}, a joint positioning and synchronization method based on centimeter wave dense 5G network was proposed, which is able to estimate the clock offsets in addition to the UE's location.

\subsection{Location-aware Communication}\label{Location-aware communication}
Accurate location information can be used not only for location-based services, like location-based advertising, autonomous driving, and so on, but also for improving the communication performance, which is known as \emph{location-aware communication} or \emph{location-aided communication}.
Location-aware communication has received fast growing attentions in recent years, where the location information can be utilized in a variety of ways to improve the communication performance in 5G-and-beyond networks, like reducing the communication overheads and delays, minimizing the energy consumption, and increasing the communication capacity.
The utilization of location information in cognitive wireless networks was studied in \cite{celebi2007utilization}, where the location-assisted network optimization was discussed, including location-assisted dynamic spectrum management, handover, as well as network planning and expansion.
In \cite{slock2012location} and \cite{dammann2013where2}, the potential of location-aware communication in multi-user and multi-cell systems was discussed, with special emphasis on utilizing location information for resource allocation.
In \cite{di2014location}, a comprehensive survey about the location-aware communication was given based on the layers of the protocol stack, including the physical, MAC, and network layers.

\subsubsection{The Physical Layer}
In the physical layer, the location information is usually used for channel estimation, beamforming, and generating the radio environment maps (REMs) to reduce the interference and signalling overhead.
For location-aided beamforming, the accurate AOD and AOA of the LoS path can be exploited to design the beam vectors of the transmitter and receiver, respectively.
Compared with the conventional beamforming methods based on full-band channel state information (CSI), the location-based beamforming schemes do not require the full-band reference signals, and the narrowband pilots are suffice \cite{koivisto2017high}, which can reduce the energy and resource consumption especially for the UE.
In \cite{kela2016location}, a beamforming method was proposed based on the AOAs or AODs of the LoS path between ANs and the UE by using the EKF to estimate and track the location of the UE.
The results showed that if the LoS path is available, with the angular error below 2 degree, the location-based beamforming scheme outperforms conventional CSI-based schemes in terms of mean user-throughput and time-frequency resources efficiency.
Another major problem of beamforming with large antenna arrays is the huge training overhead, which can be reduced by utilizing accurate location information of the transmitter and receiver \cite{garcia2016location,abdelreheem2016millimeter,abdelreheem2017location}.
Location-aided beamforming method for mmWave vehicular communication has been studied in \cite{garcia2016location}, which can reduce channel estimation overhead and speed up the initial access.
In \cite{abdelreheem2016millimeter} and \cite{abdelreheem2017location}, a compressive sensing based mmWave beamforming method was proposed, where the location information of the transmitter/receiver was used to design the sensing matrix.
Currently, the accuracy of location aided for beamforming is within meter-level, while the more accurate location information of the UE can further reduce the overhead of beamforming.

The high-precision REM \cite{yilmaz2013radio,Bi2019Engineering,zeng2020simultaneous} can be constructed by using the accurate location information combined with channel quality metric (CQM) \cite{di2014location}.
The main difference between the REM and geolocation database is that REM contains the radio elements like the knowledge of large-scale fading and location-based radio condition, which can be utilized for radio resource management (RRM) without the CSI between the BSs and the UE \cite{koivisto2017high}.
The REM construction entails a training process, where a measurement center collects the CQM from different locations of the area and performs a learning algorithm to obtain a number of radio scene parameters, like the path loss exponent and shadowing variance, tagged with the actual locations \cite{di2014location}.
In general, the number of sampling points in the area of interest, the location accuracies of these points, the dynamics of the propagation environment, and the accuracy of the propagation modelling can significantly affect the REM performance.
To compare the estimated REM with the true REM, some quality metrics are proposed, including RMSE \cite{ureten2012comparison}, correct detection zone ratio (CDZR) and false alarm zone ratio \cite{yilmaz2015location}.
In general, a REM which can predict the large-scale channel fading in the area of interest needs at least meter-level localization accuracy.
The more accurate location resolution leads to more accurate REM.
However, since the REM construction must perform online, there is a trade-off between complexity and accuracy.
An accurate REM can be used for RRM at different network layers to reduce the overhead and delay of communications.
In the physical layer, one of the best known applications is the REM-based spatial spectrum access scheme in cognitive radio (CR) \cite{nevat2012location}, where before the CR devices initiate a communication, they first query the REM database for the available frequencies depending on their locations, and select the frequency bands from a set of received unoccupied frequencies.
Furthermore, with the additional motion information of the UE, like velocity and orientation, we can predict the locations of the UE.
Combining the REM and the predicted UE locations, the proactive RRM can be achieved, which can not only maintain the communication performance at a certain time but also adapt to the upcoming events.
For example, the authors in \cite{sand2009position} outlined the adaptive mobile communication for the predicted capacity.
Moreover, if the predicted location of the UE and the accuracy of the REM are guaranteed, the proactive RRM can enable a high robustness and low latency E2E communications without the instantaneous CSI between the BS and UE \cite{hakkarainen2015high}.
Other applications of location-aided communication applications in the physical layer have also been studied, such as location-aided MIMO interference channels and coordinated multipoint (CoMP) transmission \cite{irmer2011coordinated}.

\subsubsection{The MAC Layer}
In the MAC layer, the accurate location information is beneficial for MAC layer RRM, like location-based multicasting and broadcasting, scheduling, and load balancing.
Compared with the physical layer, the location accuracy requirements on the MAC layer is lower, usually range from several meters to tens of meter.
The geographical location aided broadcasting is referred to as \emph{geocasting}.
For mobile ad hoc networks (MANET), broadcasting is an important method to quickly deliver message to the specific set of nodes, especially when the route to the destinations is still unknown.
The traditional broadcasting methods based on flooding suffer from large bandwidth consumption and the broadcast storm problem (BSP) \cite{ko2000location}.
In \cite{sun2001location}, the authors proposed a broadcasting protocol based on location-related information in terms of distance and angle between transmitter and receiver, which can achieve high reachability and bandwidth efficiency.
Recently, in \cite{yuan2017location}, the authors proposed a location aided probabilistic broadcast (LAPB) algorithm for MANET routing, which selects the more effective nodes, according to an adaptive probability based on location information, to broadcast route request.
The simulation results showed that the LAPB method is able to significantly reduce the overhead and alleviate the BSP in MANETs.
Another essential transmission scheme of ad hoc network is multicasting, which can also benefit from the accurate location information, and some location-based multicast routing protocols have been proposed in \cite{ko1999geocasting,shih2008location,zhao2010cooperative}.

Another well-known application is the location-aided radio resources scheduling, where the same resources can be shared by two different communication links if the inter-interference level is below a certain threshold.
In \cite{dammann2013where2}, the authors proposed a location-aided round robin scheduling algorithm to solve the problem introduced by fractional frequency reuse that the fractional bandwidth can meet the heavy traffic demand during rush hours, where some of the cell-central and cell-edge users are selected to share the same frequency band with minimum intra-cell interference.
Compared to alternative methods that select users based on the instantaneous channel knowledge requiring all users to feedback CSIs, the location-based scheme requires less feedback information and can achieve higher total throughput than the conventional methods.
In \cite{jiang2018passive}, a passive location resource scheduling scheme based on an improved genetic algorithm was proposed.
In \cite{girgis2016proactive}, the authors proposed proactive scheduling schemes for delay-constrained traffic, where the current user location and the priori statistics were used to predict the request arrival time slots, which can significantly reduce the transmission energy compared to the reactive methods.
For the WSNs, the energy efficiency is a critical issue due to the limited lifetime of the sensors' batteries, so the authors in \cite{zhu2014collaborative} proposed the location-based sleep scheduling schemes, which can dynamically schedule the awake or asleep status of the sensors to reduce the total energy consumption.

The future wireless networks are expected to be highly heterogenous and dense, where different types of APs, like WiFi, bluetooth, and light fidelity (LiFi), coexist in the small areas, leading to the problems of unbalanced traffic loads and inefficient resource utilization \cite{al2012survey}.
The load balancing technique is critical to solve such problems\cite{cybenko1989dynamic}, which can offload some traffic load from busy APs to idle APs.
For example, in \cite{wang2017optimization,wang2015dynamic,ma2019location}, the authors considered the utilization of load balancing techniques to optimize the  handover overhead and throughput of the LiFi and WiFi integrated hybrid networks.
The location information can help the dynamic load balancing, which is an adaptive scheme depending on the UE distribution within a cell \cite{yanmaz2005location}.
Since the distribution of UEs can be obtained based on the accurate locations of UEs, the location information can help the network APs to allocate resources more efficiently.
For instance,
in \cite{ma2019location}, the authors proposed a location-aided  load balancing scheme for hybrid LiFi and WiFi networks, which aimed to maximize the system throughput.

\subsubsection{The Network Layer}
In the network layer, location information can be used in various aspects, like geographic routing, location-aware content deliver, etc., to improve scalability and reduce network overhead and latency.
The geographic routing is usually referred to as the \emph{georouting}, which utilizes the geographic location information of the nodes in the networks to transmit the data packets to their destination \cite{di2014location}.
Different from the topology-dependent routing schemes, the georouting protocols depend on the physical location, eliminating the requirements on topology storage and reducing the associated costs, which is especially suitable for wireless ad hoc networks.
The routing performance in ad hoc networks is affected by the following factors.
First, due to the lack of centralization, the wireless nodes in ad hoc networks may not operate with the unified standards.
Furthermore, since ad hoc networks are usually dynamic and mobile (e.g. MANET), which are formed and changed depending on the particular goals, with nodes joining in and leaving the network from time to time, thus the topology of the ad hoc network is constantly changing \cite{cadger2012survey}.
All such features of ad hoc networks bring new challenges to design routing protocols.
The conventional ad hoc routing protocols can be classified into proactive and reactive protocols.
The main drawback is that both of them require the topology information for message routing, which entails high maintenance cost.
Compared to the conventional routing schemes, the georouting protocols based on the geographic location of the nodes can eliminate the need for topology information.
In the georouting scheme, when a node receives a packet, it will select the most appropriate node from its neighbours and deliver the packet in a hop-to-hop manner based on the location of the targeting nodes.
There is no specific route for a particular destination, since the selection of the nodes depends on the different network states.
One classical georouting schemes is \emph{greedy forwarding}, where the packets are forwarded to the neighboring node which is closer to the destination at each hop \cite{kuhn2008algorithmic}.
In addition to improving routing performance, other georouting protocols have been designed with particular goals.
For example, in \cite{shah2002predictive}, the authors proposed a georouting scheme which can feature quality of service (QoS) predictions based on the mobility of the UE.
In \cite{stojmenovic2004power}, an energy-efficient georouting protocol was proposed, which can guarantee delivery.
A more comprehensive overview on georouting is given in \cite{cadger2012survey}.

With the accurate location and tracking information of users, the location-aware adaptive content deliver schemes can be achieved, which include adaptive quality streaming, in-network caching, and content prefetching \cite{abou2014toward}.
Such applications require the location and trajectory information of the mobile users.
The well-known adaptive quality steaming scheme is the HTTP-based adaptive streaming (HAS) \cite{yao2011improving}, where the mobility trajectories of users were exploited to optimize HAS quality adaptation, thereby preventing data stream stalls.
In \cite{yao2011improving}, the authors developed an adaptive algorithm which can proactively switch the transmission rates based on the predicted user location and the stored REM.
The in-network caching techniques enable caching media content closer to the mobile user to reduce the delay and the prevent congestion in busy hours \cite{wang2014cache}, where the popular media content is stored at the edge of the network.
The location of users and their mobility patterns are used to predict the hot network regions where the media content will most likely be requested, so in such cases, the coarse localization method with tens or even hundreds of meter accuracy suffice.
On the other hand, the content prefetching refers to that with the help of the REMs and the predicted user's location, the network is able to deliver content proactively.
For accurate content delivery, it may require meter-level location service.
Specifically, when the networks are aware that the mobile UE will experience poor QoS based on the predicted UE's location, it can load a part or all of media content into the local storage of the UE in advance \cite{abou2014toward}.
Therefore, content prefetching has the capability to provide seamless streaming services to users.
In this case, the accurate location information of UE can improve the effectiveness
of prefetching \cite{gautam2013comparison}.

\subsection{Localization and Communication Co-design}\label{codesign}
Since radio signals can simultaneously carry data and location-related information of the transmitters, an unified study on ILAC tends to be a natural choice.
Currently, although localization systems can utilize the existing communication infrastructures like cellular networks cost-effectively, the localization and communication systems are mostly designed separately.
This is mainly due to the fact that the two lines of work in general have different design goals: one is to maximize the localization accuracy, while the other is to maximize the reliable data transmission rate through the wireless channels subject to fading and noise.
Nevertheless, there is a growing trend that the designs of the communication and localization systems can be aligned.
For instance, increasing the signal bandwidth or signal-to-noise ratio (SNR) can increase the localization accuracy as well as the channel capacity.
Hence, in this subsection, we attempt to discuss the study on ILAC for giving some insights on future networks design.

Although there are extensive work on localization in cellular networks as discussed above, the studies on localization and communication co-design are relatively rare.
In \cite{ghatak2018positioning,jeong2014beamforming,destino2017trade,destino2018impact,kumar2018trade}, some beamforming schemes for joint localization and data transmission in mmWave for 5G networks were studied, where the different trade-offs between the localization accuracy and the data rate
were derived.
In \cite{ghatak2018positioning}, the authors studied the trade-off between localization efficiency and downlink data rates by tuning the mmWave BS transmit power.
Specifically, the total transmit power of the BSs is fixed and divided into two parts: one associated with localization and the other dedicated for communication.
Hence, the optimal transmission scheme exists to select the proper power splitting factor for supporting different QoS requirements.
In \cite{jeong2014beamforming}, the authors focused on optimizing the beam vectors at the BSs to minimize the overall power consumption under certain data rate and localization accuracy requirements, where the data rate and the localization accuracy were measured as the functions of beam vectors, and the CRLB of localization accuracy for both TOA- and TDOA-based localization methods was derived.
In \cite{destino2017trade}, by assuming a finite coherence time, a trade-off between communication rate and localization accuracy for single-user LoS mmWave communication was studied.
Specifically, the total communication duration is fixed and partitioned into beam alignment and data transmission stages, and both of which are quantified as functions of the codebook size for beam alignment.
The following trade-off between localization accuracy and communication rate was revealed: more time spent for beam alignment leads to better SNR and improved localization accuracy, but it will result in less time for data transmission and hence lower data rate.
Moreover, in \cite{destino2018impact}, the impact of imperfect beam alignment on the rate-localization trade-off was considered.
In \cite{kumar2018trade}, the authors extended their prior work in \cite{destino2017trade} into a multi-user scenario, and the trade-off between the sum-rate and localization accuracy in uplink for multi-user mmWave communication system was researched.

\begin{figure} 
  \centering
\subfigure[Separated signals and receivers]{
\label{separated signal with separated receivers}
\begin{overpic}[width=0.35\textwidth]{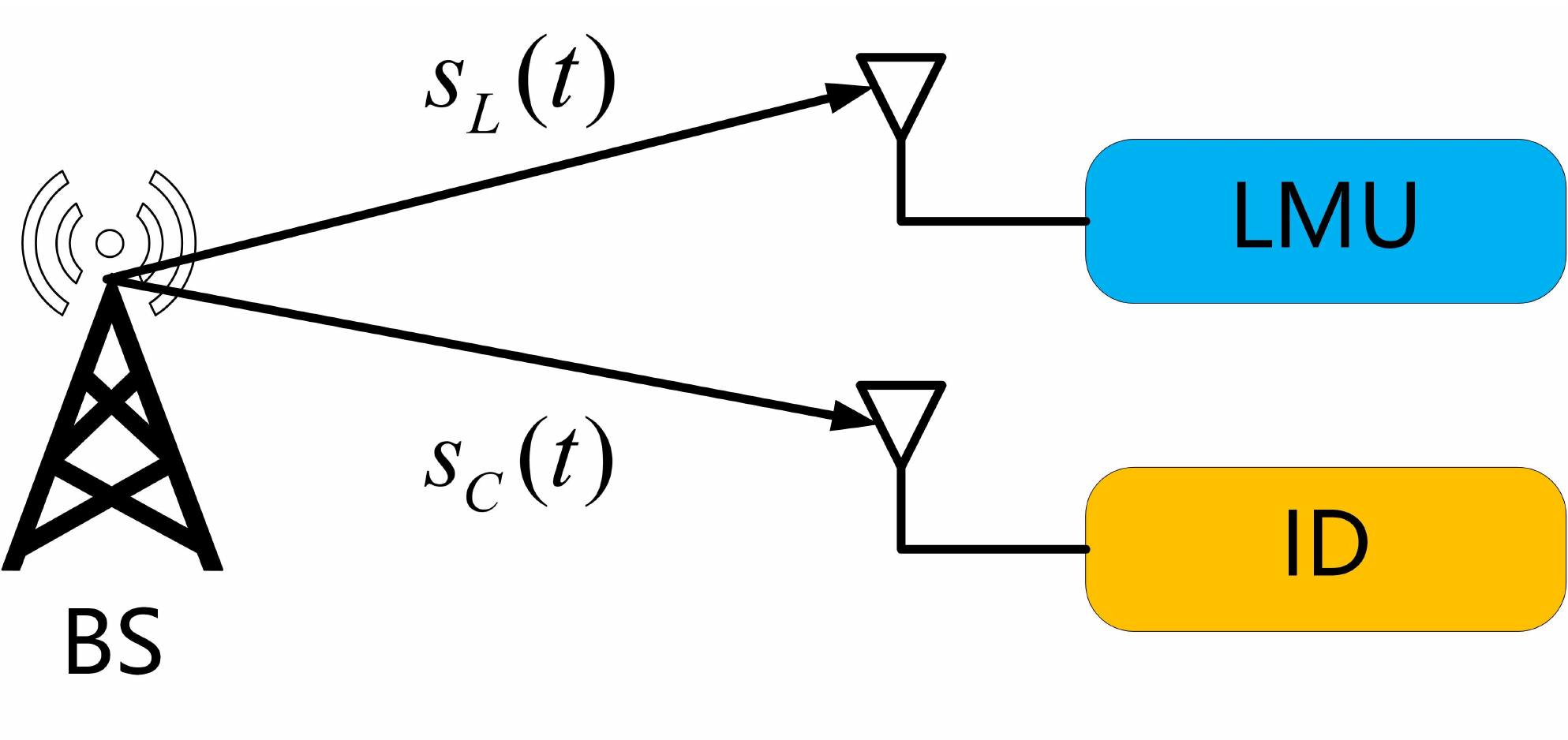}
\end{overpic}
}
\subfigure[Shared signals but separated receivers]{
\label{shared signal with separated receivers}
\begin{overpic}[width=0.35\textwidth]{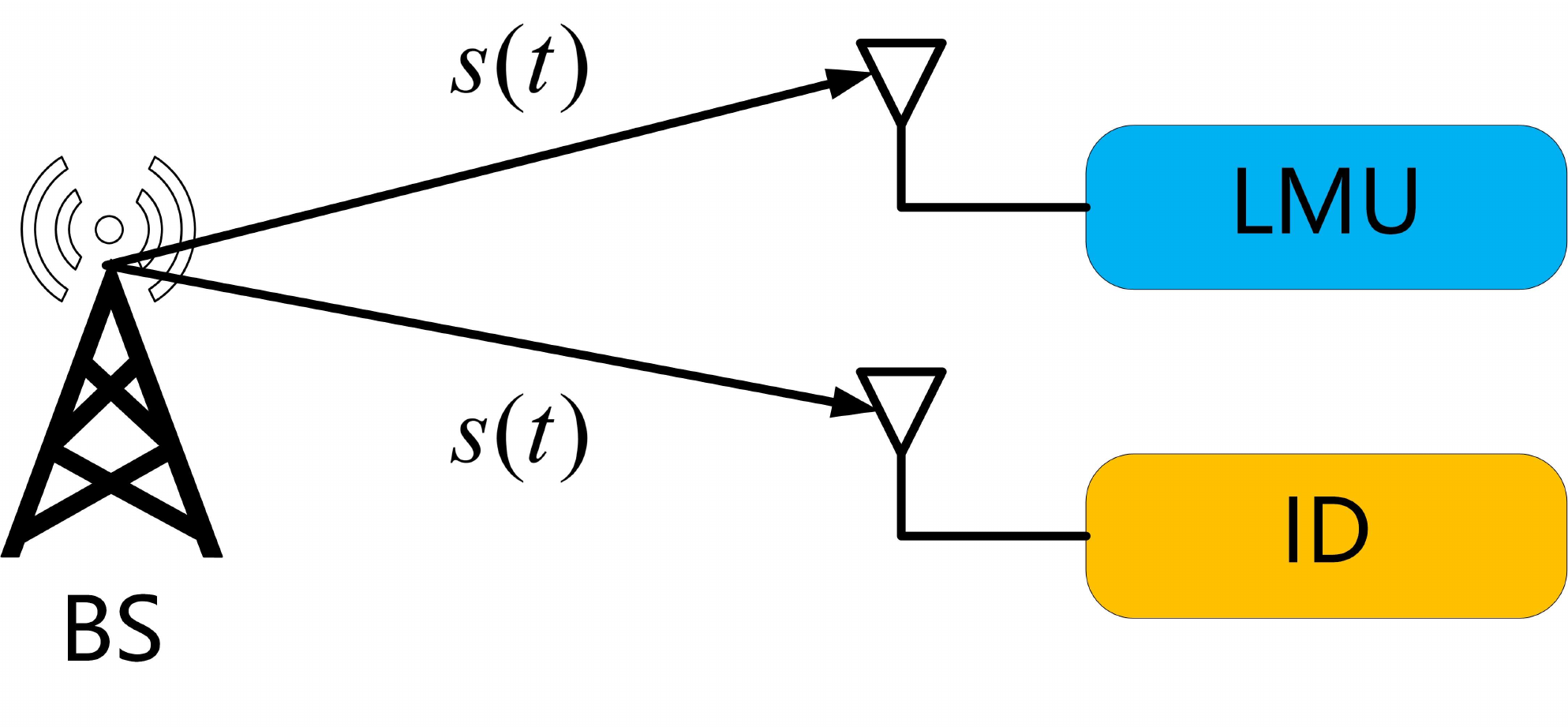}
\end{overpic}
}
\subfigure[Shared signal and receiver]{
\label{shared signal with integrated receivers}
\begin{overpic}[width=0.35\textwidth]{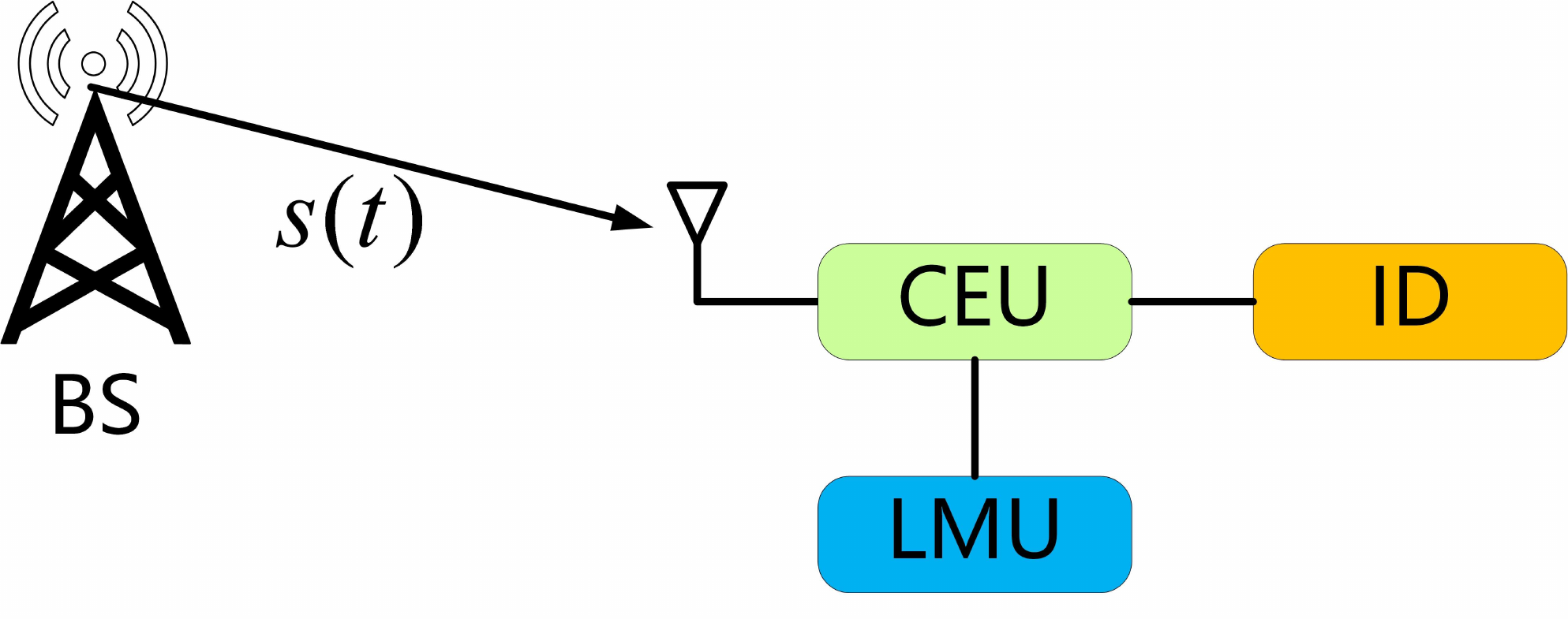}
\end{overpic}
}
\caption{An illustration of different architectures of ILAC systems.}\label{ILAC}\vspace{-10pt}
\end{figure}

In this subsection, we give the general architectures of the ILAC systems and consider their performance trade-offs.
For simplicity, we focus on the basic downlink case in the additive white Gaussian noise (AWGN) channel.
At the transmitter side, the signals for communication and localization can be designed and transmitted separately, or a common signal can be shared and reused for the both of purposes, which are referred to as \emph{separated signals} and \emph{shared signal}, respectively.
At the receiver side, the localization and communication systems can have their respective receivers that operate as an information decoder (ID) and the LMU, respectively, or share a common receiver, where the channel estimation unit (CEU) for communication can be reused for measuring location information from the common received signal, which are referred to as \emph{separated receivers} and \emph{shared receiver}, respectively.
Therefore, as illustrated in Fig. \ref{ILAC}, the architectures of the ILAC systems can be classified into three categories: \emph{separated signals and receivers}, \emph{shared signal but separated receivers}, as well as \emph{shared signal and receiver}.

\subsubsection{Separated Signals and Receivers}
As illustrated in Fig. \ref{separated signal with separated receivers}, in this case, we consider a scenario that the total bandwidth is fixed, denoted by $B_t$, which is divided into two orthogonal frequency bands for localization and communication respectively.
The signal for localization denoted by $s_L(t)$ has the bandwidth $B_L=kB_t$, where $0\le k\le1$.
Therefore, that for communication, denoted by $s_C(t)$, has the bandwidth $B_C=\left(1-k\right)B_t$, and there is no interference between them due to orthogonality in the frequency domain.
It is assumed that the two signals have the same transmission power, i.e.
$\mathbb{E}\left[\left|s(t)\right|^2\right]=\mathbb{E}\left[\left|s_L(t)\right|^2\right]=P$.
The signals received at ID and LMU are
\begin{subequations}\label{separated signal with separated receivers I}
\begin{equation}
y_C(t)=h_C(t)s_C(t)+n_C(t)
\end{equation}
\begin{equation}
y_L(t)=h_L(t)s_L(t)+n_L(t),
\end{equation}
\end{subequations}
where $h_C(t)$ and $h_L(t)$ are the communication and localization channel coefficients, respectively, where we assume $\mathbb{E}\left[\left|h_C(t)\right|^2\right]=\mathbb{E}\left[\left|h_L(t)\right|^2\right]=1$ without loss of generality; $n_C(t)$ and $n_L(t)$ denote the additive white Gaussian noise satisfied ${\cal{N}}\left(0,\sigma_C^{2}\right)$ and ${\cal{N}}\left(0,\sigma_L^{2}\right)$, respectively, where $\sigma_C^2=n_0B_C$ and $\sigma_L^2=n_0B_L$, $n_0$ is the noise power spectrum density (PSD).
According to Shannon's channel capacity theorem \cite{proakis2001digital}, the maximum data transmission rate is
\begin{equation}\label{shannon}
R=B_C\log_2\left(1+\frac{P}{n_0B_C}\right).
\end{equation}
On the other hand, considering the time-based ranging for localization, the CRLB is formulated as \cite{dardari2009ranging,gezici2005localization}
\begin{equation}\label{Time CRLB}
\mathbb{E}\left[(d-\hat{d})^2\right]\ge\text{CRLB} = \frac{c^2}{8\pi^2\beta^2\frac{P}{n_0B_L}},
\end{equation}
where
\begin{equation}\label{msb} \beta^2:=\frac{\int_{-B_L}^{B_L}f^2\left|S_L(f)\right|^2\mathrm{d}f}{\int_{-B_L}^{B_L}\left|S_L(f)\right|^2\mathrm{d}f} \end{equation}
denotes the \emph{mean square bandwidth} (MSB) of the signal $s_L(t)$, and $S_L(f)$ is the Fourier transform of $s_L(t)$, which will increase as the signal bandwidth increases.
Equations \eqref{shannon} and \eqref{Time CRLB} reveal that increasing the dedicated bandwidth can improve the data rate and decease the CRLB.
Therefore, as the total bandwidth is fixed, there is a trade-off to select the proper bandwidth splitting factor to support different QoS requirements for communication and localization, which is referred to as \emph{bandwidth splitting} scheme.
Alternatively, we can consider that the two types of signals use the same frequency band with bandwidth $B$, but with different transmission power, i.e. $\mathbb{E}\left[\left|s_C^2(t)\right|\right]=P_C$ and $\mathbb{E}\left[\left|s_L^2(t)\right|\right]=P_L$.
The total transmission power for a single-user is fixed, i.e. $P=P_C+P_L$.
For each receiver, the received signal is
\begin{equation}\label{separated signal with separated receivers I}
y(t) = h_C(t)s_C(t)+h_L(t)s_L(t)+n(t).
\end{equation}
Similarly, here we assume $\mathbb{E}\left[\left|h_C(t)\right|^2\right]=\mathbb{E}\left[\left|h_L(t)\right|^2\right]=1$ for notational simplicity.
In this case, the receiving performance is typically characterized by signal-to-interference and noise ratio (SINR), and the communication and localization performance can be obtained as
\begin{subequations}\label{SINR bound}
\begin{equation}
R = B\log_2\left(1+\frac{P_C}{n_0B+P_L}\right)
\end{equation}
\begin{equation}
\text{CRLB} = \frac{c^2}{8\pi^2\beta^2\frac{P_L}{n_0B+P_C}}.
\end{equation}
\end{subequations}
This reveals that signals with high power are beneficial for either data transmission or localization accuracy, so the
appropriately splitting the transmission power is necessary to balance the trade-off between communication and localization functionalities, which is referred to as the \emph{power splitting} scheme.

\subsubsection{Shared Signal but Separated Receivers}
As illustrated in Fig. \ref{shared signal with separated receivers}, in this case, a common signal is shared by localization and communication systems, so the study on joint waveform optimization for improving the performance of such a dual-purpose system is critical.
Although the role of localization was already highlighted in the 3G era, the dedicated reference signals for localization were not used untill the 4G-LTE, with the introduction of the PRS.
Therefore, compared with the extensive research on waveform design for communications, that for localization in cellular networks is relatively rare.
In \cite{dammann2016optimizing} and \cite{raulefs20165g}, the authors studied the impact of different signal PSD on time-based ranging accuracy.
Conventionally, the signal power is uniformly distributed over the available spectrum.
However, the authors revealed that uniform PSD is strictly sub-optimal for signal propagation delay measurement.
Actually, although the signal waveforms for localization and communications are usually designed separately, they have some similar key requirements, like low latency, high reliability, and low device complexity, rendering that their waveforms can be co-designed in the future networks.
For instance, considering the single-carrier transmission scheme, the signal that has more power concentrated at the edges of the spectrum is generally beneficial for time-based localization, which in the time domain can result in a impulse-like autocorrelation for time-based ranging.
Furthermore, in such cases the MSB of the signal can be greater according to \eqref{msb}, which leads the lower CRLB for better localization accuracy.
However, for communication, the optimal signal PSD scheme is to concentrate the signal power at the central of the mainlobes to reduce the inter-symbol-interference (ISI).
Therefore, there is a trade-off in terms of waveform design between localization and communication with different PSD requirements.
Moreover, it is important to integrate the geometrical information into waveform design for future networks, and the waveforms should be adaptive with reconfigurable features, which can flexibly configure its bandwidth, signal power, etc., depending on the real-time environment \cite{raulefs20165g}.

\subsubsection{Shared Signal and Receiver}
As illustrated in Fig. \ref{shared signal with integrated receivers}, localization and communication systems can also share a common receiver with the functionalities of LMU and ID, which can extract location-related information and decode data from a common signal.
In general, when channel parameters are unknown, the location-related information estimation in multi-path environments is closely related to channel estimation, so the CEU for communication can be also reused for localization.
For instance, the energy detector for non-coherent demodulation and the matched filter for coherent demodulation in communication systems can
be cost-effectively exploited for TOA estimation \cite{chong2007effect}.
The path amplitudes and TOAs can be jointly estimated using the ML approach, which can achieve the CRLB for large SNRs.
A primary barrier of ML estimators is the computational complexity, which results in time-consuming computation for accurate TOA estimation.
Therefore, a natural trade-off between localization and communication is that the more time spent on channel estimation can give the more accurate localization performance, but results in the higher communication latency.

\subsection{Localization and Communication in Aerial-Ground Integrated Networks}
Integrating terrestrial networks (especially the cellular networks) and aerial networks to achieve ubiquitous wireless connectivity in the 3D space is one of the visions for 6G \cite{latva2019key}.
In particular, with the high mobility and on-demand deployment capability, unmanned aerial vehicles (UAVs) have been regraded as a powerful tool to expand the wireless networks from the ground to the air space \cite{zeng2019accessing}.
In general, UAVs may be used as the low-altitude platforms (LAPs) to assist the terrestrial wireless communication from the sky, which are typically deployed at an altitude below several kilometers, while the high-altitude platforms (HAPs) consisting of floating BSs (e.g. balloons) are usually deployed in the stratosphere with tens of kilometers above the earth surface.
Compared with HAPs, UAV-based LAPs are easier and faster for deployment and more flexible for reconfiguration for critical missions, and it is able to establish strong LoS communication link with the ground UEs directly without relying on extra communication infrastructure like dish antennas \cite{zeng2016wireless}.
In general, UAV-aided communications have three main use cases, i.e. UAV-aided ubiquitous coverage, relaying, and data collection \cite{zeng2016wireless}.
On the other hand, UAVs with their own missions may also be integrated into cellular networks as new aerial users, leading to the other paradigm known as cellular-connected UAVs \cite{zeng2018cellular}.
The 3GPP started a study item to exploit the potential of LTE support for UAVs in March 2017 \cite{RP170779}, and the related technical reports TR 36.777 was released in December 2017\cite{TR36.777}, followed by a work item.
To achieve reliable data transmission for either UAV-aided communication or cellular-connected UAVs, the robust, low latency, high data rate wireless links between UAVs and terrestrial networks is necessary.
Due to the high mobility, the accurate real-time location of UAV is important for both safe operation and communication links maintenance.
In this subsection, we first focus on the localization problems in wireless networks involving UAVs, which can be classified into two main categories, i.e. \emph{localization for UAVs} and \emph{UAVs for localization}, where the UAVs play a role as agent nodes and aerial ANs, respectively.
Then, we elaborate the importance of real-time location information of UAVs for communication, which is referred to as \emph{location assisted UAV communication}.

\subsubsection{Localization for UAVs}
We first give an overview on the conventional localization and navigation methods for UAVs, and then discuss the potential localization approaches when cellular-connected UAV is enabled in future networks.
Currently, the commonly used UAV localization and navigation systems can be mainly divided into three categories, namely, GNSS, INS, and vision-based navigation.
The GNSS (e.g. GPS) can provide global coverage with meter-level localization accuracy for UAVs \cite{yoo2003low}, while it is vulnerable to disruption of satellite signals due to obstacles or signal blocking.
On the other hand, INS is a self-localization system by utilizing the motion information of the UAV for localization and navigation.
However, the INS of UAVs is expensive and unsuitable for small aircraft.
Moreover, it suffers from bias errors caused by the integral drift problem, which are continuously increasing with time, resulting in accuracy degradation \cite{nemra2010robust}.
To tackle this problem, the combination of GPS and INS was proposed, where the data fused from GPS and INS sensors through the navigation filter (e.g. EKF) can be used to improve localization performance.
For instance, in \cite{nemra2010robust}, the authors proposed a new INS/GPS sensor fusion approach based on State-Dependent Riccati Equation nonlinear filtering, which showed better UAV localization performance compared to the method based on EKF.
Nonetheless, many applications require the UAV to operate in GNSS-denied environments, like cluster urban and indoor scenarios.
To this end, the methods based on image recognition, referred to as \emph{vision-based} navigation, has emerged as a promising alternative to INS/GPS, which can be used both in outdoor and indoor environments.
The various visual sensors (e.g. cameras) are used to acquire information of surroundings, and the visual odometry and other similar methods were proposed to localize and navigate the UAV based on computer vision \cite{lu2018survey}.
The advantages of vision-based methods include that they do not rely on external signals, and the visual sensors are cheaper and easier to deploy compared to INS/GPS sensors.
However, the main limitation lies in that the UAV needs to process a large amount of sensing information in real time, especially for image processing, which greatly increase the computation complexity
Therefore, vision-based localization methods are difficult to be applied for UAVs with low power consumption and limited computing resources.
Moreover, since the vision-based methods rely on the visual information of the environment, the accuracy of these methods is usually poor in challenging environments with low-visibility conditions, like dusty or smoking environments \cite{queralta2020uwb}.
For indoor UAV localization, the radio signal based methods were also proposed,  like UWB-based \cite{tiemann2015design,queralta2020uwb,tiemann2017scalable,benini2013imu} and WiFi-based systems \cite{stojkoska2017indoor,tian2016hiquadloc}.
In \cite{tiemann2015design}, the authors proposed an UWB-based indoor localization system for UAV localization, which can achieve the RMSE under 10 cm in the horizontal plane and under 20 cm in 3D space for 95\% cases.
Furthermore, a method combined IMU, UWB, and vision-based schemes with EKF for UAV indoor localization was proposed in \cite{benini2013imu}, where the 10 cm accuracy can be achieved.
In \cite{stojkoska2017indoor}, the authors proposed an indoor UAV localization method based on WiFi RSSI, where the distances between the UAV and WiFi APs were measured to locate the UAV.
In \cite{tian2016hiquadloc}, the authors proposed a RSS fingerprint-based HiQuaLoc system for indoor UAV localization, where a RSS interpolation algorithm was proposed to reduce the overhead on training phase.

Compared with GNSS, vision-based, and short-range radio-based localization approaches, cellular-connected UAV can bring many new opportunities for UAV localization.
First, the ubiquitous accessibility of cellular networks may increase the localization coverage, which can cover outdoor and indoor environments.
Furthermore, cellular-connected UAVs can be a good complementary to improve the GNSS performance, via techniques like differential GPS (D-GPS), to achieve more robust UAV navigation.
Second, the wireless communication links between UAVs and cellular BSs, i.e. the control and non-payload communication (CNPC) link and payload data link, can be exploited and reused as the reference signals for UAV localization cost-effectively.
Third, the legacy wireless localization techniques in cellular networks from 2G to 4G, like OTDOA, UTDOA, E-CID, can be extended to the sky for UAV localization.
Finally, the new radio introduced by 5G, such as massive MIMO and mmWave communication, can be also exploited for UAV localization.
Actually, some related works like drone detection and tracking based on cellular networks are ongoing.
For instance, in \cite{solomitckii2018technologies}, the authors studied the amateur drone detection in 5G mmWave cellular networks, where the system design is outlined in terms of the density of BSs, their directional antennas, and the bandwidth, to detect the unlicensed small-sized drones.
On the other hand, massive MIMO techniques with highly directional radiation pattern have also be exploited for UAV detection and tracking in \cite{ezuma2019micro}.
In \cite{meer2020localization}, the authors studied the localizability of UAVs with cellular networks, and it was concluded that the localizability for UAVs is more favorable than that for ground users since the former has better localization performance due to the higher altitudes.

\begin{figure}
\centering
\begin{overpic}[width=0.45\textwidth]{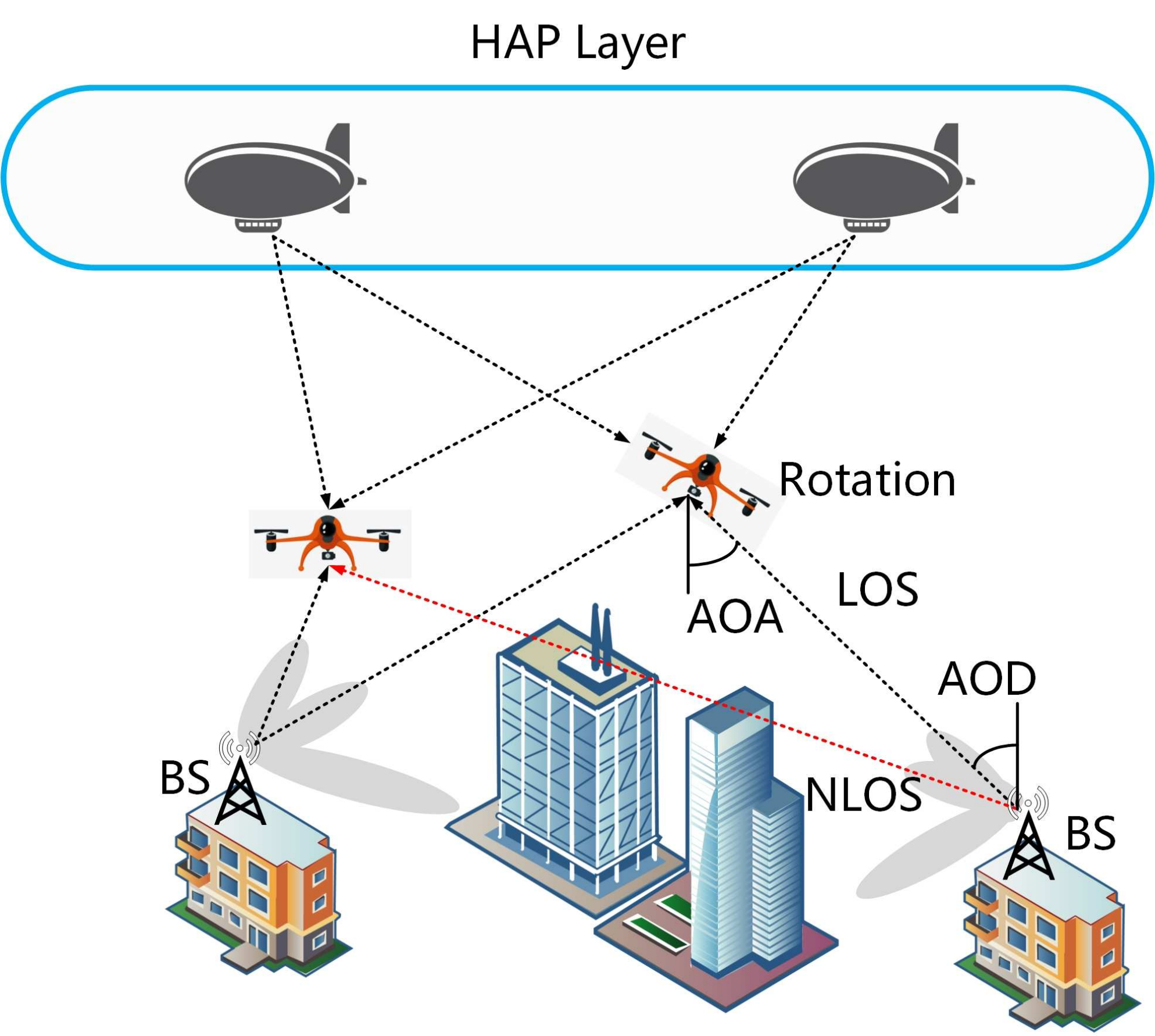}
\end{overpic}
  \caption{An illustration of 6D localization and HAP-assisted localization approaches for UAVs.}\label{6DLocalization}\vspace{-10pt}
\end{figure}

Here, we envision two promising localization techniques for UAVs in future networks, referred to as \emph{6D localization} and \emph{HAP-assisted localization}, as illustrated in Fig. \ref{6DLocalization}.
For 6D localization, the goal is to precisely estimate not only the 3D spatial location of the UAV, but also its 3D orientation in terms of roll, pitch, and roll \cite{bjornson2019massive}.
Different from the ground mobile devices for which the accurate real-time locations are typically sufficient, for the flying UAV, the accurate orientation estimation is also quite important, due to its high impact on flying gesture, power consumption, and hence the trajectory design.
This leads to the important problem of the 6D localization for UAVs.
The massive antenna arrays deployed on the BSs in 5G networks bring the opportunity to realize such a goal, by utilizing their high angular resolution to locate the cellular-connected UAV.
Furthermore, combining with mmWave technology with the high carrier frequency and wide signal bandwidth, the antennas can be squeezed into a compact form factor, which can be deployed on the UAV.
Therefore, both the AOA and AOD can be estimated in either uplink or downlink, as shown in Fig. \ref{6DLocalization}, and the location and the orientation of the UAV can be estimated simultaneously, leading to 6D localization.
However, since the BS antennas are typically downtitled for ground users, the radio coverage in the sky may not always be guaranteed, which may degrade the localization performance of UAVs.
To this end, the HAPs assisted localization approach provide a good complementary.
For HAP-assisted localization, as shown in Fig. \ref{6DLocalization}, the quasi-static floating aerial BSs can be treated as the additional ANs to assist the localization for UAVs.
Compared with terrestrial networks, HAPs can provide wide wireless coverage for very large geographic areas.
Furthermore, due to the high likelihood of LoS links in high-latitude space, the UAV localization accuracy can be improved.

\subsubsection{UAV for localization}
\begin{figure}
\centering
\begin{overpic}[width=0.5\textwidth]{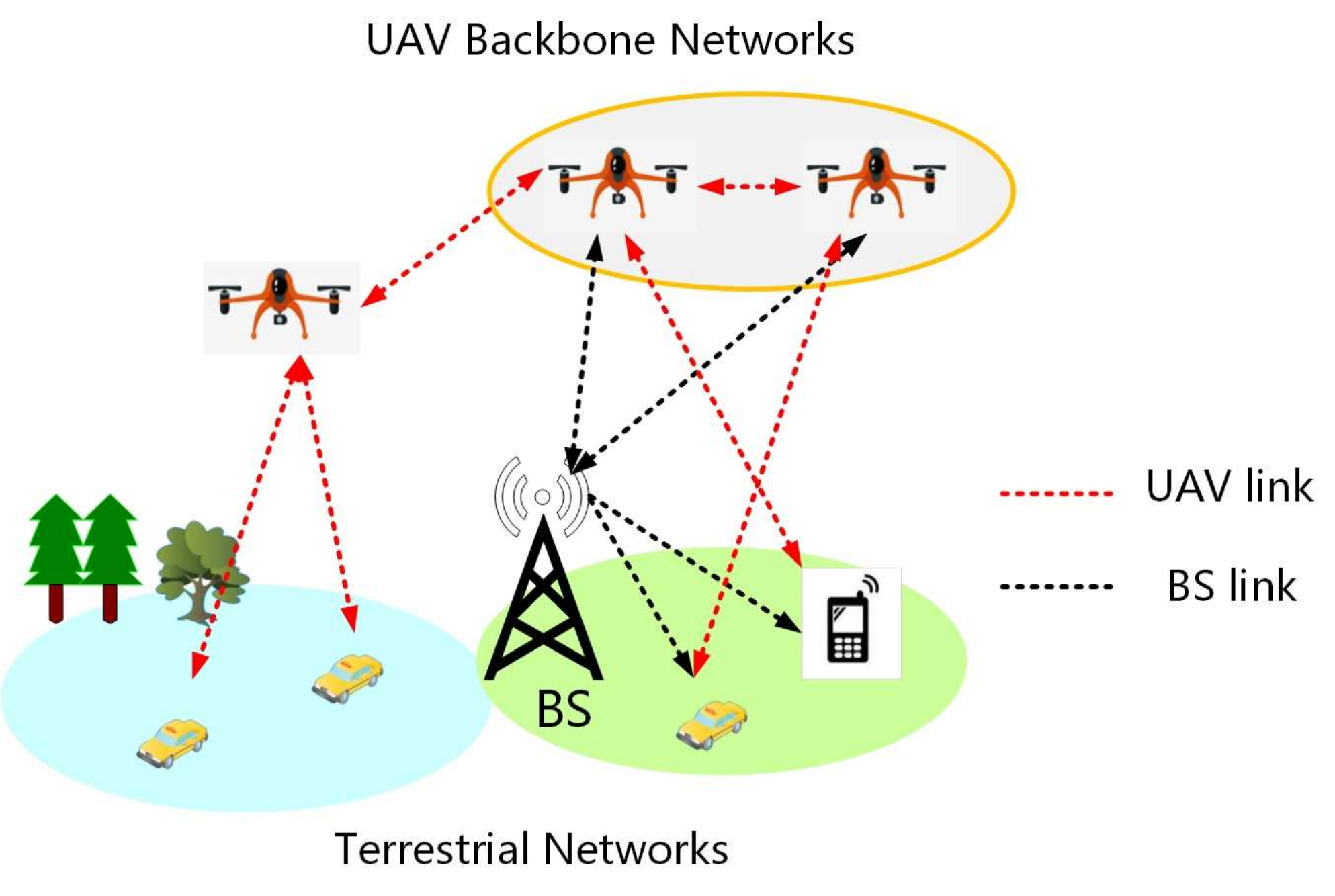}
\end{overpic}
  \caption{An illustration of FANETs and UAV-aided localization.}\label{air_ground_networks}\vspace{-10pt}
\end{figure}

Apart from estimating the locations for UAVs, the located UAVs can be also used for locating other devices in the networks, which is referred to as \emph{UAV-aided localization}.
UAV-aided localization can be mainly divided into two categories.
One is cooperative localization among UAVs, which is quite important for the flying ad hoc networks (FANETs).
The other is the UAV-assisted localization for terrestrial UEs, where the UAVs with known locations are treated as the additional ANs for providing more reference information to locate the agent node on the ground.
\paragraph{Cooperative Localization Among UAVs}
Instead of developing single-UAV systems, a group of small UAVs can form FANETs in the aerial space.
In such cases, as shown in Fig. \ref{air_ground_networks}, only a subset of UAVs can communicate with the ground BSs, which is referred to the UAV backbone networks, while other UAVs outside the coverage of the ground BSs may establish connection with the ground through the UAV backbone networks \cite{bekmezci2013flying}.
Compared to the wireless nodes in vehicular ad hoc networks (VANETs) or MANETs on the ground, FANET nodes have much higher mobility with more dynamic network topology.
Due to the performance degradation of GPS in urban and indoor scenarios and the high mobility of UAVs, multi-UAV systems require the accurate real-time location information of each UAV in FANETs, which renders the cooperative localization among UAVs important.
Intuitively, the cooperative localization approaches utilized in MANETs, VANETs, or other ad hoc networks on the ground can be similarly exploited and applied in FANETs by extending the localization schemes from 2D to 3D.
For instance, the belief propagation technique which is usually used for cooperative localization in WSNs can be utilized for multi-UAVs cooperative localization.
In \cite{wan2014cooperative}, a dynamic nonparametric belief propagation method was proposed for UAVs cooperative localization, which can locate UAVs with fault GPS successfully.
In \cite{lee2013cooperative}, the authors studied the cooperative localization between two UAVs, which were equipped with heterogeneous sensors to gather more information in a limited time.
In \cite{qu2011cooperative}, the authors proposed a cooperative localization method based on the inter-UAV relative range measurements, which can locate the UAV when the GPS is unavailable.
By assuming that all UAVs construct a ring communication topological structure, a cooperative localization method based on information synchronization was proposed in \cite{qu2010cooperative}.
Typically, since there are many LoS links among UAVs, the cooperative localization performance of UAVs should be better than that of terrestrial wireless nodes, though the practical implementation is still challenging due to the high UAV mobility.

\paragraph{UAV-Assisted Localization for Terrestrial UEs}
UAVs have higher altitude than ground BSs or UEs, which renders them easier to achieve wider coverage area on the ground with high probability of LoS links.
Therefore, when some ANs are not available to the agent node, the flying UAV with known location can be used as an additional AN to assist the terrestrial localization.
For instance, a UAV-aided localization method for ground vehicles was proposed in \cite{liu2018uav}, where each UAV first measures the RTOF of signals between the ground BSs and the UAV, and then broadcasts the measurements to the ground vehicles for localization, which can achieve decimeter-level relative position error between vehicles and meter-level absolute position accuracy.
In \cite{sorbelli2018range}, a range-based drone-aided localization method for terrestrial objects was proposed, where a flying drone was treated as a mobile anchor equipped with a GPS, and the distances between the drone and the ground objects were measured by UWB signals, then the ground objects can estimate their own position through trilateration.

\subsubsection{Location-Assisted UAV Communication}
Apart from navigation and tracking, the accurate real-time location information of a flying UAV is also beneficial for communication in the following aspects:
\begin{itemize}
\item{\textbf{3D REMs Modelling}:}
Currently, cellular networks are designed to cater for the terrestrial broadband communication, thus BS antennas are typically downtilted to reduce the inter-cell interference \cite{lin2018sky}.
With the downtilted BS antennas, the UAVs may only be served by the sidelobes, and thus the cellular coverage in the sky cannot always be guaranteed \cite{Qualcomm}.
Therefore, the coverage holes exist in the sky, and when a UAV flies over these areas, it may lose wireless connection to the cellular networks.
To overcome this problem, one useful approach is to construct 3D REMs about the area of interest to guide the design of UAVs trajectory to avoid these coverage holes.
Similar to the 2D REM as we discussed in Section \ref{Location-aware communication}, a 3D REM modelling also evolves a training process, where a UAV flies over the area of interest and collect radio measurements from the sampling points of the area to construct the 3D REM.
Therefore, the localization accuracy of the flying UAV will affect the accuracy of the REM.
For instance, in \cite{zeng2020simultaneous}, the authors proposed a method named \emph{simultaneous navigation and radio mapping} (SNARM), where the signal measurement of the UAV is used not only for optimizing the UAV trajectory, but also for creating a radio map which can predict the outage probabilities of communication at all locations in the area of interest.
\item{\textbf{Proactive Management of CNPC Links}:}
The wireless communication links of UAVs have two main categories, namely, the CNPC link and payload data link.
The CNPC link is a two-way communication link between a UAV and the ground control station (GCS), or other UAVs, which is responsible for supporting the safe control from GCS to the UAV, sending reports from the UAV to GCS, and transmitting collision alert between UAVs \cite{zeng2016wireless}.
On the other hand, the payload data link is mainly used to transmit mission-related data between the UAV and other entities, like the ground BS, mobile terminals, and other UAVs.
Different from the data link which requires higher data rate but tolerates on latency and reliability of the link, the CNPC link requires ultra-reliability, low latency, and high security to ensure safe control to the flying UAV \cite{zeng2018cellular}.
However, due to the building blockage and shadowing effects, the CNPC links may suffer from delay and attenuation, and the interference from other UAVs and ground BSs can also degrade the CNPC performance.
To enhance the reliability and robustness of the CNPC links, the combination of the 3D REMs and the predicted UAV location can provide opportunities for proactive CNPC link management.
With the predicted locations of all UAVs in the area of interest, the GCS can proactively allocate the appropriate spectrum to provide CNPC link for the UAV according to the pre-built 3D REMs, enhancing the reliability and reducing the interference.
\item{\textbf{3D Beamforming and Sub-sector Partition}:}
Another way to achieve the goal of ubiquitous wireless connectivity in 3D space is the 3D beamforming, which also relies on the accurate UAV location.
Specifically, the accurate UAV location information can be used to obtain the azimuth and elevation angles of the aerial-ground links, and further help for designing the beamforming weight vectors.
Furthermore, with the accurate 3D beamforming, the sectorization technique can be also extended to the 3D space, where the elevation angles is used to further partition the horizontal sector in current cellular systems to construct the sub-sectors for the aerial users \cite{zeng2018cellular}.
With the help of these 3D network architectures, the interference in UAV-aided communication can be significantly reduced.
Moreover, the accurate UAV location is also beneficial for RRM in these sub-sectors, like location-aided handover, multcasting, and so on, as we discussed in Section \ref{Location-aware communication}.
\end{itemize}

\section{Conclusions and Future Working Directions}
\begin{figure}
\centering
\begin{overpic}[width=0.48\textwidth]{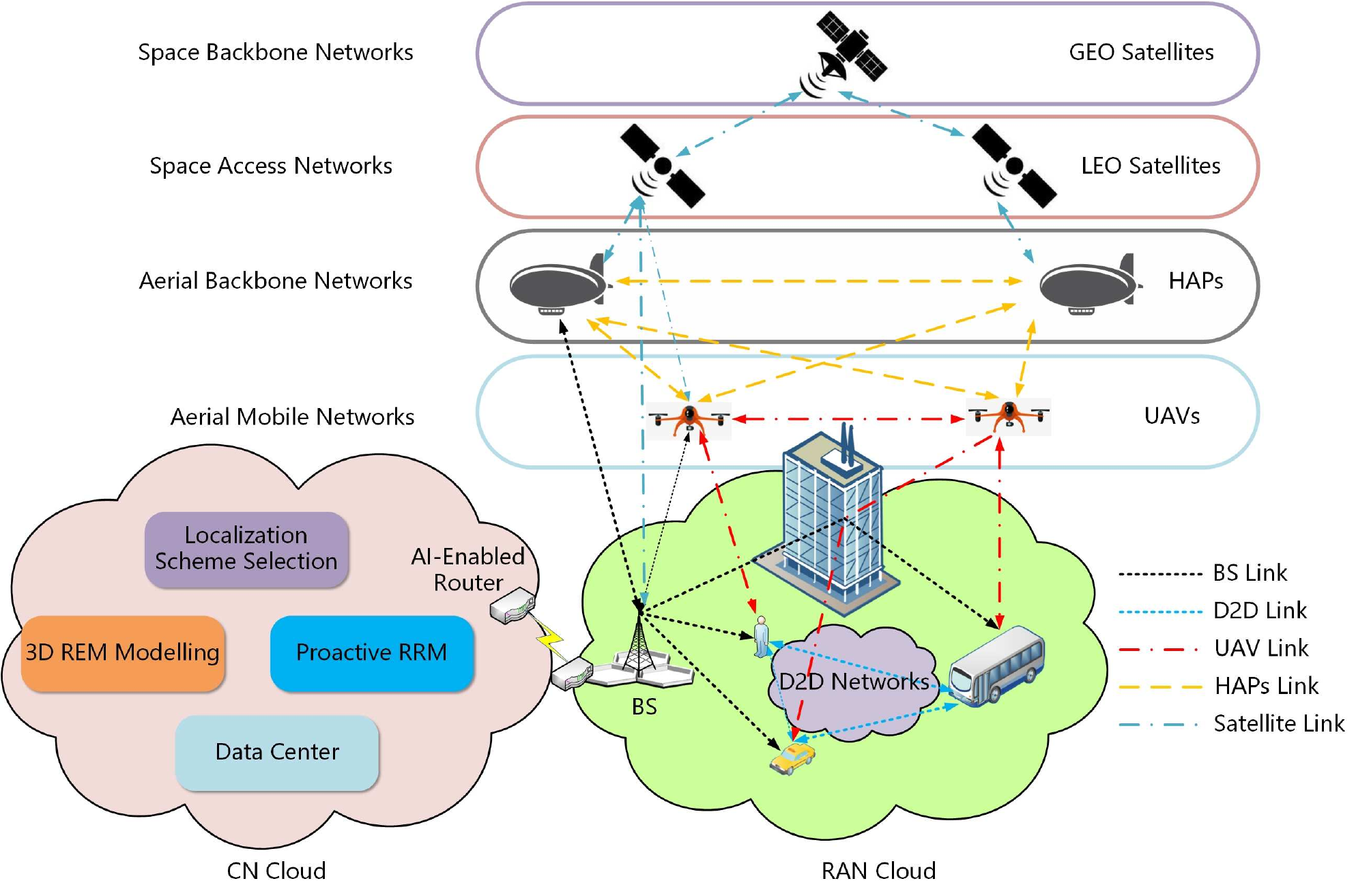}
\end{overpic}
  \caption{An illustration of future 3D wireless networks.}\label{summary}\vspace{-12pt}
\end{figure}

In this article, we have first provided an overview on the basics of wireless localization, and then discussed the vision of future network design with ILAC for 6G networks.
In summary, an envisioned architecture of future 3D wireless networks is illustrated in Fig. \ref{summary}, and some related enabling technologies and promising directions of future work are discussed as follows.

As shown in Fig. \ref{summary}, it is envisioned that the future 6G wireless networks will be artificial intelligence (AI) enabled, heterogenous, and multi-tier networks consisting of space backbone networks (SBNs), space access networks (SANs), aerial backbone networks (ABNs), aerial mobile networks (AMNs), and terrestrial networks (TNs) \cite{zhang20196G,Yao2018The,You2020The}.

The satellite networks are expected to play an important role in 6G networks due to its wide coverage, especially in remote and ocean areas.
Since various satellite systems on different earth obits are isolated and formed different autonomous systems (ASs) to provide specific services, like communications, navigation, remote sensing and so on, the real-time information sharing across different ASs is difficult \cite{Yao2018The}.
Therefore, in the future networks, the satellite networks can be divided into SBNs and SANs.
The SBNs consisting of several GEO satellites have the capabilities of data storage, routing and resource management, which can function as the control and data center of the SANs consisting of LEO satellites \cite{Yao2018The}.
LEO satellites may be equipped with high-gain large antenna arrays to provide ubiquitous coverage to enable UE direct communication \cite{You2020The}.
For localization, the GNSS provided by the geostationary satellites will be still important for mobile device localization and network synchronization due to its wide coverage.

For aerial networks, on-demand UAVs with high mobility can be deployed easily and quickly, which can form the dynamic adjustable AMNs to provide temporary communication services for special missions, like emergency communication services, UAV-aided relaying, and data collection \cite{zeng2016wireless}.
However, due to the limitations of UAVs on data storage and endurance, the ABNs consisting of floating HAPs like airships are necessary for providing reliable wireless coverage for large areas, where the airships with the computing and caching capabilities can be treated as the data center and gateways routing information to SANs and TNs.
For localization, since the locations of HAPs can be determined by the GNSS or the terrestrial BSs with dedicated antennas to transmit reference signals, they can act as the aerial ANs to locate the flying UAVs.
Furthermore, since UAVs with high mobility may provide short-distance LoS links to the ground devices, they can be used as the additional temporary ANs to improve the localization performance of ground devices.

In the TNs, mmWave, massive MIMO, and UDNs technologies can be combined to improve the communication and localization performance.
Furthermore, as a promising trend, device-centric networks will flourish the cell-free mobile communications, which will require high accurate cooperative localization among devices for resource allocation.
On the other hand, the future networks will become more heterogenous, combining with various radio access standards, like 2G/3G/LTE/5G, WiFi, as well as the Terahertz (THz) and visible light frequency bands.
Therefore, the integrated resource allocation schemes and protocols are required to leverage multiple frequencies to provide seamless wireless connectivity for mobile devices.
For localization, the system needs to select the appropriate localization approaches and reference signals depending on the surrounding radio environments of the mobile devices for better performance.

Finally, AI-enabled cloudization is a promising trend of future wireless networks, as illustrated in Fig. \ref{summary}, which can be mainly divided into RAN cloud and CN cloud.
The distributed AI units deployed at the edges of the RAN assist signal reconstruction, which is beneficial for data decoding and signal measurements to improve communication and localization performance, respectively.
In the CN cloud, the centralized AI units can be used to construct the 3D REMs, which can be used to assist the localization scheme selection and proactive RRM for communications.
As a vision of 6G, the AI-enabled networks are expected to autonomously optimize and manage the resource to dynamically maintain communication performance of the UE according to its accurate real-time location.

However, to achieve the above visions, there are still many challenges to be addressed.
Note that while some challenges have already been discussed in the previous sections, in the following, we outline some important directions for future work and highlight the promise of ILAC.

\subsection{Fundamental Performance Analysis and Design for ILAC}
Achieving ultra-high spectral efficiency is critical for future networks, especially for the IIoT application scenarios, which require massive wireless connectivity to support reliable communication and accurate localization services for massive IoT devices.
To achieve ultra-high spectral efficiency for ILAC, a common spectrum can be shared by localization and communication, or be divided into two orthogonal parts for them.
Furthermore, for shared spectrum, the signal waveform can be jointly designed for localization and communication simultaneously, or two specific waveforms can be separately designed.
Therefore, the theoretical analysis of the fundamental performance of ILAC and how to design the waveforms for ultra-high spectral efficiency is an important problem.
On the other hand, for orthogonal spectrum usage, since localization and communication both benefit from wide signal bandwidth, a critical issue is that how to split the spectrum effectively to optimize the trade-off between localization and communication according to different QoS requirements.

\subsection{Advanced Signal Processing for Multi-Tier ILAC Networks}
In multi-tier networks, the mobile terminals in different layers of networks can interact in different frequency bands with different links.
In this case, how to maintain the wireless links dynamically and effectively is a critical issue for localization and communication.
The integration of different frequency bands and dynamic resource management is a potential option to reconfigure and maintain the wireless connections, and the effective signal processing is an important factor for cross-layer information sharing.
For ILAC, since the channel estimation units can be reused for location information extraction, the localization and communication systems can partly share some hardware, and the signal processing technologies for communication can be also exploited for location-related information measurements.
Therefore, how to efficiently co-design the hardware architecture and signal processing technologies are two key questions.

\subsection{Heterogenous Networking}
The future networks are more heterogenous than before, which operate on different frequency bands with different standards.
A fundamental question is that how to switch the protocols rapidly from one to anther, while still ensuring the localization and communication performance, when the mobile terminals move quickly suffering from different radio environments.
For the distinct network architectures, each network layer employs different protocols, so a natural solution is to translate the protocols at gateways to interconnect different networks.
However, such a sequence of protocol translations is inefficient, which renders the integrated protocol design that enables cross-layer, cross-module, and cross-node data transmission critical.
Moreover, the network protocol design, especially in the physical and MAC layers, needs not only to consider the communication metrics, but also to be re-assessed from the localization perspective.

\subsection{3D REMs and Proactive RRM in Complex Environments}
The proactive RRM is beneficial for communication in terms of cell selection, channel predication, beam alignment and so on, but it will require high-accurate REMs.
Currently, the studies on REMs modelling focus on the 2D scenarios targeting to the TNs in outdoor environments.
As the networks extend to 3D space, 3D REMs modelling in clutter environment like indoor and urban city is critical and challenging, which requires more accurate 3D localization in multi-path and NLoS environments.
Although some effective multi-path and NLoS mitigation algorithms have been proposed in the literature, they are usually quite complex and only feasible for remote localization systems rather than the self-localization systems where the computation and estimation need to be done by UEs.
On the other hand, although some studies mainly focus on the localization in NLoS scenarios, by treating the signal reflectors as the virtual ANs, they are limited on the first-order reflection for analysis simplicity, whereas the higher-order reflections occur in dense multi-path environments.
Therefore, in-depth investigations on localization algorithms for multi-path environments are still needed for more accurate and low cost localization.
Furthermore, how to manage the radio resource according to the real-time location of the UE to dynamically maintain its communication performance deserves further investigations.
On the other hand, although it is well-known that the accurate location information is beneficial for communications, the related fundamental metrics are needed to reveal the relation between different location accuracy and communication performance.

\subsection{Machine Learning for ILAC}
AI, or more specifically, machine learning (ML), is one of the most promising technologies, which can bring intelligence to wireless networks with complext radio conditions.
Due to its capability to handle the accurate pattern recognition from complex raw data, it can be used to find the network dynamics and construct a user-centric intelligent networks, which can autonomously manage resources, functions, and network control to sustain the high performance according to the real-time location of the mobile users.
For ILAC, ML can be used in various perspectives, like waveform design, signal modulation/coding, resource allocation, to balance the performance trade-off between localization and communication, and create the 3D REMs to enhance the ILAC performance.

\bibliographystyle{IEEEtran}
\bibliography{RefLocalization}

\end{document}